\documentclass[a4paper,11pt]{article}
\pdfoutput=1

\usepackage[dvips]{graphicx}
\usepackage{epsfig}
\usepackage{caption,subcaption}
\usepackage{color}
\usepackage{amsmath,amssymb,amsthm,amsfonts}
\usepackage{bm,mathtools,mathrsfs}
\usepackage[noadjust]{cite}
\usepackage{url}
\usepackage{fancyhdr}
\usepackage{hyperref}
\usepackage{cases}
\usepackage{appendix}

\newcommand{\beq}{\begin{equation}}
\newcommand{\eeq}{\end{equation}}
\newcommand{\beqn}{\begin{equation*}}
\newcommand{\eeqn}{\end{equation*}}
\newcommand{\beqa}{\begin{eqnarray}}
\newcommand{\bdm}{\begin{displaymath}}
\newcommand{\edm}{\end{displaymath}}
\newcommand{\eeqa}{\end{eqnarray}}
\newcommand{\beqan}{\begin{eqnarray*}}
\newcommand{\eeqan}{\end{eqnarray*}}

\newcommand{\Real}{\mathcal{R}}

\newcommand{\hilbert}{\mathcal{H}}

\newcommand{\innprod}[2]{\left\langle{#1},{#2}\right\rangle}

\newcommand{\norm}[1]{\left\|{#1}\right\|}

\newcommand{\lev}[1]{\ensuremath{\operatorname{lev_{\leq #1}}}}

\DeclareMathOperator{\argmin}{arg\,min}
\DeclareMathOperator{\argmax}{arg\,max}
\DeclareMathOperator{\card}{card}

\DeclareMathOperator{\linspan}{span}

\DeclareMathOperator{\sign}{sgn}

\DeclareMathOperator{\sinc}{sinc}

\DeclareMathOperator{\supp}{supp}

\DeclareMathOperator{\expect}{\mathsf{E}}
\DeclareMathOperator{\prob}{Prob}

\DeclareMathOperator{\rank}{rank}
\DeclareMathOperator{\nullsp}{null}
\DeclareMathOperator{\spark}{spark}
\DeclareMathOperator{\epi}{epi}
\DeclareMathOperator{\prox}{Prox}

\theoremstyle{definition}
\newtheorem{theorem}{Theorem}
\newtheorem{algo}{Algorithm}

\newtheorem{prop}{Proposition}
\newtheorem{lemma}{Lemma}

\newtheorem{remarks}{Remarks}
\newtheorem{definition}{Definition}
\newtheorem{example}{Example}

\renewcommand{\sectionmark}[1]%
{\markright{\MakeUppercase{\thesection.\ #1}}}

\newcommand{\myfonts}{%
\fontfamily{ptm}\fontseries{b}\fontsize{9}{11}\selectfont}
\title{Sparsity-Aware Learning and Compressed
  Sensing: An Overview\footnote{This paper is based on a chapter of a new book on Machine Learning, by the first and third author, which is currently under preparation.}}
  \author{Sergios Theodoridis\\
  Dept. of Informatics {\&} Telecommunication\\
  University of Athens, Athens, Greece\\
  stheodor@di.uoa.gr\\
  \and
  Yannis Kopsinis\\
  Dept. of Informatics {\&} Telecommunication\\
  University of Athens, Athens, Greece\\
  kopsinis@ieee.org\\
  \and
  Konstantinos Slavakis\\
  Digital Technology Center\\
  University of Minnesota, Minneapolis, USA\\
  slavakis@dtc.umn.edu\\
  }

\begin{document}
\maketitle

\section{Introduction}

The notion of regularization has been widely used
as a tool to address a number of problems that are usually encountered
in Machine Learning. Improving the performance of an estimator by
shrinking the norm of the MVU estimator, guarding against overfitting,
coping with ill-conditioning, providing a solution to an underdetermined
set of equations, are some notable examples where regularization has
provided successful answers. A notable example is the ridge regression concept, where the LS loss function is
combined, in a tradeoff rationale, with the Euclidean norm of the
desired solution.

In this paper, our interest will be on alternatives to the Euclidean
norms and in particular the focus will revolve around the $\ell_1$ norm;
this is the sum of the absolute values of the components comprising a
vector. Although seeking a solution to a problem via the $\ell_1$ norm
regularization of a loss function has been known and used since the
1970s, it is only recently that has become the focus of attention of a
massive volume of research in the context of compressed sensing. At the
heart of this problem lies an underdetermined set of linear equations,
which, in general, accepts an infinite number of solutions. However, in
a number of cases, an extra piece of information is available: the true
model, whose estimate we want to obtain, is sparse; that is, only a few
of its coordinates are nonzero. It turns out that a large number of
commonly used applications can be cast under such a scenario and can be
benefited by a so-called sparse modeling.

Besides its practical significance, sparsity-aware processing has
offered to the scientific community novel theoretical tools and
solutions to problems that only a few years ago seemed to be
intractable. This is also a reason that this is an interdisciplinary field
of research encompassing scientists from, e.g., mathematics, statistics,
machine learning, signal processing. Moreover, it has already been
applied in many areas ranging from biomedicine, to communications and
astronomy.  At the time this paper is compiled, there is a ``research
happening'' in this field, which poses some difficulties in assembling
related material together. We have made an effort to put together, in a
unifying way, the basic notions and ideas that run across this new
field. Our goal is to provide the reader with an overview of the major
contributions which took place in the theoretical and algorithmic fronts
and have been consolidated over the last decade or so. Besides the methods and algorithms which are reviewed in this article, there is another path of methods based on the Bayesian learning rationale. Such techniques will be reviewed elsewhere.

\section{Parameter Estimation}
Parameter estimation is at the heart of what is known as {\it Machine Learning}; a term that is used more and more as an umbrella for a number of scientific topics that have evolved over the years within different communities, such as Signal Processing, Statistical Learning, Estimation/Detection, Control, Neurosciences, Statistical Physics, to name but a few.

In its more general and formal setting, the parameter estimation task is defined as follows. Given a set of data points $(y_n,\bm{x}_n)$, $y_n\in {\mathcal R},~\bm{x}_n\in {\mathcal R}^l,~,~n=1,2,\ldots,N$, known as the {\it training data}, and a parametric set of functions
\[
{\mathcal F}:=\{f_{\bm{\theta}},~\bm{\theta}\in {\mathcal A}\subseteq {\mathcal R}^k\},
\]
find a function in ${\mathcal F}$, which will be denoted as $f(\cdot):=f_{\bm{\theta}_*}(\cdot)$, such that given the value of $\bm{x}\in {\mathcal R}^l$, $f(\bm{x})$ best approximates the corresponding value of $y\in {\mathcal R}$. After all, the main goal of Machine Learning is {\it prediction}. In a more general setting, $y$ can also be a vector $\bm{y}\in {\mathcal R}^m$. Most of our discussion here will be limited to real valued variables. Obviously, extensions to complex valued data are readily available.

Having adopted the parametric set of functions and given the the training data set, the goal becomes that of {\it estimating} the values of the parameters $\bm{\theta}$ so that to ``fit'' the data in some (optimal) way. There are  various paths  to achieve this goal. In this paper, our approach comprises the adoption  of a {\it loss} function
\[
{\mathcal L}(\cdot,\cdot):~{\mathcal R}\times {\mathcal R}\longmapsto [0,\infty),
\]
and obtain $\bm{\theta}_*$ such that
\[
\bm{\theta}_*:=\argmin_{\bm{\theta}} J(\bm{\theta}),
\]
where
\beq
J(\bm{\theta}):=\sum_{n=1}^N{\mathcal L}(y_n,f_{\bm{\theta}}(\bm{x})).\label{papereq 1}
\eeq
In this review article, the focus will be on the Least Squares loss function, i.e.,
\[
{\mathcal L}(y,f_{\bm{\theta}}(\bm{x})):=(y-f_{\bm{\theta}}(\bm{x}))^2.
\]

\begin{figure}[!tbp]
\centering
\includegraphics[scale=1]{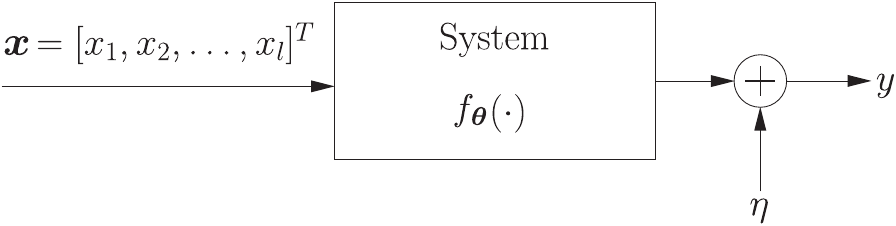}
\caption{Block diagram showing the input-output relation in a regression model.}\label{fig:system.model}
\end{figure}

Among the many parametric models, {\it regression} covers a large class of Machine Learning tasks. In linear regression, one models the relationship of a {\it dependent} variable $y$, which is considered as the output of a system, with a set {\it independent} variables, $x_1, x_2,\ldots,x_l$, which are thought as the respective inputs that activate the system in the presence of a noise (unobserved) disturbance,   $\eta$, i.e.,
\[
y=\theta_1x_1+\ldots+\theta_lx_l+\theta_0+\eta,
\]
where $\theta_0$ is known as the {\it bias} or {\it intercept}, see Figure \ref{fig:system.model}. Very often the previous input-output relationship is written as
\begin{equation}
y= \bm{x}^T \bm{\theta} +\eta \label{papereq2}
\end{equation}
where
\beq
\bm{\theta}:=[\theta_1,\ldots,\theta_0]^T,~~\mbox{and}~~\bm{x}:=[x_1,\ldots,x_l,1]^T.
\eeq
Often, $\bm{x}$ is called the {\it regressor}. Given the set of training data points, $(y_n,\bm{x}_n),~n=1,2,\ldots,N$, (\ref{papereq2}) can compactly written as
\beq
\bm{y}=X\bm{\theta}+\bm{\eta}
\eeq
where
\beq
X:=\left [\begin{array}{c} \bm{x}_1^T \\ \vdots \\ \bm{x}_N^T \end{array}\right ],~~\bm{y}=\left [\begin{array}{c} y_1\\ \vdots \\ y_N\end{array}\right],~~ \bm{\eta}=\left [\begin{array}{c} \eta_1 \\ \vdots \\ \eta_N\end{array}\right].
\eeq

For such a model, the Least Squares cost function becomes
\beq
J(\bm{\theta})=\sum_{n=1}^N(y_n-\bm{\theta}^T\bm{x}_n)^2=||\bm{y}-X\bm{\theta}||^2, \label{papereq3}
\eeq
where $||\cdot||$ denotes the Euclidean norm. Minimizing (\ref{papereq3}) with respect to $\bm{\theta}$ results to the celebrated LS estimate
\beq
\hat{\bm{\theta}}_{LS}=(X^TX)^{-1} X^T \bm{y},
\eeq
assuming the the matrix inversion is possible. However, for many practical cases, the cost function in (\ref{papereq3}) is augmented with a so called  {\it regularization} term. There are a number of reasons that justify the use of a regularization term. Guarding against overfitting, purposely introducing bias in the estimator in order to improve the overall performance, dealing with the ill conditioning of the task are examples in which the use of regularization addresses successfully. {\it Ridge regression} is a celebrated example, where the cost function is augmented as
\[
J(\bm{\theta})=||\bm{y}-X\bm{\theta}||^2+\lambda||\bm{\theta}||^2,~\lambda\ge 0
\]
leading to the estimate
\[
\hat{\bm{\theta}}_R=(X^TX+\lambda I)^{-1} X^T \bm{y},
\]
where $I$ is the identity matrix.

The major goal of this review article is to focus at alternative norms in place of the Euclidean norm, which was employed in ridge regression. As we will see, there are many good reasons in doing that.

\section{Searching for a Norm}\label{ch10:norm}

\begin{figure}[!tbp]
\centering
\includegraphics[scale=1]{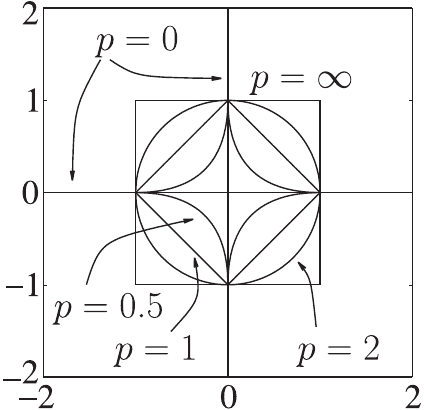}
\caption{The isovalue curves for $\norm{\theta}_p=1$ and for various
  values of $p$, in the two dimensional space. Observe that for the
  $\ell_0$ norm, the respective values cover the two axes with the exception of the point $(0,0)$. For the
  $\ell_1$ norm the isovalue curve is a rhombus and for the $\ell_2$
  (Euclidean) norm, it is a circle.}\label{fig:isovalues}
\end{figure}

Mathematicians have been very imaginative in proposing various norms in
order to equip linear spaces. Among the most popular norms used in
functional analysis are the so-called $\ell_p$ norms. To tailor things
to our needs, given a vector $\bm{\theta}\in\Real^l$, its $\ell_p$ norm
is defined as \beq \norm{\bm{\theta}}_p \coloneqq
\left(\sum_{i=1}^l|\theta_i|^p\right)^{\frac{1}{p}}. \label{ch10:norm1}
\eeq For $p=2$, the Euclidean or $\ell_2$ norm is obtained, and for
$p=1$, \eqref{ch10:norm1} results in the $\ell_1$ norm, i.e., \beq
\norm{\bm{\theta}}_1=\sum_{i=1}^l|\theta_i|.  \eeq If we let
$p\rightarrow \infty$, then we get the $\ell_{\infty}$ norm; let
$|\theta_{i_{\max}}|\coloneqq \max \left \{|\theta_1|,
|\theta_2|,\ldots,|\theta_l|\right \}$, and notice that \beq
\norm{\bm{\theta}}_\infty \coloneqq \lim_{p\rightarrow \infty}\left
(|\theta_{i_{\max}}|^p\sum_{i=1}^l\left
(\frac{|\theta_i|}{|\theta_{i_{\max}}|}\right )^p\right
)^{\frac{1}{p}}=|\theta_{i_{\max}}|, \eeq that is,
$\norm{\bm{\theta}}_\infty$ is equal to the maximum of the absolute
values of the coordinates of $\bm{\theta}$. One can show that all the
$\ell_p$ norms are true norms for $p\ge 1$; they satisfy all four
requirements that a function $\Real^l\rightarrow [0,\infty)$ must
  respect in order to be called a norm, i.e.,
\begin{enumerate}
\item $\norm{\bm{\theta}}_p\ge 0$.
\item $\norm{\bm{\theta}}_p=0 \Leftrightarrow \bm{\theta}=\bm{0}$.
\item $\norm{\alpha\bm{\theta}}_p = |\alpha| \norm{\bm{\theta}}_p$,
  $\forall \alpha \in\Real$.
\item $\norm{\bm{\theta}_1+\bm{\theta}_2}_p\le \norm{\bm{\theta}_1}_p+
  \norm{\bm{\theta}_2}_p$.
\end{enumerate}
The third condition enforces the norm function to be
({\it positively}) {\it homogeneous} and the fourth one is the
\textit{triangle inequality}. These properties also guarantee that any
function that is a norm is also a convex one. Although
strictly speaking, if we allow $p>0$ to take
values less than one in \eqref{ch10:norm1}, the resulting function is easily shown not to be a true norm, we can still call them norms, albeit knowing that this
is an abuse of the definition of a norm. An interesting case, which
will be used extensively in this paper, is the $\ell_0$ norm, which
can be obtained as the limit, for $p\rightarrow 0$, of
\beq
\norm{\bm{\theta}}_0 \coloneqq \lim _{p\rightarrow 0} \norm{\bm{\theta}}^p_p =
\lim_{p\rightarrow 0}\sum_{i=1}^l|\theta_i|^p=\sum_{i=1}^l
\chi_{(0,\infty)}(|\theta_i|),
\eeq
where $\chi_{\mathcal{A}}(\cdot)$ is the characteristic
  function  with
respect to a set $\mathcal{A}$, defined as
\begin{equation*}
\chi_{\mathcal{A}}(\tau) \coloneqq \begin{cases}
1, & \text{if}\ \tau\in \mathcal{A},\\
0, & \text{if}\ \tau\notin \mathcal{A}.
\end{cases}
\end{equation*}
That is, the $\ell_0$ norm is equal to the number of nonzero
components of the respective vector. It is very easy to check that
this function is not a true norm. Indeed, this function is not
homogeneous, i.e., $\norm{\alpha\bm{\theta}}_0\ne |\alpha|
\norm{\bm{\theta}}_0$, $\forall \alpha \neq
1$. Fig.~\ref{fig:isovalues} shows the isovalue curves, in the
two-dimensional space, that correspond to
$\norm{\bm{\theta}}_p=\rho \equiv 1$, for $p=0, 0.5, 1, 2$, and
$\infty$. Observe that for the Euclidean norm the isovalue curve has
the shape of a ``ball'' and for the $\ell_1$ norm the shape of a
rhombus.  We refer to them as the $\ell_2$ and the $\ell_1$ balls,
respectively, by slightly ``abusing'' the meaning of a
ball\footnote{Strictly speaking, a ball must also contain all the
  points in the interior.}.  Observe that in the case of the $\ell_0$
norm, the isovalue curve comprises both the horizontal and the
vertical axes, excluding the $(0,0)$ element.  If we restrict the size
of the $\ell_0$ norm to be less than one, then the corresponding set of points
becomes a singleton, i.e., $(0,0)$. Also, the set of all the points
that have $\ell_0$ norm less than or equal to two, is the $\Real^2$
space. This, slightly ``strange'' behavior, is a consequence of the
discrete nature of this ``norm''.

\begin{figure}[!tbp]
\centering
\includegraphics[scale=1]{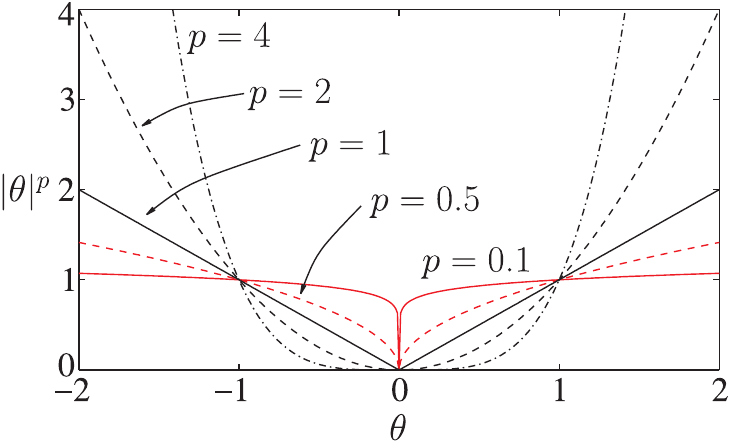}
\caption{Observe that the epigraph, that is, the region above the
  graph, is nonconvex for values $p<1$, indicating the nonconvexity of
  the respective $|\cdot|^p$ function. The value $p=1$ is the
  smallest one for which convexity is retained. Also note that, for
  large values of $p>1$, the contribution of small values of $\theta$
  to the respective norm becomes insignificant.}
\label{fig:lp.function}
\end{figure}

Fig.~\ref{fig:lp.function} shows the graph of $|\cdot|^p$, which is
the individual contribution of each component of a vector to the
$\ell_p$ norm, for different values of $p$. Observe that a) for $p<1$,
the region which is formed above the graph (known as epigraph)
is not a convex one, which verifies what we have already said; i.e,
the respective function is not a true norm, b) for values of the
argument $|\theta|>1$, the larger the value of $p\ge 1$ and the larger
the value of $|\theta|$ the higher its respective contribution to the
norm. Hence, if $\ell_p$ norms, $p\ge 1$, are used to
regularize a loss function, such large values become the dominant ones and
the optimization algorithm will concentrate on these by penalizing
them to get smaller, so that the overall cost to be reduced. On the
other hand, for values of the argument $|\theta|<1$ and closer to
zero, the $\ell_1$ norm is the only one (among $p\ge 1$) that retains
relatively large values and, hence, the respective components can
still have a say in the optimization process and can be penalized by
being pushed to smaller values.  Hence, if the $\ell_1$ norm is used
to replace the $\ell_2$ one in the regularization equation, \textit{only those
  components of the vector, that are really significant in reducing the model misfit
  measuring term in the regularized
  cost function, will be kept and the rest will be forced to
  zero}. The same tendency, yet more aggressive, is true for $0\le
p<1$.  The extreme case is when one considers the $\ell_0$ norm. Even
a small increase of a component from zero, makes its contribution to
the norm large, so the optimizing algorithm has to be very
``cautious'' in making an element nonzero.

From all the true norms ($p\ge 1$), the $\ell_1$ is the only
one that shows respect to small values. The rest of the $\ell_p$
norms, $p>1$, just squeeze them, to make their values even smaller
and care, mainly, for the large values.  We will return to this point
very soon.

\section[The LASSO]{The Least Absolute Shrinkage and Selection
  Operator (LASSO)}\index{LASSO}\label{ch10:lasso}

We have already discussed some of the benefits in
adopting the regularization method for enhancing the performance of an
estimator. However, in this paper, we are going to see and study more reasons
that justify the use of regularization. The first one refers to what
is known as the {\it interpretation} power of an estimator. For
example, in the regression task, we want to select those components,
$\theta_i$, of $\bm{\theta}$ that have the most important say in the
formation of the output variable. This is very important if the number
of parameters, $l$, is large and we want to concentrate on the most
important of them. In a classification task \cite{Theod-10}, not all features are
informative, hence one would like to keep the most informative of them
and make the less informative ones equal to zero. Another related
problem refers to those cases where we know, a-priori, that a number
of the components of a parameter vector are zero but we do not know which
ones. The discussion we had at the end of the previous section starts
now to become more meaningful. Can we use, while regularizing, an
appropriate norm that can assist the optimization process a) in
unveiling such zeros or b) to put more emphasis on the most significant of its
components, those that play a decisive role in reducing the misfit measuring term
in the regularized cost function, and set the rest of them equal to
zero? Although the $\ell_p$ norms, with $p<1$, seem to be the natural
choice for such a regularization, the fact that they are not convex
makes the optimization process hard. The $\ell_1$ norm is the one that
is ``closest'' to them yet it retains the computationally attractive
property of convexity.

The $\ell_1$ norm has been used for such problems for a long time. In the
seventies, it was used in seismology \cite{Taylor-10, Claer-10}, where the
reflected signal, that indicates changes in the various earth substrates,
is a sparse one, i.e., very few values are relatively large and the rest
are small and insignificant. Since then, it has been used to tackle similar
problems in different applications, e.g., \cite{Santo-10,
  Dono-10}. However, one can trace two papers that were really catalytic in
providing the spark for the current strong interest around the $\ell_1$
norm. One came from statistics, \cite{Tibsh-10}, which addressed the
\textit{LASSO} task (first formulated, to our knowledge, in
\cite{Santo-10}), to be discussed next, and one from the signal analysis
community, \cite{Chen-10}, which formulated the \textit{Basis Pursuit}, to
be discussed in a later section.

We first address our familiar regression task
\[
\bm{y} = X\bm{\theta}+\bm{\eta}, \quad \bm{y},\bm{\eta} \in {\cal
  R}^N, \bm{\theta}\in {\cal R}^l,
\]
and obtain the estimate of the unknown parameter $\bm{\theta}$ via
the LS loss, regularized by the $\ell_1$ norm, i.e., for $\lambda \geq
0$,
\beqa
\hat{\bm{\theta}} & \coloneqq & \argmin_{\bm{\theta}\in \Real^l}
L(\bm{\theta},\lambda) \\
& \coloneqq & \argmin_{\bm{\theta}\in \Real^l} \left(
\sum_{n=1}^N(y_n-\bm{x}_n^T\bm{\theta})^2+\lambda \norm{\bm{\theta}}_1 \right)
\nonumber \\
& =& \argmin_{\bm{\theta}\in \Real^l}
\left((\bm{y}-X\bm{\theta})^T(\bm{y}-X\bm{\theta})+ \lambda
\norm{\bm{\theta}}_1\right). \label{ch10lassos1}
\eeqa
In order to simplify the analysis, we will assume
hereafter, without harming generality, that the data are centered. If this
is not the case, the data can be centered by subtracting the sample mean
$\bar{y}$ from each one of the output values. The estimate of the
bias term will be equal to the sample mean $\bar{y}$. The task in
\eqref{ch10lassos1} can be \textit{equivalently} written in the following two
formulations
\begin{eqnarray}
\hat{\bm{\theta}}: \min_{\bm{\theta}\in \Real^l} &&
(\bm{y}-X\bm{\theta})^T(\bm{y}-X\bm{\theta}), \nonumber \\
\text{s.t.} && \norm{\bm{\theta}}_1 \le \rho, \label{ch10lassos2}
\end{eqnarray}
or
\begin{eqnarray}
\hat{\bm{\theta}}: \min_{\bm{\theta}\in \Real^l} &&
\norm{\bm{\theta}}_1, \nonumber \\
\text{s.t.} && (\bm{y}-X\bm{\theta})^T(\bm{y}-X\bm{\theta}) \le
\epsilon,  \label{ch10lassos3}
\end{eqnarray}
given the user-defined parameters $\rho, \epsilon \geq 0$. The formulation
in \eqref{ch10lassos2} is known as the LASSO and the one in
\eqref{ch10lassos3} as the {\it Basis Pursuit Denoising} (BPDN)\index{Basis
  Pursuit Denoising}, e.g.,  \cite{Fred-10}. All three formulations can be shown to be equivalent
for specific choices of $\lambda, \epsilon$, and $\rho$. The minimized cost function in \eqref{ch10lassos1} corresponds to the Lagrangian of the formulations in \eqref{ch10lassos2} and \eqref{ch10lassos3}. However, this
functional dependence is hard to compute, unless the columns of $X$ are
mutually orthogonal. Moreover, this equivalence does not necessarily imply
that all three formulations are equally easy or difficult to solve. As we
will see later on, algorithms have been developed along
each one of the previous formulations. From now on, we will refer to all
three formulations as the LASSO task, in a slight abuse of the standard
terminology, and the specific formulation will be apparent from the
context, if not stated explicitly.

As it was discussed before, the Ridge regression admits a closed form solution, i.e,
\[
\hat{\bm{\theta}}_R=\left (X^TX+\lambda I\right )^{-1}X^T\bm{y}.
\]
In contrast, this is not the case for LASSO and its solution requires
iterative techniques. It is straightforward to see that LASSO can be
formulated as a standard convex quadratic problem with linear
inequalities. Indeed, we can rewrite \eqref{ch10lassos1} as
\begin{eqnarray*}
\min_{\{\theta_i, u_i\}_{i=1}^l} &&
(\bm{y}-X\bm{\theta})^T(\bm{y}-X\bm{\theta})+\lambda \sum_{i=1}^lu_i
\\
\text{s.t.} && \left\{\begin{aligned}
& -u_i\le \theta_i\le u_i,\\
& u_i\ge  0,
\end{aligned} \quad i=1,2, \ldots, l,\right.
\end{eqnarray*}
which can be solved by any standard convex
optimization method, e.g., \cite{Ye-10, Boyd-10}. The reason that developing algorithms for the LASSO
has been a hot research topic
is due to the emphasis  in obtaining {\it efficient} algorithms by
exploiting the specific nature of this task, especially for cases
where $l$ is very large, as it is often the case in practice.

In order to get a better insight of the nature of the solution
that is obtained by LASSO, let us assume that the regressors are
mutually orthogonal and of unit norm, hence $X^TX=I$. Orthogonality of
the input matrix helps to decouple the coordinates and results to $l$
one-dimensional problems, that can be solved analytically. For this
case, the LS estimator becomes
\[
\hat{\bm{\theta}}_{\text{LS}} = (X^TX)^{-1}X^T\bm{y} = X^T\bm{y},
\]
and the ridge regression gives
\beq
\hat{\bm{\theta}}_{R} = \frac{1}{1+\lambda}\hat{\bm{\theta}}_{\text{LS}},
\eeq
that is, every component of the LS estimator is simply shrunk by the
\textit{same} factor, $\frac{1}{1+\lambda}$.

\begin{figure}[!tbp]
\centering
\includegraphics[scale=1]{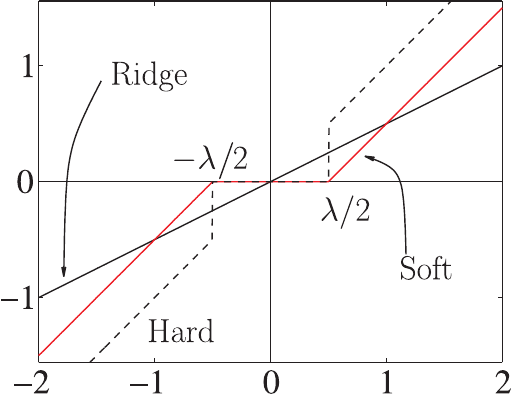}
\caption{Output-input curves for the hard thresholding, soft thresholding
  operators together with the linear operator associated with the ridge
  regression, for the same value of $\lambda = 1$. } \label{fig:hard.soft}
\end{figure}

In the case of the $\ell_1$ regularization, the minimized Lagrangian function is no more
differentiable, due to the presence of the absolute values in the $\ell_1$
norm. So, in this case, we have to consider the notion of the
subdifferential (see Appendix). It is known that if the zero vector
belongs to the subdifferential set of a convex function at a point, this
means that this point corresponds to a minimum of the function. Taking the
subdifferential of the Lagrangian defined in \eqref{ch10lassos1} and recalling
that the subdifferential of a differentiable function includes
\textit{only} the respective gradient, we obtain that
\begin{equation*}
\bm{0} \in -2X^T\bm{y} + 2X^TX\bm{\theta}+\lambda\partial
 \norm{\bm{\theta}}_1,\label{ch10-l1min1}
 \end{equation*}
where $\partial$ stands for the subdifferential operator (see Appendix). If $X$ has orthonormal columns, the previous equation can be written
component-wise as follows
\begin{equation}
0\in -\hat{\theta}_{\text{LS},i} + \hat{\theta}_{1,i} + \frac{\lambda}{2}
\partial \left|\hat{\theta}_{1,i} \right|, \quad \forall
i, \label{componentwise.lasso.task}
\end{equation}
where the subdifferential of the function $|\cdot|$, derived in Appendix, is given as
\begin{equation*}
\partial|\theta| = \begin{cases}
\{1\}, & \text{if}\ \theta>0,\\
\{-1\}, & \text{if}\ \theta<0,\\
[-1,1], & \text{if}\ \theta=0.
\end{cases}
\end{equation*}
Thus, we can now write
\begin{numcases}{\hat{\theta}_{1,i} =}
\hat{\theta}_{\text{LS},i}-\frac{\lambda}{2}, &
if $\hat{\theta}_{1,i} > 0$, \label{ch10:las1}\\
\hat{\theta}_{\text{LS},i} + \frac{\lambda}{2}, &
if $\hat{\theta}_{1,i} < 0$. \label{ch10:las2}
\end{numcases}
Notice that \eqref{ch10:las1} can only be true if
$\hat{\theta}_{\text{LS},i} > \frac{\lambda}{2}$, and \eqref{ch10:las2}
only if $\hat{\theta}_{\text{LS},i} < -\frac{\lambda}{2}$. Moreover, in the
case where $\hat{\theta}_{1,i}=0$, then \eqref{componentwise.lasso.task}
and the subdifferential of $|\cdot|$ suggest that necessarily
$\left|\hat{\theta}_{\text{LS},i} \right| \leq
\frac{\lambda}{2}$. Concluding, we can write in a more compact way that
\begin{equation}
\hat{\theta}_{1,i}=\sign(\hat{\theta}_{\text{LS},i})\left
(\left|\hat{\theta}_{\text{LS},i}\right|-\frac{\lambda}{2} \right
)_+,\label{ch10-l1min2}
\end{equation}
where $(\cdot)_+$ denotes the ``positive part'' of the respective argument;
it is equal to the argument if this is non-negative, and zero
otherwise. This is very interesting indeed. In contrast to the ridge regression
that shrinks all coordinates of the unregularized LS solution by the same
factor, LASSO forces all coordinates, whose absolute value is less than or
equal to $\lambda/2$, to zero, and the rest of the coordinates are reduced,
in absolute value, by the same amount $\lambda/2$. This is known as {\it
  soft thresholding}\index{Soft thresholding}, to distinguish it from the
\textit{hard thresholding} operation; the latter is defined as $\theta\cdot
\chi_{(0,\infty)}\left(|\theta|-\frac{\lambda}{2}\right)$, $\theta\in
\Real$, where $\chi_{(0,\infty)}(\cdot)$ stands for the characteristic
function with respect to the set $(0,\infty)$. Fig.~\ref{fig:hard.soft}
shows the graphs illustrating the effect that the ridge regression, LASSO
and hard thresholding have on the unregularized LS solution, as a function
of its value (horizontal axis).  Note that our discussion here, simplified
via the orthonormal input matrix case, has quantified what we had said
before about the tendency of the $\ell_1$ norm to push small values to
become {\it exactly zero}. This will be further strengthened, via a more
rigorous mathematical formulation, in Section \ref{Ch10:thesparsest}.

\begin{example}
Assume that the unregularized LS solution, for a given regression
task, $\bm{y}=X\bm{\theta}+\bm{\eta}$, is given by:
\[
\hat{\bm{\theta}}_{\text{LS}}=[0.2, -0.7, 0.8, -0.1, 1.0]^T.
\]
Derive the solutions for the corresponding ridge regression and
$\ell_1$ norm regularization tasks. Assume that the input matrix $X$
has orthonormal columns and that the regularization parameter is
$\lambda=1$. Also, what is the result of hard thresholding the vector
$\hat{\bm{\theta}}_{\text{LS}}$ with threshold equal to $0.5$?

We know that the corresponding solution for the ridge regression is
\[
\hat{\bm{\theta}}_R= \frac{1}{1+\lambda}
\hat{\bm{\theta}}_{\text{LS}}=[0.1, -0.35, 0.4, -0.05, 0.5]^T.
\]
The solution for the $\ell_1$ norm regularization is given by soft
thresholding, with threshold equal to $\lambda/2=0.5$, hence the
corresponding vector is
\[
\hat{\bm{\theta}}_1=[0, -0.2, 0.3, 0, 0.5]^T.
\]
The result of the hard thresholding operation  is the vector $[0,
  -0.7, 0.8, 0, 1.0]^T$.
\end{example}

\begin{figure}[!tbp]
\centering
\includegraphics[scale=1]{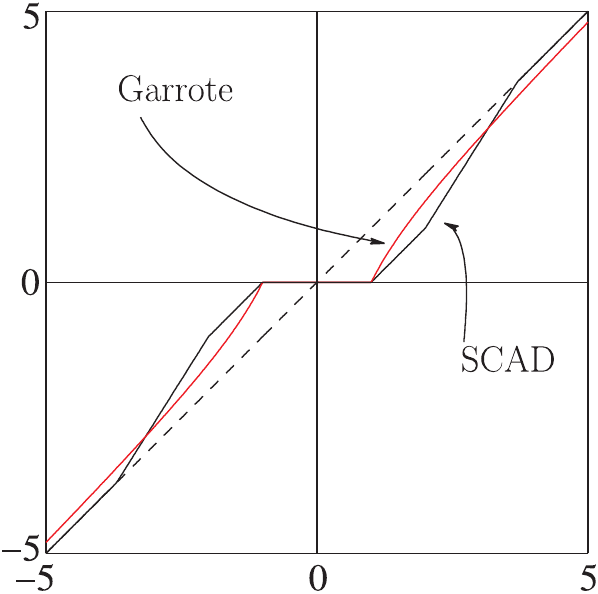}
\caption{Output-input graph for the SCAD and nonnegative garotte rules with
  parameters $\alpha= 3.7$, and $\lambda_{\text{SCAD}}=
  \lambda_{\text{garr}} = 1$. Observe that both rules smooth out the
  discontinuity associated with the hard thresholding rule. Notice, also,
  that the SCAD rule removes the bias, associated with the soft
  thresholding rule, for large values of the input variable. On the
  contrary, the garrote thresholding rule allows some bias for large input
  values, which diminishes as $\lambda_{\text{garr}}$ gets smaller and
  smaller.} \label{fig:garrote}
\end{figure}

\begin{remarks}\mbox{}
\begin{itemize}

\item The hard and soft thresholding rules are only two possibilities
  out of a larger number of alternatives. Note that the hard
  thresholding operation is defined via a discontinuous function and
  this makes this rule to be unstable, in the sense of being very
  sensitive to small changes of the input. Moreover, this shrinking
  rule  tends to exhibit large variance in the resulting
  estimates. The soft thresholding rule is a continuous function, but,
  as it is readily seen from the graph in Fig.~\ref{fig:hard.soft}, it
  introduces bias even for the large values of the input argument.  In
  order to ameliorate  such shortcomings,  a number of alternative
  thresholding operators have been introduced and studied both
  theoretically and experimentally. Although these are not within the
  mainstream of our interest, we provide two popular examples for the
  sake of completeness; the {\it Smoothly Clipped Absolute Deviation}
  (SCAD)\index{SCAD thresholding rule}:
\begin{equation*}
\label{eq:SCAD}
\hat{\theta}_{\text{SCAD}} =
\begin{dcases}
\sign(\theta)\left(|\theta|-\lambda_{\text{SCAD}}\right )_+, & |\theta|\le
2\lambda_{\text{SCAD}},\\
\frac{(\alpha-1)\theta- \alpha\lambda_{\text{SCAD}} \sign(\theta)}{\alpha-2}, &
2\lambda_{\text{SCAD}} < |\theta|\leq \alpha\lambda_{\text{SCAD}},\\
\theta, & |\theta|>\alpha\lambda_{\text{SCAD}},
\end{dcases}
\end{equation*}
and the {\it nonnegative garrote} thresholding rule \index{Nonnegative
  garrote thresholding rule}:
\begin{equation*}
\label{eq:nngarrote}
\hat{\theta}_{\text{garr}} =
\begin{dcases}
0, & |\theta|\le \lambda_{\text{garr}},\\
\theta-\frac{\lambda_{\text{garr}}^2}{\theta}, & |\theta|>\lambda_{\text{garr}}.
\end{dcases}
\end{equation*}
Fig.~\ref{fig:garrote} shows the respective graphs. Observe that, in both
cases, an effort has been made to remove the discontinuity (associated with
the hard thresholding) and to remove/reduce the bias for large values of
the input argument. The parameter $\alpha$ is a user-defined one. For a
more detailed discussion on this topic, the interested reader can refer,
for example, to \cite{Anto-10}.
\end{itemize}
\end{remarks}

\section{Sparse Signal Representation}\index{Sparse Signal Representation}
\label{ch10:sparse}

In the previous section, we brought into our discussion the need for taking
special care for zeros. Sparsity is an attribute that is met in a plethora
of natural signals, since nature tends to be parsimonious.  In this section,
we will briefly present a number of application cases, where the existence
of zeros in a mathematical expansion is of paramount importance, hence it
justifies to further strengthen our search for and developing related
analysis tools.

\begin{figure}[!tbp]
\centering
\includegraphics[scale=1]{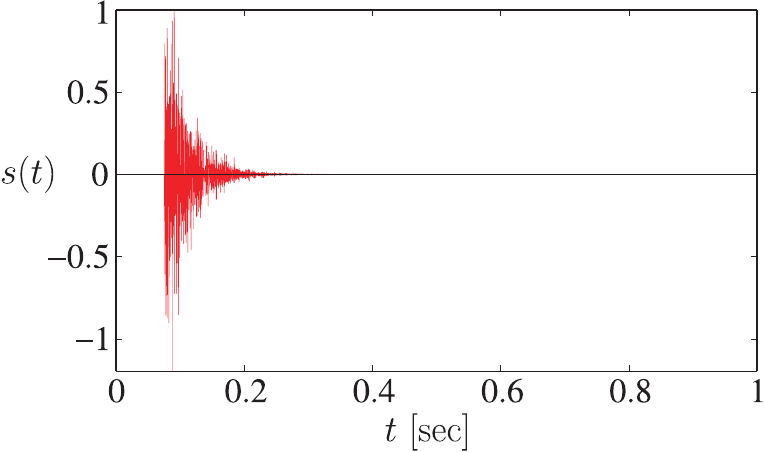}
\caption{The impulse response function of an echo-path in a telephone
  network. Observe that although it is of relatively short duration, it is
  not a-priori known where exactly in time will
  occur.} \label{fig:echo.path}
\end{figure}

Echo cancelation is a major task in Communications. In a number
of cases, the echo path, represented by a vector comprising the values of the impulse
response samples, is a sparse one.  This is the case, for example, in
internet telephony and in acoustic and network environments, e.g.,
\cite{Nayl-10, Bene-10, Aren-10}. Fig.~\ref{fig:echo.path} shows the
impulse response of such an echo path. The impulse response of the echo
path is of short duration; however, the delay with which it appears is not
known.  So, in order to model it, one has to use a long impulse response,
yet only a relatively small number of the coefficients will be significant
and the rest will be close to zero.  Of course, one could ask why not use
an LMS or an RLS \cite{Haykin, SayedBook}and eventually the significant coefficients will be
identified. The answer is that this turns out not to be the most efficient
way to tackle such problems, since the convergence of the algorithm can be
very slow. In contrast, if one embeds, somehow, into the problem the
a-priori information concerning the existence of (almost) zero
coefficients, then the convergence speed can be significantly increased and
also better error floors can be attained.

A similar situation, as in the previous case, occurs in wireless
communication systems, which involve multipath channels.  A typical
application is in high definition television (HDTV) systems, where the
involved communications channels consist of a \textit{few}
non-negligible echoes, some of which may have quite large time delays
with respect to the main signal, see, e.g. \cite{Ghosh-10, Cott-10,
  Ariy-10, Rond-10}.  If the information signal is transmitted at high
symbol rates through such a dispersive channel, then the introduced
intersymbol interference (ISI) has a span of several tens up to
hundreds of symbol intervals. This in turn implies that quite long
channel estimators are required at the receiver's end in order to
reduce effectively the ISI component of the received signal, although
only a small part of it has values substantially different to zero.
The situation is even more demanding whenever the channel frequency
response exhibits deep nulls. More recently, sparsity has been
exploited in channel estimation for multicarrier systems, both for
single antenna as well as for MIMO systems \cite{Eiwen1-10,
  Eiwen2-10}. A thorough and in depth treatment related to sparsity in
multipath communication systems is provided in \cite{Bajwa-10}.

\begin{figure}[!tbp]
  \centering
  \includegraphics[scale=1]{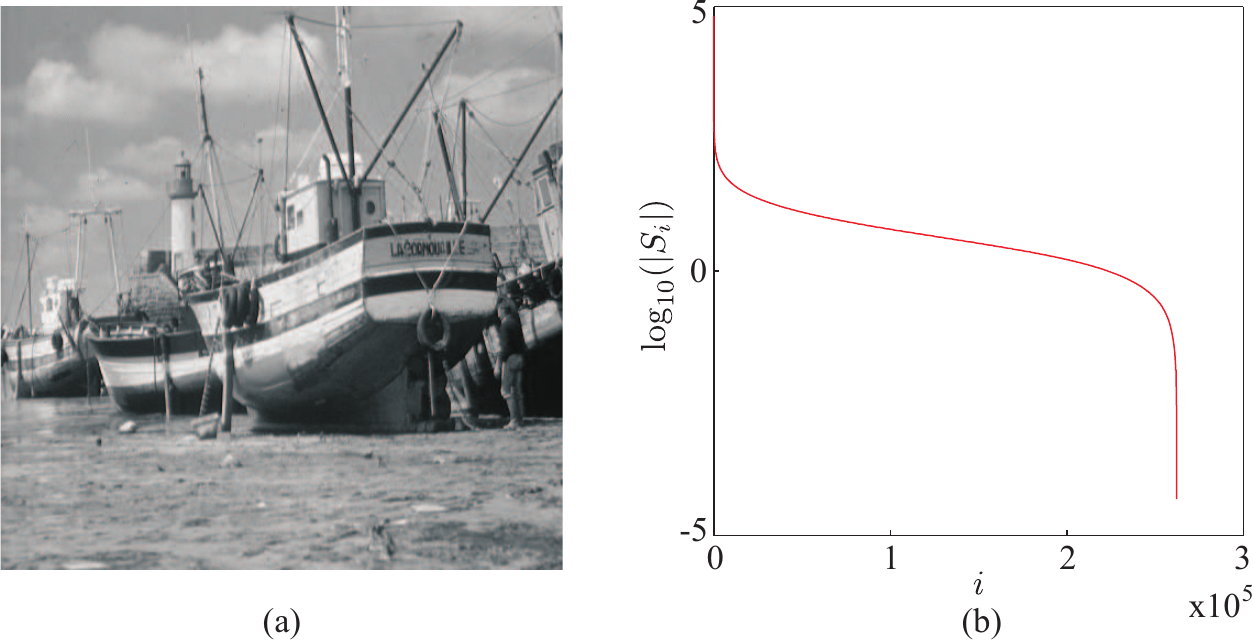}
  \caption{(a) A $512 \times 512$ pixel image and (b) The magnitude of its Discrete Cosine Transform components in descending order and logarithmic scale. Note that more than $95\%$ of the total energy is contributed by only the $5\%$ of the largest components}\label{fig:boatwavelets}
\end{figure}
Another example, which might be more widely known, is that of signal
compression. It turns out that if the signal modalities, with which we
communicate, e.g., speech, and also we sense the world, e.g., images,
audio, are transformed into a suitably chosen domain then they are sparsely
represented; only a relatively small number of the signal components in
this domain are large and the rest are close to zero. As an example, Fig. \ref{fig:boatwavelets}a shows an image and Fig. \ref{fig:boatwavelets}b the plot of the magnitude of the obtained Discrete Cosine Transform (DCT) components, which are computed by writing the corresponding image array as a vector in lexicographic order. Note that more than $95\%$ of the total energy is contributed by only the $5\%$ of the largest components. This is at the heart
of any compression technique. Only the large coefficients are chosen to be
coded and the rest are considered to be zero. Hence, significant gains are
obtained in memory/bandwidth requirements while storing/transmitting such
signals, without much perceptual loss. Depending on the modality, different
transforms are used. For example, in JPEG-2000, an image array, represented
in terms of a vector that contains the intensity of the gray levels of the
image pixels, is transformed via the discrete wavelet transform (DWT) and
results to a transform vector that comprises only a few large
components. Such an operation is of the form
\begin{equation}
\bm{S} =  \Phi^H \bm{s}, \quad \bm{s},
\bm{S}\in\mathcal{C}^l, \label{ch10-sparsre}
\end{equation}
where $\bm{s}$ is the vector of the ``raw'' signal samples, $\bm{S}$ the
vector of the transformed ones, and $\Phi$ is the $l\times l$
transformation matrix. Often, this is an orthonormal matrix,
$\Phi^H\Phi=I$. Basically, a transform is nothing else than a projection of
a vector on a new set of coordinate axes, which comprise the columns of the
transformation matrix $\Phi$.  Celebrated examples of such transforms are
the wavelet, the discrete Fourier (DFT) and the discrete cosine (DCT)
transforms, e.g., \cite{Theod-10}.  In such cases, where the transformation
matrix is orthonormal, one can write that \beq
\bm{s}=\Psi\bm{S},\label{ch10-sparsre-1} \eeq where $\Psi=\Phi$. Equation
\eqref{ch10-sparsre} is known as the \textit{analysis} and
\eqref{ch10-sparsre-1} as the \textit{synthesis} equation.

Compression via such transforms exploit the fact that many signals in
nature, which are rich in context, can be {\it compactly} represented in an
appropriately chosen basis, depending on the modality of the signal. Very
often, the construction of such bases tries to ``imitate'' the sensory
systems that the human (and not only) brain has developed in order to sense
these signals; and we know that nature (in contrast to modern humans) does
not like to waste resources. A standard compression task comprises the
following stages: a) Obtain the $l$ components of $\bm{S}$, via the
analysis step \eqref{ch10-sparsre}, b) keep the, say, $k$ most significant
of them, c) code these values, as well as their respective locations in the
transform vector $\bm{S}$, and d) obtain the (approximate) original signal $\bm{s}$, when
needed (after storage or transmission), via the synthesis
equation \eqref{ch10-sparsre-1}, where in place of $\bm{S}$ only its $k$
most significant components are used, which are the ones that were coded,
while the rest are set equal to zero.
However, there is something
unorthodox in this process of compression, as it has been practised till
very recently. One processes (transforms) large signal vectors of $l$
coordinates, where $l$ in practice can be quite large, and then uses only a
small percentage of the transformed coefficients and the rest are simply
ignored. Moreover, one has to store/transmit the location of the respective
large coefficients that were finally coded. A natural question that is now
raised is the following: Since $\bm{S}$ in the synthesis equation is
(approximately) sparse, can one compute it via an alternative path than the
analysis equation in \eqref{ch10-sparsre}? The issue here is to investigate
whether one could use a more informative way of obtaining measurements from
the available raw data, so that less than $l$  measurements are
sufficient to recover all the necessary information. The ideal case would
be to be able to recover it via a set of $k$ such measurement samples, since this is
the number of the significant free parameters.
On the other hand, if this sounds a bit extreme, can one obtain $N$
($k<N<l$) such signal-related measurements, from which one can obtain the
$k$ needed components of $\bm{S}$? It turns out that such an approach is
possible and it leads to the solution of an \textit{underdetermined} system
of linear equations, under the constraint that the unknown target vector is
a sparse one. The importance of such techniques becomes even more apparent
when, instead of an orthonormal basis, as discussed before, a more general
type of expansion is adopted, in terms of what is known as
\textit{overcomplete dictionaries}\index{Overcomplete dictionaries}.

A dictionary \cite{Mall-10} is a collection of parameterized waveforms,
which are discrete-time signal samples, represented as vectors
$\bm{\psi}_i\in\mathcal{C}^l$, $i\in \mathcal{I}$. For example, the columns of a
DFT or a DWT matrix comprise a dictionary. These are two examples of what
is known as {\it complete} dictionaries, which consist of $l$ (orthonormal)
vectors, i.e., a number equal to the length of the signal vector. However,
in many cases in practice, using such dictionaries is very restrictive. Let
us take, for example, a segment of audio signal, from a news media or a
video, that needs to be processed. This consists, in general, of different
types of signals, namely speech, music, environmental sounds. For each type
of these signals, different signal vectors (dictionaries) may be more
appropriate in the expansion for the analysis. For example, music signals
are characterized by a strong harmonic content and the use of sinusoids
seems to be best for compression, while for speech signals a Gabor type
signal expansion (sinusoids of various frequencies weighted by sufficiently narrow pulses
at different locations in time, \cite{Coifm-10, Theod-10}), may be a better
choice. The same applies when one deals with an image. Different parts of
an image, e.g., parts which are smooth or contain sharp edges, may demand a
different expansion vector set, for obtaining the best overall
performance. The more recent tendency, in order to satisfy such needs, is
to use {\it overcomplete} dictionaries. Such dictionaries can be obtained,
for example, by concatenating different dictionaries together, e.g., a DFT
and a DWT matrix to result in a combined $l\times 2l$ transformation
matrix. Alternatively, a dictionary can be ``trained'' in order to effectively represent a set of available signal exemblars, a task which is often referred to as dictionary learning \cite{tosic_dictionary_2011,rubinstein_dictionaries_2010,yaghoobi_parametric_2009}.  While using such overcomplete dictionaries, the synthesis equation
takes the form
\begin{equation}
\bm{s} =\sum_{i\in\mathcal{I}}
\theta_i\bm{\psi}_i. \label{ch-10signrepr}
\end{equation}
Note that, now, the analysis is an ill-posed problem, since the elements
$\{\bm{\psi}_i\}_{i\in\mathcal{I}}$ (usually called \textit{atoms}) of the
dictionary are not linearly independent, and there is not a unique set of
coefficients $\{\theta_i\}_{i\in\mathcal{I}}$ which generates
$\bm{s}$. Moreover, we expect most of these coefficients to be (nearly)
zero.  Note that, in such cases, the cardinality of ${\cal I}$ is larger
than $l$. This necessarily leads to underdetermined systems of equations
with infinite many solutions. The question that is now raised is whether we
can exploit the fact that most of these coefficients are known to be zero,
in order to come up with a unique solution, and if yes, under which
conditions such a solution is possible?

Besides the previous examples, there is a number of cases where an
underdetermined system of equations is the result of our inability to
obtain a sufficiently large number of measurements, due to physical and
technical constraints. This is for example the case in MRI imaging, which
will be presented in more detail later on.

\section{In Quest for the Sparsest Solution}\label{Ch10:thesparsest}

Inspired by the discussion in the previous section, we now turn our
attention to the task of solving underdetermined systems of equations,
by imposing the sparsity constraint on the solution \cite{eladbook}. We will develop
the theoretical set up in the context of the regression task and we
will adopt the notation that has been adopted for this task. Moreover,
we will adhere to the real data case, in order to simplify the presentation.
The theory can be readily extended to the more general complex data case, see, e.g., \cite{Wright-10,maleki_complexlasso_2011}.
We assume that we are given a set of measurements, $\bm{y} \coloneqq [y_1,
  y_2, \ldots, y_N]^T \in \Real^N$, according to the linear model
\begin{equation}
\bm{y}=X\bm{\theta}, \quad\bm{y}\in\Real^N, \bm{\theta}\in\Real^l,
l>N, \label{ch10-thesp-1}
\end{equation}
where $X$ is the $N\times l$ input matrix, which is assumed to be of
full row rank, i.e., $\rank(X) = N$. Our starting point is the
noiseless case.  The system in \eqref{ch10-thesp-1} is an
underdetermined one and accepts an infinite number of solutions. The
set of possible solutions lies in the intersection of the $N$
hyperplanes\footnote{In $\Real^l$, a hyperplane is of dimension
  $l-1$. A plane has dimension lower than $l-1$.} in the
$l$-dimensional space,
\[
\bigl\{\bm{\theta}\in \Real^l: y_n=\bm{x}^T_n\bm{\theta} \bigr\}, \quad
n=1,2,\ldots,N.
\]
We know from geometry, that the intersection of $N$ non-parallel hyperplanes
(which in our case is guaranteed by the fact that $X$ has been assumed to
be full row rank, hence $\bm{x}_n$ are mutually independent) is a plane of
dimensionality $l-N$ (e.g., the intersection of two (non-parallel)
(hyper)planes in the 3-dimensional space is a straight line; that is, a
plane of dimensionality equal to one). In a more formal way, the set of all
possible solutions, to be denoted as $\Theta$, is an {\it affine} set. An
affine set is the translation of a linear subspace by a constant
vector. Let us pursue this a bit further, since we will need it later on.

Let the null space of $X$ be the set $\nullsp(X)$, defined as the
linear subspace
\[
\nullsp(X)=\left \{\bm{z}\in\Real^l: X\bm{z}=\bm{0}\right\}.
\]
Obviously, if $\bm{\theta}_0$ is a solution to \eqref{ch10-thesp-1},
i.e., $\bm{\theta}_0 \in \Theta$, then it is easy to verify that
$\forall \bm{\theta}\in \Theta$,
$X(\bm{\theta}-\bm{\theta}_0)=\bm{0}$, or $\bm{\theta}- \bm{\theta}_0
\in\nullsp(X)$. As a result,
\begin{equation*}
\Theta= \bm{\theta}_0 + \nullsp(X), \label{ch10-thesp-2}
\end{equation*}
and $\Theta$ is an affine set. We also know from linear algebra basics, that the null space of a full row
rank matrix, $N\times l$, $l>N$, is a subspace of dimensionality
$l-N$. Fig.~\ref{fig:AffineSetAndl2vsl1min} illustrates the case for one
measurement sample in the $2$-dimensional space, $l=2$ and $N=1$. The set
of solutions $\Theta$ is a line, which is the translation of the linear
subspace crossing the origin (the $\nullsp(X)$). Therefore, if one wants to
determine a \textit{single} point that lies in the affine set of solutions,
$\Theta$, then an extra constraint/a-priori knowledge has to be imposed

In the sequel, three such possibilities are examined.

\subsubsection{The $\ell_2$ Norm Minimizer}

Our goal now becomes to pick a point in (the affine set)
$\Theta$, that corresponds to the minimum $\ell_2$ norm. This is
equivalent to solving the following constrained task
\begin{eqnarray}
\min_{\bm{\theta}\in\Real^l} & & \norm{\bm{\theta}}_2^2 \nonumber \\
\text{s.t.} & &\bm{x}^T_n\bm{\theta}=y_n, \quad
n=1,2,\ldots,N. \label{l2.min.task}
\end{eqnarray}
The previous
optimization task accepts a \textit{unique} solution given in closed
form  as
\begin{equation}
\hat{\bm{\theta}}=X^T\left (XX^T\right )
^{-1}\bm{y}. \label{ch10l2min1}
\end{equation}
The geometric interpretation of this solution is provided in
Fig.~\ref{fig:AffineSetAndl2vsl1min}a, for the case of $l=2$ and $N=1$. The
radius of the Euclidean norm ball keeps increasing, till it touches the
plane that contains the solutions. This point is the one with the minimum
$\ell_2$ norm or, equivalently, the point that lies closest to the
origin. Equivalently, the point $\hat{\bm{\theta}}$ can be seen as the
(metric) projection of $\bm{0}$ onto $\Theta$.

Minimizing the $\ell_2$ norm, in order to solve a linear set of
underdetermined equations, has been used in various applications. The
closest to us is in the context of determining the unknown
coefficients in an expansion using an overcomplete dictionary of
functions (vectors) \cite{Daub-10}. A main drawback of this method is
that it is not sparsity preserving. There is no guarantee that the
solution in (\ref{ch10l2min1}) will give zeros even if the true model vector
$\bm{\theta}$ has zeros. Moreover, the method is {\it resolution
  limited} \cite{Chen-10}. This means that, even if there may be a
sharp contribution of specific atoms in the dictionary, this is not
portrayed in the obtained solution. This is a consequence of the fact
that the information provided by  $XX^T$  is a global one, containing
all atoms of the dictionary in an ``averaging'' fashion, and the final
result tends to smooth out the individual contributions, especially
when the dictionary is overcomplete.

\subsubsection{The $\ell_0$ Norm Minimizer}


\begin{figure}[!tbp]
  \centering
  \includegraphics[scale=1]{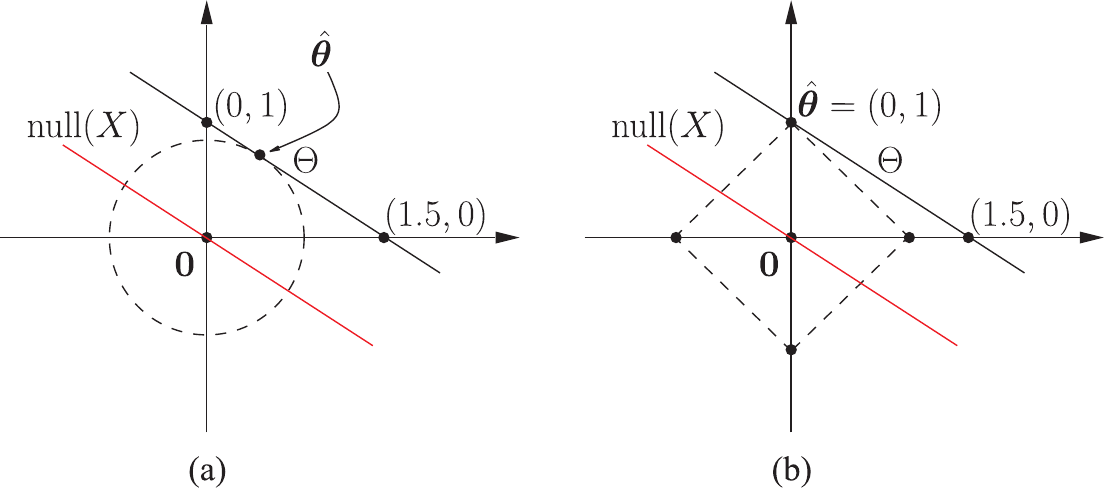}
  \caption{(a) The $\ell_2$ norm minimizer. The dotted circle corresponds to the smallest $\ell_2$
  ball that intersects the set $\Theta$. As such, the intersection point,
  $\hat{\bm{\theta}}$, is the $\ell_2$ norm minimizer of the task
  \eqref{l2.min.task}. Notice that the vector $\hat{\bm{\theta}}$ contains
  no zero component. (b) The $\ell_1$ norm minimizer. The dotted rhombus corresponds to the smallest
  $\ell_1$ ball that intersects $\Theta$. Hence, the intersection point,
  $\hat{\bm{\theta}}$, is the solution of the constrained $\ell_1$
  minimization task of \eqref{ch10:liminim}. Notice that the obtained
  estimate $\hat{\bm{\theta}}=(0,1)$ contains a zero.}\label{fig:AffineSetAndl2vsl1min}
\end{figure}

Now we turn our attention to the $\ell_0$ norm (once more, it is
pointed out that this is an abuse of the definition of the norm, as
stated before), and we make sparsity our new flag under which a
solution will be obtained. Recall from Section \ref{ch10:sparse} that
such a constraint is in line with the natural structure
that underlies a number of applications. The task now becomes
\beqa
\min_{\bm{\theta}\in\Real^l} & & \norm{\bm{\theta}}_0 \nonumber \\
\text{s.t.} & &\bm{x}^T_n\bm{\theta}=y_n, \quad
n=1,2,\ldots,N, \label{ch10:l0minim}
\eeqa
that is, from all the points that lie on the plane of all possible
solutions find the \textit{sparsest} one; i.e., the one with the least
number of nonzero elements. As a matter of fact, such an approach is
within the spirit of \textit{Occam's razor} rule. It corresponds to the
smallest number of parameters that can explain the obtained
measurements. The points that are now raised are:
\begin{itemize}
\item Is a solution to this problem unique and under which conditions?
\item Can a solution be obtained with low enough complexity in realistic time?
\end{itemize}
We postpone the answer to the first question later on. As for the second one, the
news is no good. Minimizing the $\ell_0$ norm under a set of linear
constraints is a task of combinatorial nature and as a matter of fact the
problem is, in general, NP-hard \cite{Nata-10}. The way to approach the
problem is to consider all possible combinations of zeros in $\bm{\theta}$,
removing the respective columns of $X$ in \eqref{ch10-thesp-1} and check
whether the system of equations is satisfied; keep as solutions the ones
with the smallest number of nonzero elements. Such a searching technique
exhibits complexity of an exponential dependence on
$l$. Fig.~\ref{fig:AffineSetAndl2vsl1min}a illustrates the two points
($(1.5,0)$ and $(0,1)$) that comprise the solution set of minimizing the
$\ell_0$ norm for the single measurement (constraint) case.

\subsubsection{The $\ell_1$ Norm Minimizer}

The current task is now given by
\beqa
\min_{\bm{\theta}\in\Real^l} & &
\norm{\bm{\theta}}_1 \nonumber \\ \text{s.t.} & &\bm{x}^T_n\bm{\theta}=y_n,
\quad n=1,2,\ldots,N. \label{ch10:liminim}
\eeqa
Fig.~\ref{fig:AffineSetAndl2vsl1min}b illustrates the geometry. The $\ell_1$ ball is
increased till it touches the affine set of the possible solutions.  For
this specific geometry, the solution is the point $(0,1)$. In our discussion
in Section \ref{ch10:norm}, we saw that the $\ell_1$ norm is the one, out
of all $\ell_p$, $p\ge 1$ norms, that bears some similarity with the
sparsity favoring (nonconvex) $\ell_p$, $p<1$ ``norms''. Also, we have
commented that the $\ell_1$ norm encourages zeros, when the respective
values are small. In the sequel, we will state one lemma, that establishes
this zero-favoring property in a more formal way.  The $\ell_1$ norm
minimizer is also known as \textit{Basis Pursuit}\index{Basis Pursuit} and
it was suggested for decomposing a vector signal in terms of the atoms of
an overcomplete dictionary \cite{Chen-10}.

The $\ell_1$ minimizer can be brought into the standard Linear
Programming (LP) form and then can be solved by recalling any related
method; the simplex method or the more recent interior point methods
are two possibilities, see, e.g., \cite{Boyd-10, Dantz-10}. Indeed,
consider the (LP) task
\beqan
\min_{\bm{x}} && \bm{c}^T\bm{x} \\
\text{s.t.} && A\bm{x}=\bm{b} \\
 && \bm{x}\ge \bm{0}.
 \eeqan
To verify that our $\ell_1$ minimizer can be cast in the previous
form, notice first that any $l$-dimensional vector $\bm{\theta}$ can
be decomposed as
\begin{equation*}
\bm{\theta}= \bm{u}-\bm{v}, \quad \bm{u}\geq \bm{0}, \bm{v}\geq \bm{0}.
\end{equation*}
Indeed, this holds true if, for example,
\begin{equation*}
\bm{u} \coloneqq \bm{\theta}_+, \quad \bm{v} \coloneqq (-\bm{\theta})_+,
\end{equation*}
where $\bm{x}_+$ stands for the vector obtained after taking the
positive parts of the components of $\bm{x}$. Moreover, notice
that
\begin{equation*}
\norm{\bm{\theta}}_1 = [1,1,\ldots,1] \begin{bmatrix}  \bm{\theta}_+
  \\ (-\bm{\theta})_+
\end{bmatrix} = [1,1,\ldots,1] \begin{bmatrix}  \bm{u}
  \\ \bm{v}\end{bmatrix}.
\end{equation*}
Hence, our $\ell_1$ minimization task can be recast in the LP form, if
\begin{alignat*}{2}
\bm{c} & \coloneqq [1,1,\ldots,1]^T,\quad & \bm{x} & \coloneqq [\bm{u}^T,
  \bm{v}^T]^T,\\
A & \coloneqq [X,-X], & \bm{b} & \coloneqq \bm{y}.
\end{alignat*}

\subsubsection{Characterization of the $\ell_1$ norm minimizer}

\begin{lemma}\label{lemma:charact.l1.min}
An element $\bm{\theta}$ in the affine set, $\Theta$, of the solutions
of the underdetermined linear system (\ref{ch10-thesp-1}), has minimal
$\ell_1$ norm {\it if and only if} the following condition is
satisfied:
\beq
 \left |\sum_{i:~\theta_i\ne 0}\sign(\theta_i) z_i\right |\le
 \sum_{i:~\theta_i=0}|z_i|, \quad\forall
 \bm{z}\in\nullsp(X).\label{ch10:chara-1}
 \eeq
 Moreover, the $\ell_1$ minimizer is {\it unique if and only if}
 the inequality in \eqref{ch10:chara-1} is a {\it strict} one for all
 $\bm{z}\ne \bm{0}$ (see, e.g., \cite{Pink-10}).
\end{lemma}

\begin{remarks}\label{rem:zeroes.dim.null}
The previous lemma has a very interesting and
 important consequence. If $\hat{\bm{\theta}}$ is the {\it unique}
 minimizer of \eqref{ch10:liminim}, then
 \beq
 \card\{i: \hat{\theta}_i=0\} \ge \dim(\nullsp(X)),
 \eeq
where $\card\{\cdot\}$ denotes the cardinality of a set. In words, the number
of zero coordinates of the unique minimizer cannot be smaller than the
dimension of the null space of $X$. Indeed, if this is not the case, then
the unique minimizer could have less zeros than the dimensionality of
$\nullsp(X)$. As it can easily be shown, this means that we can always find a
$\bm{z}\in\nullsp(X)$, which has zeros in the same locations where the
coordinates of the unique minimizer are zero, and at the same time it
is not identically zero, i.e., $\bm{z}\ne \bm{0}$.
However, this would
violate \eqref{ch10:chara-1}, which
in the case of uniqueness holds as a strict inequality.
\end{remarks}

\begin{definition}\label{def:k.sparse}
A vector $\bm{\theta}$ is called $k$-sparse if it has {\it at most}
$k$ nonzero components.
\end{definition}

\begin{remarks}\label{rem:k.leq.N}
If the minimizer of (\ref{ch10:liminim}) is {\it unique}, then it is a
$k$-sparse vector with
 \[
 k\le N.
 \]
This is a direct consequence of the Remark~\ref{rem:zeroes.dim.null},
and the fact that for the matrix $X$,
\begin{equation*}
\dim(\nullsp(X))= l-\rank(X) = l-N.
\end{equation*}
Hence, the number of the nonzero elements of the unique minimizer must
be at most equal to $N$.

If one resorts to geometry, all the previously stated results become
crystal clear.
\end{remarks}

\subsubsection{Geometric interpretation}

\begin{figure}[!tbp]
                \centering
                \includegraphics[scale=1]{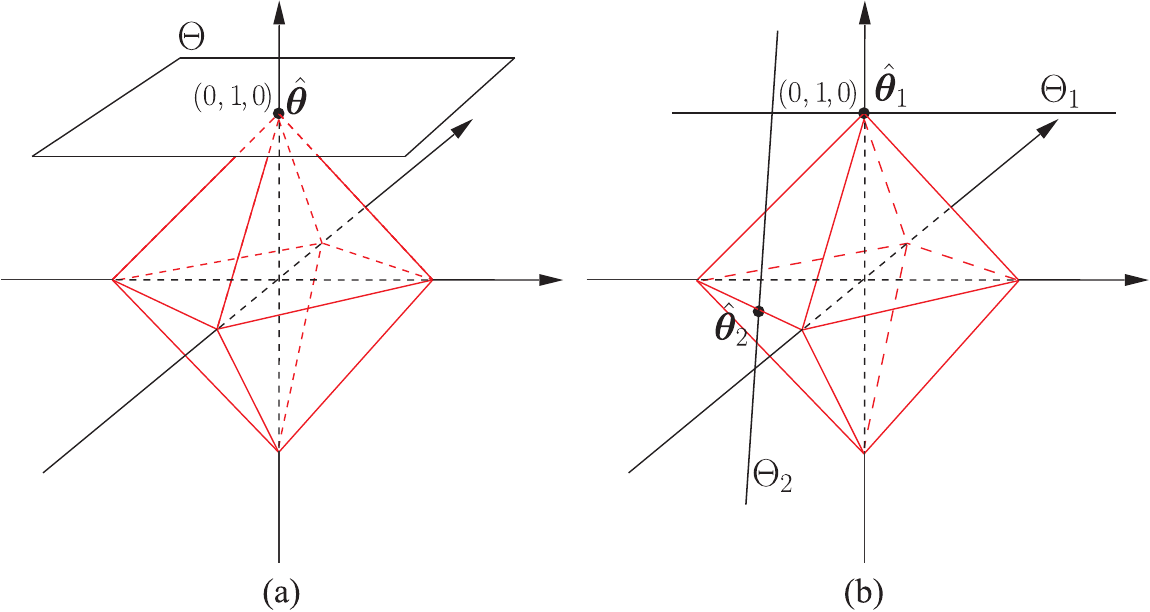}
        \caption{(a) The $\ell_1$ ball intersecting with a plane. The only possible scenario, for the existence of a unique
  common intersecting point of the $\ell_1$ ball with a plane in the
  Euclidean $\Real^3$ space, is for the point to be located at one of the
  corners of the $\ell_1$ ball, i.e., to be an $1$-sparse vector. (b) The $\ell_1$ ball intersecting with lines.
In this case, the sparsity level of the unique intersecting point is relaxed; it
could be an $1$- or a $2$-sparse vector.}\label{fig:l1.ball.and.Theta}
\end{figure}

Assume that our target solution resides in the $3$-dimensional space
and that we are given one measurement
\[
y_1=\bm{x}_1^T\bm{\theta}=x_{11}\theta_1+x_{12}\theta_2+x_{13}\theta_3.
\]
Then the solution lies in the $2$-dimensional (hyper)plane, which is
described by the previous equation. To get the minimal $\ell_1$
solution we keep increasing the size of the $\ell_1$ ball\footnote{Observe that in the $3$-dimensional space the $\ell_1$ ball looks
  like a diamond.} (the set of all points that have equal $\ell_1$ norm)
till it touches this plane. The only way that these two geometric objects
have a single point in common (unique solution) is when they meet at a
corner of the diamond. This is shown in Fig.~\ref{fig:l1.ball.and.Theta}a. In
other words, the resulting solution is $1$-sparse, having two of its
components equal to zero. This complies with the finding stated in
Remark~\ref{rem:k.leq.N}, since now $N=1$. For any other orientation of the
plane, this will either cut across the $\ell_1$ ball or will share with the
diamond an edge or a side. In both cases, there will be infinite many
solutions.


Let us now assume that we are given an extra measurement,
\[
y_2=x_{21}\theta_1+x_{22}\theta_2+x_{23}\theta_3.
\]
The solution now lies in the intersection of the two previous planes, which
is a straight line. However, now, we have more alternatives for a unique
solution. A line, e.g., $\Theta_1$, can either touch the $\ell_1$ ball at a
corner ($1$-sparse solution) or, as it is shown in
Fig.~\ref{fig:l1.ball.and.Theta}b, it can touch the $\ell_1$ ball at one of its
edges, e.g., $\Theta_2$. The latter case, corresponds to a solution that
lies on a $2$-dimensional subspace, hence it will be a $2$-sparse
vector. This also complies with the findings stated in
Remark~\ref{rem:k.leq.N}, since in this case, we have $N=2$, $l=3$ and the
sparsity level for a unique solution can be either $1$ or $2$.

Note that uniqueness is associated with the particular geometry and
orientation of the affine set, which is the set of all possible solutions
of the underdetermined system of equations. For the case of the square
$\ell_2$ norm, the solution was always unique. This is a consequence of the
(hyper)spherical shape formed by the Euclidean norm. From a mathematical
point of view, the square $\ell_2$ norm is a strict convex function. This
is not the case for the $\ell_1$ norm, which is convex, albeit not a strict
convex function.

\begin{figure}[!tbp]
   \centering
   \includegraphics[scale=1]{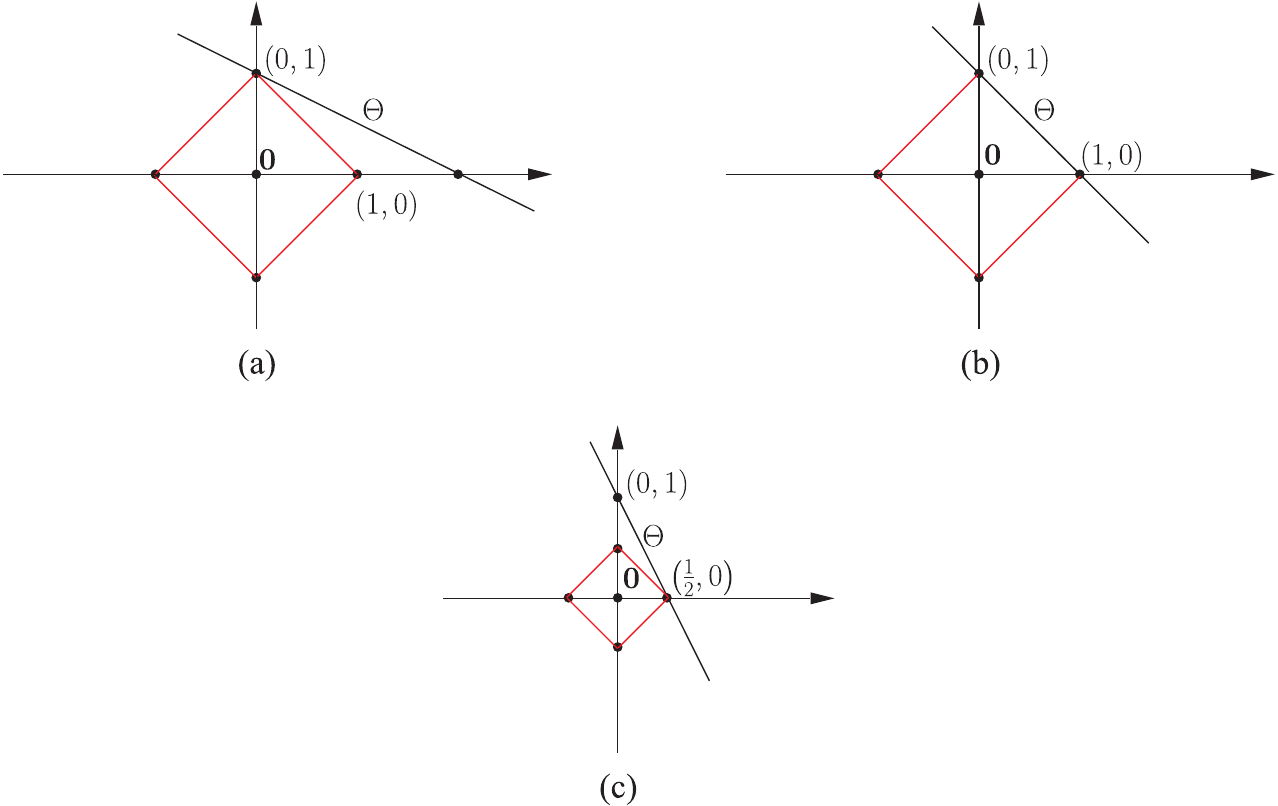}
   \caption{(a) Sensing with $\bm{x}
       = [\frac{1}{2},1]^T$, (b) sensing with $\bm{x}
       = [1, 1]^T$, (c)
     sensing with $\bm{x} = [2, 1]^T$. The choice of the sensing
     vector $\bm{x}$ is crucial to unveiling the true sparse solution
     $(0,1)$. Only the sensing vector $\bm{x} = [\frac{1}{2},1]^T$
     identifies uniquely the desired $(0,1)$.}\label{fig:example.l1.min}
\end{figure}

\begin{example}\label{example:l1.min}
Consider a sparse vector parameter $[0,1]^T$, which we assume to be
unknown. We will use one measurement to {\it sense} it. Based on this
single measurement, we will use the $\ell_1$ minimizer of
\eqref{ch10:liminim} to recover its true value. Let us see what happens. We
will consider three different values of the ``sensing'' (input) vector
$\bm{x}$ in order to obtain the measurement $y=\bm{x}^T\bm{\theta}$: a)
$\bm{x}=[\frac{1}{2}, 1]^T$, b) $\bm{x}=[1, 1]^T$, and c) $\bm{x}=[2,
  1]^T$. The resulting measurement, after sensing $\bm{\theta}$ by
$\bm{x}$, is $y=1$ for all the three previous cases.

Case a): The solution will lie on the straight line
\[
\Theta = \left\{[\theta_1, \theta_2]^T\in \Real^2 :
\frac{1}{2}\theta_1+\theta_2=1\right\},
\]
which is shown in Fig.~\ref{fig:example.l1.min}a. For this setting,
expanding the $\ell_1$ ball, this will touch the line (our solutions' affine
set) at the corner $[0,1]^T$. This is a unique solution, hence it is
sparse, and it coincides with the true value.

Case b): The solutions lies on the straight line
\begin{equation*}
\Theta = \left\{[\theta_1, \theta_2]^T\in \Real^2 :
\theta_1+\theta_2=1\right\},
\end{equation*}
which is shown in Fig.~\ref{fig:example.l1.min}b. For this set up,
there is an infinite number of solutions, including two sparse ones.

Case c): The affine set of solutions is described by
\[
\Theta= \left\{[\theta_1, \theta_2]^T\in \Real^2:
2\theta_1+\theta_2=1\right\},
\]
which is sketched in Fig.~\ref{fig:example.l1.min}c. The solution in
this case is sparse, but it is not the correct one.

This example is quite informative. \textit{If we sense (measure) our
  unknown parameter vector with appropriate sensing (input) data, the use
  of the $\ell_1$ norm can unveil the true value of the parameter vector,
  even if the system of equations is underdetermined, provided that the
  true parameter is sparse}. This now becomes our new goal. To investigate
whether what we have just said can be generalized, and under which
conditions holds true, if it does. In such a case, the choice of the
regressors (which we just called them sensing vectors) and hence the input matrix
(which, from now on, we will refer to, more and more frequently, as the
sensing matrix) acquire an extra significance. It is not enough for the
designer to care only for the rank of the matrix, i.e., the linear
independence of the sensing vectors. One has to make sure that the
corresponding affine set of the solutions has such an orientation, so that
the touch with the $\ell_1$ ball, as this increases from zero to meet this
plane, is a ``gentle'' one, i.e., they meet at a single point, and more
important at the correct one; that is, at the point that represents the
true value of the sparse parameter, which we are searching for.
\end{example}


\begin{remarks}\label{rem:normalize.X}\mbox{}
\begin{itemize}
\item Often in practice, the columns of the input matrix, $X$, are
  normalized to unit $\ell_2$ norm. Although $\ell_0$ norm is insensitive
  to the values of the nonzero components of $\bm{\theta}$, this is not the
  case with the $\ell_1$ and $\ell_2$ norms. Hence, while trying to
  minimize the respective norms, and at the same time to fulfill the
  constraints, components that correspond to columns of $X$ with high
  energy (norm) are favored more than the rest. Hence, the latter become
  more popular candidates to be pushed to zero. In order to avoid such situations, the
  columns of $X$ are normalized to unity, by dividing each element of the
  column vector by the respective (Euclidean) norm.
\end{itemize}
\end{remarks}

\section{Uniqueness of the $\ell_0$ Minimizer}\label{ch10:uniql0}

Our first goal is to derive {\it sufficient} conditions that guarantee
uniqueness of the $\ell_0$ minimizer, which has been defined in Section
\ref{Ch10:thesparsest}.

\begin{definition}\label{def:spark}
The \textit{spark}\index{Spark of a matrix} of a full rank $N\times l$ ($l\ge
N$) matrix, $X$, denoted as $\spark(X)$, is the \textit{smallest} number of
its linearly dependent columns.
\end{definition}

According to the previous definition, \textit{any} $m<\spark(X)$ columns of
$X$ are, necessarily, \textit{linearly independent}. The spark of a square,
$N\times N$, full rank matrix is equal to $N+1$.

\begin{remarks}\mbox{}
\begin{itemize}
\item In contrast to the rank of a matrix, which can be easily
determined, its spark can only be obtained by resorting to a
combinatorial search over all possible combinations of the columns of
the respective matrix, see, e.g., \cite{Fred-10, DonoEl-10}. The
notion of the spark was used in the context of sparse representation,
under the name of {\it Uniqueness Representation Property}, in
\cite{GorodRao-10}. The name ``spark'' was coined in
\cite{DonoEl-10}. An interesting discussion relating this matrix index
with other indices,  used in other disciplines, is given in
\cite{Fred-10}.
\end{itemize}
\end{remarks}

\begin{example}\label{example:spark}
Consider the following matrix
\[
X=\left [\begin{array}{cccccc}
1 & 0& 0& 0& 1& 0 \\
0 & 1& 0& 0& 1& 1 \\
0 & 0& 1& 0& 0& 1 \\
0 &0 & 0 &1& 0& 0 \end{array}\right ].
\]
The matrix has rank equal to 4 and spark equal to 3. Indeed, any pair of
columns are linearly independent. On the other hand, the
  first, the second and the fifth columns are linearly dependent. The same
  is also true for the combination of the second, third and sixth columns.
\end{example}

\begin{lemma}\label{lem:spark.geq.l0.norm}
If $\nullsp(X)$ is the null space of $X$, then
\begin{equation*}
\norm{\bm{\theta}}_0\ge \spark(X), \quad \forall \bm{\theta}\in \nullsp(X),
\bm{\theta}\ne \bm{0}.
\end{equation*}
\end{lemma}

\noindent\textit{Proof:} To derive a contradiction, assume that there
exists a $\bm{0}\neq \bm{\theta}\in\nullsp(X)$ such that
$\norm{\bm{\theta}}_0< \spark(X)$. Since by definition
$X\bm{\theta}=\bm{0}$, there exists a number of $\norm{\bm{\theta}}_0$
columns of $X$ that are linearly dependent. However, this contradicts the
minimality of $\spark(X)$, and the claim of
Lemma~\ref{lem:spark.geq.l0.norm} is established.

\begin{lemma}\label{lem:suff.cond.sparse.solution}
If a linear system of equations, $X\bm{\theta}=\bm{y}$, has a solution that
satisfies
\begin{equation*}
\norm{\bm{\theta}}_0 < \frac{1}{2}\spark(X),
\end{equation*}
then this is the
\textit{sparsest} possible solution. In other words, this is,
necessarily, the \textit{unique} solution of the $\ell_0$ minimizer.
\end{lemma}

\noindent\textit{Proof:} Consider any other solution $\bm{h} \neq
\bm{\theta}$. Then, $\bm{\theta}-\bm{h}\in\nullsp(X)$, i.e.,
\[
X(\bm{\theta}-\bm{h})=\bm{0}.
\]
Thus, according to Lemma~\ref{lem:spark.geq.l0.norm},
\beq \spark(X)\le
\norm{\bm{\theta}-\bm{h}}_0 \le \norm{\bm{\theta}}_0 +
\norm{\bm{h}}_0. \label{ch10:sprk2}
\eeq
Observe that although the $\ell_0$ ``norm'' is not a true norm, it can be
readily verified by simple inspection and reasoning that the triangular
property is satisfied. Indeed, by adding two vectors together, the
resulting number of nonzero elements will always be at most equal to the
total number of nonzero elements of the two vectors. Therefore, if
$\norm{\bm{\theta}}_0 < \frac{1}{2}\spark(X)$, then \eqref{ch10:sprk2}
suggests that
\begin{equation*}
\norm{\bm{h}}_0  >  \frac{1}{2}\spark(X) > \norm{\bm{\theta}}_0.
\end{equation*}

\begin{remarks}\label{rem:spark}\mbox{}
\begin{itemize}
\item Lemma~\ref{lem:suff.cond.sparse.solution} is a very interesting
  result. We have a sufficient condition to check whether a solution
  is the unique optimal in a, generally, NP-hard problem. Of course,
  although this is nice from a theoretical point of view, is not of
  much use by itself, since the related bound (the $\spark$) can only be
  obtained after a combinatorial search. Well, in the next section, we
  will see that we can relax the bound by involving another index, in
  place of the spark, which can be easily computed.
\item An obvious consequence of the previous lemma is that if the
  unknown parameter vector is a sparse one with $k$ nonzero elements,
  then if matrix $X$ is chosen so that to
  have $\spark(X)>2k$, then the true parameter vector is necessarily
  the sparsest one that satisfies the set of equations, and the
  (unique) solution to the $\ell_0$ minimizer.
\item In practice, the goal is to sense the unknown parameter vector
  by a matrix that has as high a spark as possible, so that the
  previously stated sufficiency condition to cover a wide range of
  cases. For example, if the spark of the input matrix is, say, equal
  to three, then one can check for optimal sparse solutions up to a
  sparsity level of $k=1$. From the respective definition, it is
  easily seen  that the values of the spark are in the range $1<
  \mbox{spark}(X)\le N+1$.
\item Constructing an $N\times l$ matrix $X$ in a random manner, by
  generating i.i.d entries, guarantees, with high probability, that
  $\spark(X)=N+1$; that is, any $N$ columns of the matrix are
  linearly independent.
\end{itemize}
\end{remarks}

\subsection{Mutual Coherence}\index{Mutual coherence of a
  matrix}\label{ch10mutcoh}

Since the spark of a matrix is a number that is difficult to compute,
our interest shifts to another index, which can be derived easier and at
the same time can offer a useful bound on the spark. The {\it mutual
  coherence} of an $N\times l$ matrix $X$ \cite{Mall-10}, denoted as
$\mu(X)$, is defined as
\begin{equation}
\mu(X) \coloneqq \max_{1\le i< j\le l} \frac{|\mathbf{x}_i^T
  \mathbf{x}_j|}{\norm{\mathbf{x}_i}
  \norm{\mathbf{x}_j}}, \label{ch10mutcoh-1}
\end{equation}
where $\mathbf{x}_i$, $i=1,2,\ldots, l$, denote the columns of $X$ (notice
the difference in notation between a row $\bm{x}_i^T$ and a column
$\mathbf{x}_i$ of the matrix $X$). This number reminds us of the
correlation coefficient between two random variables. Mutual coherence is
bounded as $0\le \mu(X)\le 1$.  For a square orthogonal matrix, $X$,
$\mu(X)=0$. For general matrices, with $l>N$, $\mu(X)$ satisfies
\begin{equation*}
\sqrt {\frac{l-N}{N(l-1)}}\le \mu(X) \le 1,
\end{equation*}
which is known as the {\it Welch bound}\index{Welch bound} \cite{Welch-10}. For large values of $l$, the lower bound becomes,
approximately, $\mu(X)\ge \frac{1}{\sqrt{N}}$. Common sense reasoning
guides us to construct input (sensing) matrices of mutual coherence as
small as possible. Indeed, the purpose of the sensing matrix is to
``measure'' the components of the unknown vector and ``store'' this
information in the measurement vector $\bm{y}$.  Thus, this should be done
in such a way so that $\bm{y}$ to retain as much information about the
components of $\bm{\theta}$ as possible.  This can be achieved if the
columns of the sensing matrix, $X$, are as ``independent'' as
possible. Indeed, $\bm{y}$ is the result of a combination of the columns of
$X$, each one weighted by a different component of $\bm{\theta}$. Thus, if
the columns are as much ``independent'' as possible then the information
regarding each component of $\bm{\theta}$ is contributed by a different
direction making its recovery easier.  This is easier understood if $X$ is
a square orthogonal matrix.  In the more general case of a non-square
matrix, the columns should be made as ``orthogonal'' as possible.

\begin{example}\label{example:mu.DFT}
Assume that $X$ is an $N\times 2N$ matrix, formed by concatenating two
orthonormal bases together,
\[
X=[I,W],
\]
where $I$ is the identity matrix, having as columns the vectors
$\bm{e}_i$, $i=1,2,\ldots, N$, with elements equal to
\[
 \delta_{ir}=
\begin{cases}
1, & \text{if}\ i=r,\\
0, & \text{if}\ i\ne r,
\end{cases}
 \]
for $r=1,2,\ldots,N$. The matrix $W$ is the orthonormal DFT matrix,
defined as
\[
W=\frac{1}{\sqrt{N}}
\begin{bmatrix}
1 & 1 & \ldots & 1 \\
1 & W_N & \ldots & W_N^{N-1} \\
\vdots & \vdots & \ddots & \vdots \\
1 & W_N^{N-1} & \ldots & W_N^{(N-1)(N-1)}
\end{bmatrix},
\]
where
\[
W_N \coloneqq \exp \left(-j\frac{2\pi}{N}\right).
\]
Such an overcomplete dictionary could be used to
represent signal vectors in terms of the expansion in
(\ref{ch-10signrepr}), that comprise the sum of sinusoids with very
narrow spiky-like pulses.  The inner products between
any two columns of $I$ and between any two columns of $W$ are zero,
due to orthogonality. On the other hand, it is easy to see that the
inner product between any column of $I$ and any column of $W$ has
absolute value equal to $\frac{1}{\sqrt{N}}$. Hence, the mutual
coherence of this matrix is $\mu(X)=\frac{1}{\sqrt{N}}$. Moreover,
observe that the spark of this matrix is $\spark(X)=
N+1$.
\end{example}

\begin{lemma}\label{lem:bound.spark.mu}
For any  $N\times l$ matrix $X$, the following inequality holds
\beq
\spark(X) \ge 1+\frac{1}{\mu(X)} \label{ch10spmu}.
\eeq
\end{lemma}

The proof is given in \cite{DonoEl-10} and it is based on arguments that
stem from matrix theory applied on the Gram matrix, $X^TX$, of $X$. A ``superficial'' look at the previous bound is that for very small
values of $\mu(X)$ the spark can be larger than $N+1$! Looking at the
proof, it is seen that in such cases the spark of the matrix attains its
maximum value $N+1$.

The result complies with a common sense reasoning. The smaller the value of
$\mu(X)$ the more independent are the columns of $X$, hence the higher the
value of its spark is expected to be.  Based on this lemma, we can now
state the following theorem, first given in \cite{DonoEl-10}. Combining the
way that Lemma \ref{lem:suff.cond.sparse.solution} is proved and
\eqref{ch10spmu}, we come to the following important theorem.

\begin{theorem}\label{thm:mu.bound.unique.l0}
If the linear system of equations in \eqref{ch10-thesp-1} has a
solution that satisfies the condition
\beq
\norm{\bm{\theta}}_0 < \frac{1}{2}\left (1+\frac{1}{\mu(X)}\right ), \label{ch10theor}
\eeq
then this solution is the \textit{sparsest one}.
\end{theorem}

\begin{remarks}\label{rem:mu}\mbox{}
\begin{itemize}
\item The bound in \eqref{ch10theor} is ``psychologically'' important. It
  relates an easily computed bound to check whether the solution to a
  NP-hard task is the optimal one. However, it is not a particularly good
  bound and it restricts the range of values in which it can be applied.
  As we saw in Example~\ref{example:mu.DFT}, while the maximum possible
  value of the spark of a matrix was equal to $N+1$, the minimum possible
  value of the mutual coherence was $\frac{1}{\sqrt{N}}$. Therefore, the
  bound based on the mutual coherence restricts the range of sparsity,
  i.e., $\norm{\bm{\theta}}_0$, where one can check optimality, to around
  $\frac{1}{2}\sqrt{N}$.  Moreover, as the previously stated Welch bound
  suggests, this $\mathcal{O}(\frac{1}{\sqrt{N}})$ dependence of the mutual
  coherence seems to be a more general trend and not only the case for
  Example~\ref{example:mu.DFT}, see, e.g., \cite{DonoHu-10}. On the other
  hand, as we have already stated in the Remarks~\ref{rem:spark} , one can
  construct random matrices with spark equal to $N+1$; hence, using the
  bound based on the spark, one could expand the range of sparse vectors up
  to $\frac{1}{2}N$.
\end{itemize}
\end{remarks}

\section[Equivalence of $\ell_0$ and $\ell_1$ Minimizers]{Equivalence
  of $\ell_0$ and $\ell_1$ Minimizers: Sufficiency
  Conditions} \label{ch10:equivl0l1}

We have now come to the crucial point and we will establish the
conditions that guarantee  the equivalence between the $\ell_1$ and the  $\ell_0$
minimizers.  Hence, under such conditions, a problem, that
is in general NP-hard problem, \textit{can be solved via a tractable
  convex optimization task}. Under these conditions, the zero value
encouraging nature of the $\ell_1$ norm, that has already been discussed,
obtains a much higher stature; it provides the sparsest solution.

\subsection{Condition Implied by the Mutual Coherence Number}

\begin{theorem}\label{thm:l0.l1.coincide}
Let the underdetermined system of equations
\begin{equation*}
\bm{y}=X\bm{\theta}, \label{ch10-10.3}
\end{equation*}
where $X$ is an $N\times l$ $(N<l)$ full row rank matrix. If a
solution exists and satisfies the condition
\beq
\norm{\bm{\theta}}_0<\frac{1}{2}\left (1+\frac{1}{\mu(X)}\right ),
\eeq
then this  is  the {\it unique} solution  of both, the $\ell_0$ as
well the $\ell_1$ minimizers.
\end{theorem}

This is a very important theorem and it was shown independently in
\cite{DonoEl-10, Grib-10}. Earlier versions of the theorem addressed the
special case of a dictionary comprising two orthonormal bases,
\cite{DonoHu-10, EladBr-10}. A proof is also summarized in \cite{Fred-10}. This theorem established, for a first time, what it was
till then empirically known: often, the $\ell_1$ and $\ell_0$ minimizers
result in the same solution.

\begin{remarks}\label{rem:l0.l1.coincide}\mbox{}
\begin{itemize}
\item The theory that we have presented so far is very satisfying, since it
  offers the theoretical framework and conditions that guarantee uniqueness
  of a sparse solution to an underdetermined system of equations. Now we
  know that, under certain conditions, the solution, which we obtain by
  solving the convex $\ell_1$ minimization task, is the (unique) sparsest
  one. However, from a practical point of view, the theory, which is based
  on mutual coherence, does not say the whole story and falls short to
  predict what happens in practice. Experimental evidence suggests that the
  range of sparsity levels, for which the $\ell_0$ and $\ell_1$ tasks give
  the same solution, is much wider than the range guaranteed by the mutual
  coherence bound. Hence, there is a lot of theoretical happening in order
  to improve this bound. A detailed discussion is beyond the scope of this paper. In the sequel, we will present one of these bounds, since it is the
  one that currently dominates the scene. For more details and a related discussion
  the interested reader may consult, e.g., \cite{DonoTan-10}.
\end{itemize}
\end{remarks}

\subsection{The Restricted Isometry Property (RIP)}\index{Restricted
  Isometry Property}\label{ch10isom}

\begin{definition}\label{def:rip}
For each integer $k=1,2,\ldots$, define the \textit{isometry
  constant}\index{Isometry constant} $\delta_k$ of an $N\times l$
matrix $X$ as the \textit{smallest} number such that
\begin{equation}
(1-\delta_k) \norm{\bm{\theta}}^2_2\le \norm{X\bm{\theta}}^2_2\le
(1+\delta_k) \norm{\bm{\theta}}_2^2, \label{ch10iso-1}
\end{equation}
holds true for \textit{all} $k$-sparse vectors $\bm{\theta}$.
\end{definition}

This definition was introduced in \cite{Cand-10}. We loosely say that
matrix $X$ obeys the RIP of order $k$ if $\delta_k$ is not too close to
one. When this property holds true, it implies that the Euclidean norm of
$\bm{\theta}$ is approximately {\it preserved}, after projecting it on the
rows of $X$. Obviously, if matrix $X$ were orthonormal then
$\delta_k=0$. Of course, since we are dealing with non-square matrices this
is not possible. However, the closer $\delta_k$ is to zero, the closer to
orthonormal {\it all} subsets of $k$ columns of $X$ are. Another view point
of \eqref{ch10iso-1} is that it preserves Euclidean distances between
$k$-sparse vectors. Let us consider two $k$-sparse vectors,
$\bm{\theta}_1$, $\bm{\theta}_2$ and apply \eqref{ch10iso-1} to their
difference $\bm{\theta}_1-\bm{\theta}_2$, which, in general, is a
$2k$-sparse vector. Then we obtain
\begin{equation}
 (1-\delta_{2k}) \norm{\bm{\theta}_1-\bm{\theta}_2}^2_2\le
\norm{X(\bm{\theta}_1-\bm{\theta}_2)}^2_2\le
(1+\delta_{2k}) \norm{\bm{\theta}_1-\bm{\theta}_2}_2^2. \label{ch10iso-1b}
 \end{equation}
Thus, when $\delta_{2k}$ is small enough, the Euclidean distance is
preserved after projection in the lower dimensional measurements' space. In
words, if the RIP holds true, this means that searching for a sparse vector
in the lower dimensional subspace formed by the measurements, $\Real^N$,
and not in the original $l$-dimensional space, one can still recover the
vector since distances are preserved and the target vector is not
``confused'' with others. After projection on the rows of $X$, the
discriminatory power of the method is retained. It is interesting to point
out that the RIP is also related to the condition number of the Grammian
matrix. In \cite{Cand-10,Bara-10}, it is pointed out that if $X_r$ denotes
the matrix that results by considering only $r$ of the columns of $X$, then
the RIP in (\ref{ch10iso-1}) is equivalent with requiring the respective
Grammian, $X^T_rX_r$, $r\le k$, to have its eigenvalues within the interval
$[1-\delta_k,1+\delta_k$]. Hence, the more well conditioned the matrix is,
the better is for us to dig out the information hidden in the lower
dimensional measurements space.

\begin{theorem}\label{thm:cond.l1.equiv.l0}
Assume that for some $k$,  $\delta_{2k}<\sqrt{2}-1$. Then the solution to the
$\ell_1$ minimizer of \eqref{ch10:liminim}, denoted as
$\bm{\theta}_*$, satisfies the following two conditions
\begin{equation}
\norm{\bm{\theta}-\bm{\theta}_*}_1\le C_0
\norm{\bm{\theta}-\bm{\theta}_k}_1, \label{ch10isom2a}
\end{equation}
and
\begin{equation}
\norm{\bm{\theta}-\bm{\theta}_*}_2\le
C_0k^{-\frac{1}{2}}\norm{\bm{\theta}-\bm{\theta}_k}_1, \label{ch10isom2}
\end{equation}
for some constant $C_0$. In the previously stated formulas, $\bm{\theta}$
is the true (target) vector that generates the measurements in
\eqref{ch10:liminim} and $\bm{\theta}_k$ is the vector that results from
$\bm{\theta}$ if we keep its $k$ largest components and set the rest equal
to zero, \cite{Cand-10, Cand1-10, Cand2-10, Cand3-10}.
\end{theorem}

Hence, if the true vector is a sparse one, i.e.,
$\bm{\theta}=\bm{\theta}_k$, then the $\ell_1$ minimizer recovers the
(unique) exact value. On the other hand, if the true vector is not a sparse
one, then the minimizer results in a solution whose accuracy is dictated by
a genie-aided procedure that knew in advance the locations of the $k$
largest components of $\bm{\theta}$.  This is a groundbreaking result.
Moreover, it is deterministic, it is always true and not with high
probability. Note that the isometry property of order $2k$ is used, since
at the heart of the method lies our desire to preserve the norm of the
differences between vectors.

Let us now focus on the case where there is a $k$-sparse vector that
generates the measurements, i.e., $\bm{\theta}= \bm{\theta}_k$. Then it is
shown in \cite{Cand3-10} that the condition $\delta_{2k}<1$ guarantees that
the $\ell_0$ minimizer has a unique $k$-sparse solution. In other words, in
order to get the equivalence between the $\ell_1$ and $\ell_0$ minimizers,
the range of values for $\delta_{2k}$ has to be decreased to
$\delta_{2k}<\sqrt{2}-1$, according to
Theorem~\ref{thm:cond.l1.equiv.l0}. This sounds reasonable. If we relax the
criterion and use $\ell_1$ instead of $\ell_0$, then the sensing matrix has
to be more carefully constructed. Although we are not going to provide the
proofs of these theorems here, since their formulation is well beyond the
scope of this paper, it is interesting to follow what happens if
$\delta_{2k}=1$. This will give us a flavor of the essence behind the
proofs. If $\delta_{2k}=1$, the left hand side term in \eqref{ch10iso-1b}
becomes zero. In this case, there may exist two $k$-sparse vectors
$\bm{\theta}_1, \bm{\theta}_2$ such that $X(\bm{\theta}_1 - \bm{\theta}_2)=
\bm{0}$, or $X\bm{\theta}_1=X\bm{\theta}_2$. Thus, it is not possible to
recover all $k$-sparse vectors, after projecting them in the measurements
space, by any method.

The previous argument also establishes a connection between RIP and the
spark of a matrix. Indeed, if $\delta_{2k}<1$, this guarantees that any
number of columns of $X$ up to $2k$ are linearly independent, since for any
$2k$-sparse $\bm{\theta}$, \eqref{ch10iso-1} guarantees that
$\norm{X\bm{\theta}}_2>0$. This implies that $\spark(X)>2k$. A connection
between RIP and the coherence is established in \cite{Cai-10}, where it is
shown that if $X$ has coherence $\mu(X)$, and unit norm columns, then $X$
satisfies the RIP of order $k$ with $\delta_k$, where
$\delta_k\le(k-1)\mu(X)$.

\subsubsection{Constructing Matrices that Obey the RIP of order
  $k$}\label{ch10:randommatr}

It is apparent from our previous discussion, that the higher the value
of $k$, for which the RIP property of a matrix, $X$, holds true, the
better, since a larger range of sparsity levels can be handled. Hence,
a main goal towards this direction is to construct such matrices. It
turns out that verifying the RIP for a matrix of a general structure
is a difficult task. This reminds us of the spark of the matrix, which
is also a difficult task to compute.  However, it turns out that for a
certain class of random matrices, the RIP follows fairly easy. Thus,
constructing such sensing matrices has dominated the scene of related
research.  We will present a few examples of such matrices, which are
also very popular in practice, without going into details of the
proofs, since this is out of our scope and the interested
reader may dig this information from the related references.

Perhaps, the most well known example of a random matrix is the
Gaussian one, where the entries $X(i,j)$ of the sensing matrix are
i.i.d.\ realizations from a Gaussian pdf
$\mathcal{N}(0,\frac{1}{N})$. Another popular example of such matrices
is constructed by sampling i.i.d.\ entries from a Bernoulli, or related,
distributions

\[
X(i,j)=\left\{
\begin{alignedat}{3}
 &\frac{1}{\sqrt{N}}, & \quad \text{with probability}\ \frac{1}{2}, \\
 -&\frac{1}{\sqrt{N}}, & \quad \text{with probability}\ \frac{1}{2},
\end{alignedat} \right.
\]

or
\[
X(i,j) =\left\{
\begin{alignedat}{3}
+\sqrt{\frac{3}{N}}&, & \quad \text{with probability}\ \frac{1}{6},\\
0&, & \quad \text{with probability}\ \frac{2}{3},\\
-\sqrt{\frac{3}{N}}&, & \quad \text{with probability}\ \frac{1}{6}.
\end{alignedat} \right.
\]

Finally, one can adopt the uniform distribution and construct the
columns of $X$ by sampling uniformly at random on the unit sphere in
$\Real^N$. It turns out,  that such
matrices obey the RIP of order $k$, with overwhelming probability, provided that the number of
measurements, $N$, satisfy the following inequality
\begin{equation}
N\ge Ck\ln(l/k), \label{ch10ripra1}
\end{equation}
where $C$ is some constant, which depends on the isometry constant
$\delta_k$. In words, having such a matrix at our disposal, one can
recover a $k$-sparse vector from $N<l$ measurements, where $N$ is
larger than the sparsity level by an amount controlled by the
inequality \eqref{ch10ripra1}. More on these issues can be
obtained from, e.g., \cite{Bara-10, Mend-10}.

Besides random matrices, one can construct other matrices that obey
the RIP. One such example includes the partial Fourier matrices,
 which are formed by selecting uniformly at random $N$
  rows drawn from the $l\times l$ DFT matrix. Although the required
number of samples for the RIP to be satisfied may be larger than the bound
in \eqref{ch10ripra1} (see, \cite{Rud-10}), Fourier-based sensing
matrices offer certain computational advantages, when it comes to
storage ($\mathcal{O}(N\ln l$)) and matrix-vector products
($\mathcal{O}(l\ln l)$), \cite{Cand1b}. In \cite{Haup-10}, the case of
random Toeplitz sensing matrices, containing statistical dependencies
across rows, is considered and it is shown that they can also satisfy
the RIP with high probability. This is of particular importance in
signal processing and communications applications, where it is very
common for a system to be excited in its input via a time series,
hence independence between successive input rows cannot be assumed. In
\cite{Rive-10, Duartesep-10}, the case of separable matrices is considered where the
sensing matrix is the result of a Kronecker product of matrices, which
satisfy the RIP individually.  Such matrices are of interest for
multidimensional signals, in order to exploit the sparsity structure
along each one of the involved dimensions. For example, such signals
may occur while trying to ``encode'' information associated with an
event whose activity spreads across the temporal, spectral, spatial,
etc., domains.

In spite of their theoretical elegance, the derived bounds, that
determine the number of the required measurements for certain sparsity
levels, fall short of what is the experimental evidence, e.g., \cite{DonoTan-10}. In practice,
a rule of thumb is to use $N$ of the order of $3k$-$5k$, e.g.,
\cite{Cand3-10}. For large values of $l$, compared to the sparsity
level, the analysis in \cite{Donotu-10} suggests that we can recover
most sparse signals when  $N\approx 2k\ln (l/N)$. In an effort to overcome the shortcomings associated with the RIP, a number of other techniques have been proposed, e.g. \cite{Cohe-09,Bick-10,Tang-11,DonoTan-10}. Furthermore, in specific
applications, the use of an empirical study may be a more appropriate
path.

Note that, in principle, the minimum number of measurements that are
required to recover a $k$ sparse vector from $N<l$ measurements is $N\ge
2k$. Indeed, in the spirit of the discussion after
Theorem~\ref{thm:cond.l1.equiv.l0}, the main requirement that a sensing
matrix must fulfil is the following: not to map two different $k$-sparse
vectors to the same measurement vector $\bm{y}$. Otherwise, one can never
recover both vectors from their (common) measurements.  If we have $2k$
measurements and a sensing matrix that guarantees that any $2k$ columns are
linearly independent, then the previously stated requirement is readily
seen that it is satisfied. However, the bounds on the number of
measurements set in order the respective matrices to satisfy the RIP are
higher. This is because RIP accounts also for the stability of the recovery
process. We will come to this issue soon, in Section \ref{ch10compressive},
where we talk about \textit{stable} embeddings.

\section[Recovery from Noisy Measurements]{Robust Sparse Signal
  Recovery from Noisy Measurements}\label{ch10-robrecovery}

In the previous section, our focus was on recovering a sparse solution
from an underdetermined system of equations. In the formulation of the
problem, we assumed that there is no noise in the obtained
measurements. Having acquired a lot of experience and insight from a
simpler problem, we now turn our attention to the more realistic task,
where uncertainties come into the scene. One type of uncertainty may
be due to the presence of noise and our measurements' model comes back
to the standard regression form
\beq
\bm{y}=X\bm{\theta}+\bm{\eta}, \label{ch10:robrec1}
\eeq
where $X$ is our familiar  non-square $N\times l$ matrix. A sparsity-aware
formulation for recovering $\bm{\theta}$ from
\eqref{ch10:robrec1} can be cast as
\beqa
\min_{\bm{\theta}\in \Real^l} && \norm{\bm{\theta}}_1 \nonumber\\
\text{s.t.} && \norm{\bm{y}-X\bm{\theta}}_2^2\le
\epsilon, \label{ch10lassorip}
\eeqa
 which coincides with the LASSO task given in
\eqref{ch10lassos3}.  Such a formulation implicitly assumes that the
noise is bounded and the respective range of values is controlled by
$\epsilon$.  One can consider a number of different variants. For
example, one possibility would be to minimize the $\norm{\cdot}_0$
norm instead of the $\norm{\cdot}_1$, albeit loosing the computational
elegance of the latter. An alternative route would be to replace
the Euclidean norm in the constraints with another one.

Besides the
presence of noise, one could see the previous formulation from a
different perspective. The unknown parameter vector, $\bm{\theta}$,
may not be exactly sparse, but it may consist of a few large
components, while the rest are small and close to, yet not necessarily
equal to, zero. Such a model misfit can be accommodated by allowing a
deviation of $\bm{y}$ from $X\bm{\theta}$.

In this relaxed setting of a sparse solution recovery, the notions of
uniqueness and equivalence, concerning the $\ell_0$ and $\ell_1$ solutions,
no longer apply. Instead, the issue that now gains in importance is that of
\textit{stability} of the solution.  To this end, we focus on the
computationally attractive $\ell_1$ task.  The counterpart of Theorem
\ref{thm:cond.l1.equiv.l0} is now expressed as follows.

\begin{theorem}\label{thm:bound.approxim.robust.sparse}
Assume that the sensing matrix, $X$, obeys the RIP with
$\delta_{2k}<\sqrt{2}-1$, for some $k$.  Then the solution $\bm{\theta}_*$ of
\eqref{ch10lassorip} satisfies the following (\cite{Cand1-10,
  Cand2-10}),
\beq
\norm{\bm{\theta}-\bm{\theta}_*}_2\le
C_0k^{-\frac{1}{2}} \norm{\bm{\theta}-\bm{\theta}_k}_1 +
C_1 \sqrt{\epsilon}, \label{ch10-lassperfrip}
\eeq
for some constants $C_1$, $C_0$.
\end{theorem}

This is also an elegant result. If the model is exact and $\epsilon=0$
we obtain (\ref{ch10isom2}). If not, the higher the uncertainty
(noise) term in the model, the higher our ambiguity about the
solution. Note, also, that the ambiguity about the solution depends on
how far the true model is from $\bm{\theta}_k$. If the true model is
$k$-sparse, the first term on the right hand side of the inequality is
zero. The values of $C_1, C_0$ depend on $\delta_{2k}$ but they are
small, e.g., close to five or six, \cite{Cand2-10}.

The important conclusion, here, is that \textit{the LASSO formulation for
  solving inverse problems (which in general tend to be ill-conditioned) is a stable one and the noise
  is not amplified excessively during the recovery process}.

\section{Compressed Sensing: The Glory of
  Randomness}{\index{Compressed sensing}\label{ch10compressive}

The way in which this paper was deplored followed, more or less, the
sequence of developments that took place during the evolution of the sparsity-aware parameter estimation field. We intentionally made an
effort to follow such a path, since this is also indicative of how science
evolves in most cases. The starting point had a rather strong mathematical
flavour: to develop conditions for the solution of an underdetermined
linear system of equations, under the sparsity constraint and in a
mathematically tractable way, i.e., using convex optimization. In the end,
the accumulation of a sequence of individual contributions revealed that
the solution can be (uniquely) recovered if the unknown quantity is sensed
via randomly chosen data samples.  This development has, in turn, given
birth to a new field with strong theoretical interest as well as with an
enormous impact on practical applications. This new emerged area is known
as {\it compressed sensing} or {\it compressive sampling} (CS). Although CS
builds around the LASSO and Basis Pursuit (and variants of them, as we will
soon see), it has changed our view on how to sense and process signals
efficiently.

\subsubsection{Compressed Sensing}

In compressed sensing, the goal is to directly acquire as few samples as
possible that encode the minimum information, which is needed to obtain a
compressed signal representation.  In order to demonstrate this, let us
return to the data compression example, which was discussed in Section
\ref{ch10:sparse}. There, it was commented that the ``classical'' approach
to compression was rather unorthodox, in the sense that first all (i.e., a
number of $l$) samples of the signal are used, then they are processed to obtain
$l$ transformed values, from which only a small subset is used for
coding. In the CS setting, the procedure changes to the following one.

Let $X$ be an $N\times l$ sensing matrix, which is applied to the (unknown)
signal vector, $\bm{s}$, in order to obtain the measurements, $\bm{y}$, and $\Psi$ be
the dictionary matrix that describes the domain where the signal $\bm{s}$
accepts a sparse representation, i.e.,
\beqa
\bm{s}&=&\Psi\bm{\theta}, \nonumber \\
\bm{y}&=&X\bm{s}. \label{ch10sens}
\eeqa
Assuming that at most $k$ of the components of $\bm{\theta}$ are
nonzero, this can be obtained by the following optimization task
\beqa
\min_{\bm{\theta}\in \Real^l} & & \norm{\bm{\theta}}_1 \nonumber\\
\text{s.t.}& & \bm{y}=X\Psi\bm{\theta}, \label{ch10sens-1a}
\eeqa
\textit{provided that the combined matrix} $X\Psi$ {\it complies} with
the RIP and the number of measurements, $N$, satisfies the associated
bound given in \eqref{ch10ripra1}. Note that $\bm{s}$ needs not to be
stored and can be obtained any time, once $\bm{\theta}$ is
known. Moreover, as we will soon discuss, the measurements, $y_n$,
$n=1,2,\ldots,N$, can be acquired directly from an analogue signal
$s(t)$, prior to obtaining
its sample (vector) version, $\bm{s}$! Thus, from such a perspective,
CS fuses the data acquisition and the compression steps together.

There are different ways to obtain a sensing matrix, $X$, that leads
to a product $X\Psi$, which satisfies the RIP. It can be shown,
that if $\Psi$ is orthonormal and $X$ is a random
matrix, which is constructed as discussed at the end of Section
\ref{ch10isom}, then the product $X\Psi$ obeys the RIP, provided that
\eqref{ch10ripra1} is satisfied, \cite{Cand2-10}. An alternative way to obtain a
combined matrix, that respects the RIP, is to consider another
orthonormal matrix $\Phi$, whose columns have low coherence with the
columns of $\Psi$ (coherence between two matrices is defined in
\eqref{ch10mutcoh-1}, where, now, the pace of $\mathbf{x}_i$ is taken by a column of
$\Phi$ and that of $\mathbf{x}_j$ by a column of $\Psi$). For example, $\Phi$ could be the DFT matrix and $\Psi=I$ or vice versa. Then choose $N$ rows of
$\Phi$ uniformly at random to form $X$ in \eqref{ch10sens}. In other
words, for such a case, the sensing matrix can be written as $R\Phi$,
where $R$ is a $N\times l$ matrix that extracts $N$ coordinates
uniformly at random. The notion of incoherence (low coherence) between
the sensing and the basis matrices is closely related to RIP. The more
incoherent the two matrices are, the less the number of the required
measurements for the RIP to hold,
e.g., \cite{Cand1a-10, Rud-10}. Another way to view incoherence is
that the rows of $\Phi$ cannot be sparsely represented in terms of the
columns of $\Psi$. It turns out that if the sensing matrix $X$ is a
random one, formed as it has already been described in Section
\ref{ch10:randommatr}, then RIP and the incoherence with any $\Psi$
are satisfied with high probability.

The news get even better to say that all the previously stated
philosophy can be extended to the more general type of signals, which
are not, necessarily, sparse or sparsely represented in terms of the
atoms of a dictionary, and
they are known as {\it compressible}\index{Compressible signals}. A
signal vector is said to be compressible if its expansion in terms of
a basis consists of just a
few large coefficients $\theta_i$ and the rest are small. In other
words, the signal vector is {\it approximately} sparse in some
basis. Obviously, this is the most interesting case in practice, where
exact sparsity is scarcely (if ever) met. Reformulating the arguments
used in Section \ref{ch10-robrecovery}, the CS task for this case can
be cast as:
\beqa
\min_{\bm{\theta}\in \Real^l} & & \norm{\bm{\theta}}_1 \nonumber \\
\text{s.t.} & & \norm{\bm{y}-X\Psi\bm{\theta}}_2^2\le \epsilon, \label{ch10sens-1}
\eeqa
and everything that has been said in Section \ref{ch10-robrecovery} is
also valid for this case, if in place of $X$ we consider the product
$X\Psi$.

\begin{remarks}\label{rem:cs}
\begin{itemize}\mbox{}
\item An important property in compressed sensing is that the sensing
  matrix, which provides the measurements, may be chosen independently on
  the matrix $\Psi$; that is, the basis/dictionary in which the signal is
  sparsely represented. In other words, the sensing matrix can be
  ``universal'' and can be used to provide the measurements for
  reconstructing any sparse or sparsely represented signal in any
  dictionary, provided RIP is not violated.
\item Each measurement, $y_n$, is the result of an inner product
  (projection) of the signal vector with a row, $\bm{x}_n^T$, of the
  sensing matrix, $X$. Assuming that the signal vector, $\bm{s}$, is the
  result of a sampling process on an analogue signal, $s(t)$, then $y_n$
  can be directly obtained, to a good approximation, by taking the inner
  product (integral) of $s(t)$ with a sensing waveform, $x_n(t)$, that
  corresponds to $\bm{x}_n$. For example, if $X$ is formed by $\pm 1$, as
  described in Section \ref{ch10:randommatr}, then the configuration shown
  in Fig.~\ref{fig:analogue.sampling} can result to $y_n$. An important
  aspect of this approach, besides avoiding to compute and store the $l$
  components of $\bm{s}$, is that multiplying by $\pm 1$ is a relatively
  easy operation. It is equivalent with changing the polarity of the signal
  and it can be implemented by employing inverters and mixers. It is a
  process that can be performed, in practice, at much higher rates than
  sampling (we will come to it soon). If such a scenario is adopted, one
  could obtain measurements of an analogue signal at much lower rates than
  required for classical sampling, since $N$ is much lower than $l$. The
  only condition is that the signal vector must be sparse, in some
  dictionary, which may not necessarily be known during the data
  acquisition phase. Knowledge of the dictionary is required only during
  the reconstruction of $\bm{s}$. Thus, in the CS rationale, processing
  complexity is removed from the ``front end'' and is transferred to the
  ``back end'', by exploiting $\ell_1$ optimization.

\begin{figure}[!tbp]
\centering
\includegraphics[scale=1]{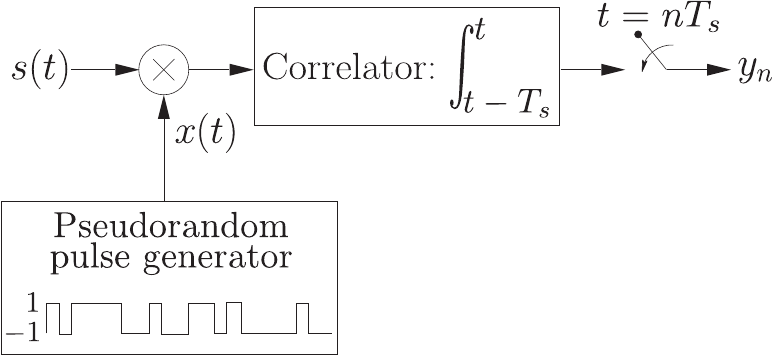}
\caption{Sampling an analogue signal $s(t)$ in order to generate the
  measurement $y_n$ at the time instant $n$. The sampling period $T_s$ is much lower than that required by the Nyquist sampling.}\label{fig:analogue.sampling}
\end{figure}

One of the very first applications that were inspired by the previous
approach, is the so-called {\it one pixel camera}\index{One pixel
  camera} \cite{Takh-10}. This was one among the most catalytic
examples, that spread the rumour about the practical power of CS. CS is
an example of what is commonly said: ``There is nothing more practical
than a good theory''!
\end{itemize}

\end{remarks}

\subsubsection{Dimensionality Reduction and Stable Embeddings}

We are now going to shed light to what we have said so far in this paper
from a different view. In both cases, either when the unknown quantity was
a $k$-sparse vector in a high dimensional space, $\Real^l$, or if the
signal $\bm{s}$ was (approximately) sparsely represented in some dictionary
($\bm{s}=\Psi\bm{\theta}$), we chose to work in a lower dimensional space
($\Real^N$); that is, the space of the measurements, $\bm{y}$. This is a
typical task of dimensionality reduction. The main task in any (linear)
dimensionality reduction technique is to choose the proper matrix $X$, that
dictates the projection to the lower dimensional space. In general, there
is always a loss of information by projecting from $\Real^l$ to $\Real^N$,
with $N<l$, in the sense that we cannot recover any vector,
$\bm{\theta}_l\in \Real^l$, from its projection $\bm{\theta}_N\in
\Real^N$. Indeed, take any vector $\bm{\theta}_{l-N}\in \nullsp(X)$, that
lies in the $(l-N)$-dimensional null space of the (full rank) $X$ (see
Section \ref{Ch10:thesparsest}).  Then, all vectors
$\bm{\theta}_l+\bm{\theta}_{l-N}\in \Real^l$ share the same projection in
$\Real^N$. However, what we have discovered in our tour in this paper is
that if the original vector is sparse then we can recover it exactly. This
is because all the $k$-sparse vectors do not lie anywhere in $\Real^l$, but
rather in a subset of it; that is, in the \textit{union of subspaces},
each one having dimensionality $k$.  If the signal $\bm{s}$ is sparse in
some dictionary $\Psi$, then one has to search for it in the union of all
possible $k$-dimensional subspaces of $\Real^l$, which are spanned by $k$
column vectors from $\Psi$, \cite{Bara1-10, Lu-10}. Of course, even in this
case, where sparse vectors are involved, not any projection can guarantee
unique recovery. The guarantee is provided if the projection in the lower
dimensional space is a \textit{stable embedding}\index{Stable
  embeddings}. A stable embedding in a lower dimensional space must
guarantee that if $\bm{\theta}_1\ne \bm{\theta}_2$, then their projections
remain also different. Yet this is not enough. A stable embedding must
guarantee that distances are (approximately) preserved; that is, vectors
that lie far apart in the high dimensional space, have projections that
also lie far apart. Such a property guarantees robustness to noise. Well,
the sufficient conditions, which have been derived and discussed throughout
this paper, and guarantee the recovery of a sparse vector lying in
$\Real^l$ from its projections in $\Real^N$, are conditions that guarantee
stable embeddings.  The RIP and the associated bound on $N$ provides a
condition on $X$ that leads to stable embeddings. We commented on this
norm-preserving property of RIP in the related section. The interesting
fact that came out from the theory is that we can achieve such stable
embeddings via random projection matrices.

Random projections for dimensionality reduction are not new and have
extensively been used in pattern recognition, clustering and data
mining, see, e.g., \cite{Achli-10, Blum-10, Dasg-10, Theod-10}. More
recently, the spirit underlying compressed sensing has been exploited
in the context of pattern recognition, too. In this application, one
needs not to return to the original high dimensional space, after the
information-digging activity in the low dimensional measurements
subspace. Since the focus in pattern recognition is to identify the
class of an object/pattern, this can be performed in the measurements
subspace, provided that there is no class-related information loss.
In \cite{Cald-10}, it is shown, using compressed sensing arguments,
that if the data is approximately linearly separable in the original
high dimensional space and the data has a sparse
representation, even in an unknown basis, then projecting randomly in
the measurements subspace retains the structure of linear
separability.

\textit{Manifold learning}\index{manifold learning} is another area where
random projections have been recently applied. A manifold is, in general, a nonlinear $k$-dimensional surface, embedded in a higher dimensional
(ambient) space. For example, the surface of a sphere is a two-dimensional
manifold in a three-dimensional space. More on linear and nonlinear
techniques for manifold learning can be found in, e.g., \cite{Theod-10}. In
\cite{Wakin-10, Baramanif-10}, the compressed sensing rationale is extended
to signal vectors that live along a $k$-dimensional submanifold of the
space $\Real^l$. It is shown that choosing a matrix, $X$, to project and a
sufficient number, $N$, of measurements, then the corresponding submanifold
has a stable embedding in the measurements subspace, under the projection
matrix, $X$; that is, pairwise Euclidean and geodesic distances are
approximately preserved after the projection mapping.  More on these issues
can be found in the given references and in, e.g., \cite{Bara1-10}.

\subsubsection{Sub-Nyquist Sampling: Analog-to-Information Conversion}

In our discussion in the Remarks presented before, we touched a very
important issue; that of going from the analogue domain to the
discrete one.  The topic of analog-to-digital (A/D) conversion has
been at the forefront of research and technology since the seminal
works of Shannon, Nyquist, Whittaker and Kotelnikof were published,
see, for example, \cite{Unse-10} for a thorough related review. We all
know that if the highest frequency of an analog signal, $s(t)$, is
less than $F/2$, then Shannon's theorem suggests that no loss of
information is achieved if the signal is sampled, at least, at the
Nyquist rate of $F=1/T$, where $T$ is the corresponding sampling
period, and the signal can be perfectly recovered by its samples
\[
s(t)=\sum_{n}s(nT)\sinc(Ft-n),
\]
where $\sinc(\cdot)$ is the sampling function
\[
\sinc(t)=\frac{\sin(\pi t)}{\pi t}.
\]
While this has been the driving force behind the development of signal
acquisition devices, the increasing complexity of emerging applications
demands increasingly higher sampling rates, that cannot be accommodated by
today's hardware technology. This is the case, for example, in wideband
communications, where conversion speeds, as dictated by Shannon's bound,
have become more and more difficult to obtain. Consequently, alternatives
to high rate sampling are attracting a strong interest with the goal to
reduce the sampling rate by exploiting the {\it underlying structure} of
the signals at hand. In many applications, the signal comprises a few
frequencies or bands, see Fig.~\ref{fig:few.freq.bands} for an
illustration. In such cases, sampling at the Nyquist rate is
inefficient. This is an old problem and it has been addressed by a number
of authors, in the case where the locations of the non-zero bands in the
frequency spectrum are known, see, e.g., \cite{Vaug-10,Lin-10,Venk-10}.  CS
theory has inspired research to study cases where the locations
(carrier frequencies) of the bands are not known a-priori. A typical
application of this kind, of high practical interest, lies within the field
of Cognitive radio, e.g., \cite{Yu-10,Tian-10,Mish-10}. In contrast to what
we have studied so far in this paper, the sparsity now characterizes the
analog signal, and this poses a number of challenges that need to be
addressed. In other words, one can consider that
\begin{equation*}
s(t)=\sum_{i\in \mathcal{I}}\theta_i\psi_i(t),
\end{equation*}
where $\psi_i(t),~i\in \mathcal{I}$, are the functions that comprise the
dictionary, and only a small subset of the coefficients $\theta_i$ are
nonzero. Note that although each one of the dictionary functions can be of
high bandwidth, the true number of degrees of freedom of the signal is
low. Hence, one would like to sample the signal not at the Nyquist rate but
at a rate determined by the sparsity level of the coefficients' set. We
refer to such a scenario as {\it
  Analog-to-Information}\index{Analog-to-Information} sampling or {\it
  sub-Nyquist}\index{sub-Nyquist sampling} sampling.

\begin{figure}[!tbp]
\centering
\includegraphics[scale=1]{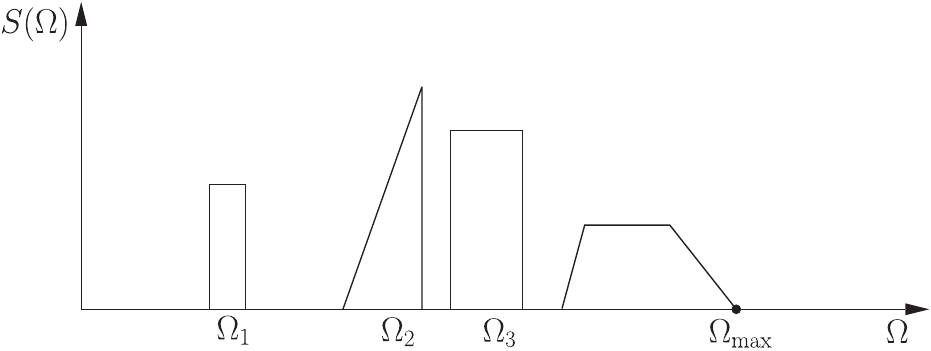}
\caption{The Fourier transform of an analogue signal, $s(t)$, which is sparse in the frequency domain; only a
  limited number of frequency bands contribute to its spectrum content
  $S(\Omega)$, where $\Omega$ stands for the angular frequency. Nyquist's
  theory guarantees that sampling at a frequency larger than or equal to twice the maximum
  $\Omega_{\max}$ is sufficient to recover the original analogue
  signal. However, this theory does not exploit information related to the sparse structure of
  the signal in the frequency domain.}\label{fig:few.freq.bands}
\end{figure}

An approach inspired directly by the theory of CS was first presented in
\cite{Kiro-10} and later improved and theoretically developed in
\cite{Tropp-10}. The approach builds around the assumption that the signal
consists of a sum of sinusoids and the \textit{random demodulator} of
Figure \ref{fig:analogue.sampling} is
adopted. In \cite{Mish-10, Mish11-10}, the more general
case of a signal consisting of a number of frequency bands, instead of
tones, was treated. In addition, the task of extracting each band of the
signal from the compressed measurements, that enables (low rate) baseband
processing, is addressed. In principle, CS related theory would enable far
fewer data samples than traditionally required when capturing signals with
relatively high bandwidth, but with a low information rate. However, from a
practical point of view, there are still a number of hardware
implementation related issues, such as random jittering, to be solved
first, e.g., \cite{ChB:Chenhard,ChB:Maetcher}.

An alternative path to sub-Nyquist sampling embraces a different class of
analog signals known as {\it multipulse} signals; that is, signals that
consist of a stream of short pulses. Sparsity now refers to the time
domain, and such signals may not even be bandlimited. Signals of this type
can be met in a number of applications, such as in radar, ultrasound,
bioimaging and neuronal signal processing, see, e.g., \cite{Drag-10}.  An
approach, known as \textit{finite rate of innovation sampling}, passes an
analogue signal, having $k$ degrees of freedom per second, through a linear
time invariant filter and then samples at a rate of $2k$ samples per
second. Reconstruction is performed via rooting a high-order polynomial,
see, e.g., \cite{Vett-10, Blu-10} and the references therein. In
\cite{Matou-10}, the task of sub-Nyquist sampling is treated using CS
theory arguments and an expansion in terms of Gabor functions; the signal
is assumed to consist of a sum of a few pulses of finite duration, yet of
unknown shape and time positions.

The task of sparsity-aware learning in the analogue domain is still in its
early stages and there is currently a lot of on-going activity; more on
this topic can be obtained in \cite{Duart2eld-10, MashaliYonina-10} and the
references there in.

\section{Sparsity-Promoting Algorithms}\label{ch10-algorithms}

In the previous sections, our emphasis was to highlight the most important
aspects underlying the theory of sparse signal/vector recovery from an
underdetermined set of linear equations. We now turn our attention to the
algorithmic aspects of the problem \cite{eladbook}. The issue now becomes that of
discussing {\it efficient} algorithmic schemes, which can achieve the
recovery of the unknown set of parameters. In Sections \ref{ch10:lasso} and
\ref{Ch10:thesparsest}, we saw that the constrained $\ell_1$ norm
minimization (Basis Pursuit) can be solved via Linear Programming
techniques and the LASSO task via convex optimization schemes. However,
such general purpose techniques tend to be inefficient, since, often, they
require many iterations to converge and the respective computational
resources can be excessive for practical applications, especially in high
dimensional spaces, $\Real^l$. As a consequence, a
huge research effort has been invested with the goal to develop efficient
algorithms, that are tailored-made to these specific tasks. This is still a
hot on-going area of research and definite conclusions are still risky to
be drawn. Our aim here is to provide the reader with some general trends
and philosophies that characterize the related activity. We will focus on
the most commonly used and cited algorithms, which at the same time are
structurally simple and the reader can follow them, without requiring a
deeper knowledge on optimization. Moreover, these algorithms involve, in
one way or another, arguments that are directly related to points and
notions that we have already used while presenting the theory; thus, they
can also be exploited from a pedagogical point of view, in order to
strengthen the reader's understanding of the topic. We start our review
with the class of batch algorithms, where all data are assumed to be
available prior to the application of the algorithm, and then we will move
on to online/time-adaptive schemes. Furthermore, our emphasis is on
algorithms that are appropriate for any sensing matrix. This is stated in
order to point out that in the literature efficient algorithms have also
been developed for specific forms of highly structured sensing matrices;
exploiting their particular structure can lead to reduced computational
demands, \cite{Gilb-10, Needl-10}.

There are currently three rough types of families along which this
algorithmic activity is growing: a) greedy algorithms, b) iterative
shrinkage schemes, and c) convex optimization techniques. We have used the
word rough, since, in some cases, it may be difficult to assign an
algorithm to a specific family.

\subsection{Greedy Algorithms}\index{Greedy algorithms}\label{ch10:greedy}

Greedy algorithms have a long history, see, for example, \cite{Tem-10} for
a comprehensive list of references. In the context of dictionary learning,
a greedy algorithm known as \textit{Matching Pursuit}\index{Matching
  Pursuit} was introduced in \cite{Mall-10}. A greedy algorithm is built
upon a series of {\it locally} optimal {\it single-term} updates. In our
context, the goals are: a) to unveil the ``active'' columns of the sensing
matrix $X$; that is, those columns that correspond to the nonzero locations
of the unknown parameters and b) to estimate the respective sparse
parameter vector. The set of indices which correspond to the nonzero vector
components is also known as the {\it support}. To this end, the set of
active columns of $X$ (and the support) is increased by one at each
iteration step. In the sequel, an updated estimate of the unknown sparse
vector is obtained. Let us assume that, at the $(i-1)$th iteration step,
the algorithm has selected the columns denoted as
$\mathbf{x}_{j_1},\mathbf{x}_{j_2},\ldots, \mathbf{x}_{j_{i-1}}$, with
$j_1,j_2,\ldots,j_{i-1} \in \{1,2,\ldots,l\}$. These indices are the
elements of the currently available support, $S^{(i-1)}$.  Let $X^{(i-1)}$
be the $N\times (i-1)$ matrix having $\mathbf{x}_{j_1},
\mathbf{x}_{j_2},\ldots, \mathbf{x}_{j_{i-1}}$ as its columns. Let, also,
the current estimate of the solution be $\bm{\theta}^{(i-1)}$, which is a
$(i-1)$-sparse vector, with zeros at all locations with index outside the
support.

\begin{algo}[\textbf{Orthogonal Matching Pursuit (OMP)}]\index{Orthogonal
  Matching Pursuit}\mbox{}

\noindent The algorithm is initialized with $\bm{\theta}^{(0)} \coloneqq
\bm{0}$, $\bm{e}^{(0)} \coloneqq \bm{y}$ and $S^{(0)} \coloneqq \emptyset
$. At iteration step $i$, the following computational steps are performed:

\begin{enumerate}

\item Select the column $\mathbf{x}_{j_i}$ of $X$, which is
  \textit{maximally} correlated to (forms the least angle with) the
  respective error vector,
  $\bm{e}^{(i-1)} \coloneqq \bm{y}-X\bm{\theta}^{(i-1)}$, i.e.,
\begin{equation*}
\mathbf{x}_{j_i}:\ j_i \coloneqq \argmax_{j=1,2,\ldots,l} \frac{\left|\mathbf{x}_j^T
  \bm{e}^{(i-1)}\right|}{\norm{\mathbf{x}_j}_2}.
\end{equation*}

\item Update the support and the corresponding set of active columns:
  $S^{(i)} = S^{(i-1)}\cup \{j_i\}$, and
  $X^{(i)} = [X^{(i-1)},\mathbf{x}_{j_i}]$.

\item\label{omp.ls.step} Update the estimate of the parameter vector: Solve
  the Least-Squares (LS) problem that minimizes the norm of the error,
  using the active columns of $X$ only, i.e.,
\[
\tilde{\bm{\theta}} \coloneqq \argmin_{\bm{z}\in\Real^i}
\norm{\bm{y}-X^{(i)}\bm{z}}_2^2.
\]
Obtain $\bm{\theta}^{(i)}$ by inserting the elements of
$\tilde{\bm{\theta}}$ in the respective locations
$(j_1,j_2,\ldots,j_i)$, which comprise the support (the rest of the
elements of $\bm{\theta}^{(i)}$ retain their zero values).

 \item Update the error vector
 \[
 \bm{e}^{(i)} \coloneqq \bm{y}-X\bm{\theta}^{(i)}.
 \]
 \end{enumerate}
\end{algo}

The algorithm terminates if the norm of the error becomes less than a
preselected user-defined constant, $\epsilon_0$. The following observations
are in order.

\begin{figure}[!tbp]
\centering
\includegraphics[scale=1]{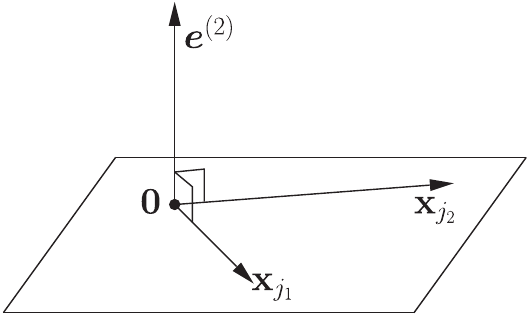}
\caption{The error vector at the $i$th iteration is orthogonal to the
  subspace spanned by the currently available set of active
  columns. Here is an illustration for the case of the $3$-dimensional
  Euclidean space $\Real^3$, and for $i=2$.} \label{fig:error.perp.span}
\end{figure}

\begin{remarks}\label{rem:omp}\mbox{}
\begin{itemize}
\item Since $\bm{\theta}^{(i)}$, in Step 3, is the result of a LS
  task, it is known that the error vector is
  orthogonal to the subspace spanned by the active columns involved,
  i.e.,
\[
\bm{e}^{(i)}\bot \linspan \left\{\mathbf{x}_{j_1}, \ldots,
\mathbf{x}_{j_i} \right\}.
\]
This guarantees that taking the correlation, in the next step, of the
columns of $X$ with $\bm{e}^{(i)}$ none of the previously selected columns
will be reselected; they result to zero correlation, being orthogonal to
$\bm{e}^{(i)}$, see Fig.~\ref{fig:error.perp.span}.

\item It can be shown that the column, which has maximal correlation (maximum absolute value of
  the inner product) with the currently available error vector, is the one
  that maximally reduces (compared to any other column) the $\ell_2$ norm
  of the error, when $\bm{y}$ is approximated by linearly combining the
  currently available active columns. This is the point where the heart of
  the greedy strategy beats. This minimization is with respect to a
  \textit{single term}, keeping the rest fixed, as they have been obtained
  from the previous iteration steps \cite{Fred-10}.

\item Starting with all the components being zero, if the algorithm stops
  after $k_0$ iteration steps, the result will be a $k_0$-sparse solution.

\item Note that there is no optimality in this searching strategy. The
  only guarantee is that the $\ell_2$ norm of the error vector is
   decreased at every iteration step.  In general, there is no guarantee
  that the algorithm can obtain a solution close to the true one, see,
  e.g., \cite{Teml1-10}. However, under certain constraints on the
  structure of $X$, performance bounds can be obtained, see, e.g.,
  \cite{Tropp1-10, DaveOMP-10, ZangOMP-10}.

\item The complexity of the algorithm amounts to $\mathcal{O}(k_0lN)$
  operations, which are contributed by the computations of the
  correlations, plus the demands raised by the solution of the LS task, in
  Step \ref{omp.ls.step}, whose complexity depends on the specific
  algorithm used. The $k_0$ is the sparsity level of the delivered solution
  and, hence, the total number of iteration steps that are performed.
\end{itemize}
\end{remarks}

Another more qualitative argument, that justifies the selection of the
columns based on their correlation with the error vector, is the
following. Assume that the matrix $X$ is orthonormal. Let also
$\bm{y}=X\bm{\theta}$. Then, $\bm{y}$ lies in the subspace spanned by the
active columns of X, i.e., those which correspond to the non-zero
components of $\bm{\theta}$. Hence, the rest of the columns are orthogonal
to $\bm{y}$, since $X$ is assumed to be orthonormal. Taking the correlation
of $\bm{y}$, at the first iteration step, with all the columns, it is
certain that one among the active columns will be chosen. The inactive
columns result in zero correlation. A similar argument holds true for all
subsequent steps, since all the activity takes place in a subspace that is
orthogonal to all the inactive columns of $X$. In the more general case,
where $X$ is not orthonormal, we can still use the correlation as a measure
that quantifies geometric similarity. The smaller the correlation/the
magnitude of the inner product is, the more orthogonal two vectors
are. This brings us back to the notion of mutual coherence, which is a
measure of the maximum correlation (least angle) among the columns of $X$.

\subsubsection{OMP Can Recover Optimal Sparse Solutions: Sufficiency
  Condition}

We have already stated that, in general, there are no guarantees that OMP
will recover optimal solutions. However, when the unknown vector is
sufficiently sparse, with respect to the structure of the sensing matrix
$X$, then OMP can exactly solve the $\ell_0$ minimization task in
\eqref{ch10:l0minim} and recover the solution in $k_0$ steps, where $k_0$
is the sparsest solution that satisfies the associated linear set of
equations.

\begin{theorem}
Let the mutual coherence (Section \ref{ch10mutcoh}) of the sensing matrix,
$X$, be $\mu(X)$. Assume, also, that the linear system,
$\bm{y}=X\bm{\theta}$, accepts a solution such as
\beq \norm{\bm{\theta}}_0
< \frac{1}{2}\left (1+\frac{1}{\mu(X)}\right ). \label{ch10:10.6equ}
\eeq
Then, OMP guarantees to recover the sparsest solution in
$k_0 = \norm{\bm{\theta}}_0$ steps.
\end{theorem}

We know from Section \ref{ch10mutcoh} that, under the previous
condition, any other solution will be necessarily less sparse. Hence,
there is a unique way to represent $\bm{y}$ in terms of $k_0$ columns
of $X$. Without harming generality, let us assume that the true
support corresponds to the first $k_0$ columns of $X$, i.e.,
\[
\bm{y}=\sum_{j=1}^{k_0}\theta_j\mathbf{x}_j, \quad \theta_j\neq 0,
\forall j\in\{1,\ldots,k_0\}.
\]
The theorem is a direct consequence of the following proposition:

\begin{prop}
If the condition \eqref{ch10:10.6equ} holds true, then the OMP
algorithm will never select a column with index outside the true
support, see, e.g., \cite{Tropp1-10}. In a more
formal way, this is expressed as
\[
j_i=\argmax_{j=1,2, \ldots,l} \frac{\left|\mathbf{x}_j^T\bm{e}^{(i-1)}
  \right|}{\norm{\mathbf{x}_j}_2}\in \{1,\ldots,k_0\}.
\]
\end{prop}


\begin{figure}[!tbp]
                 \centering
                \includegraphics[scale=0.9]{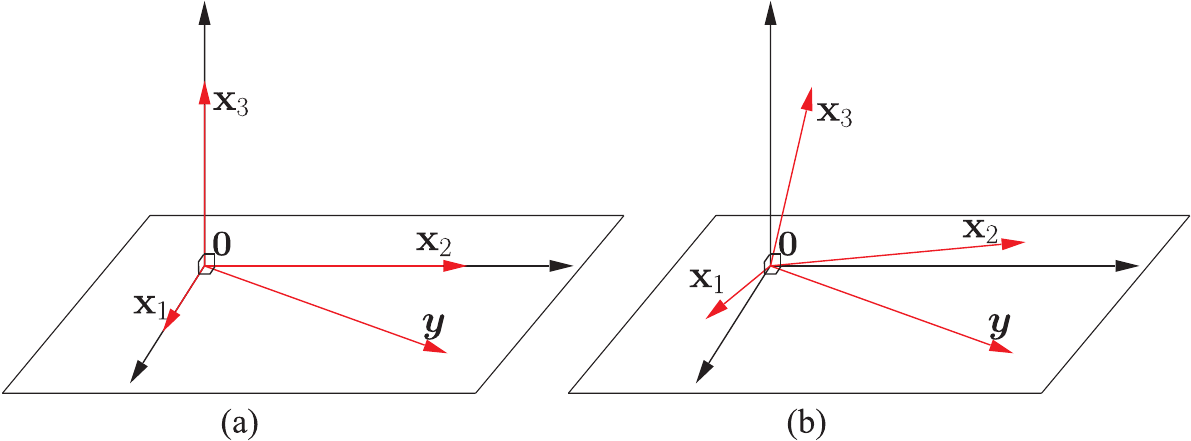}
        \caption{(a) In the case of an orthogonal matrix, the measurement
          vector $\bm{y}$ will be orthogonal to any inactive column; here,
          $\mathbf{x}_3$. (b) In the more general case, it is expected to
          ``lean" closer (form smaller angles) to the active than to the
          inactive columns.} \label{fig:prop.omp}
\end{figure}
A geometric interpretation of this proposition is the following: if
the angles formed between all the possible pairs among the columns of
$X$ are large enough in the $\Real^l$ space, which guarantees that
$\mu(X)$ is small enough, then $\bm{y}$ will lean more (form smaller
angle) towards any one of the active columns, which contribute to its
formation, compared to the rest, which are inactive and do not
participate in the linear combination that generates $\bm{y}$.
Fig.~\ref{fig:prop.omp} illustrates the geometry, for the extreme case
of mutually orthogonal vectors (Fig.~\ref{fig:prop.omp}a) and for
the more general case, where the vectors are not orthogonal, yet the
angle between any pair of columns is large enough
(Fig.~\ref{fig:prop.omp}b).

In a nutshell, the previous proposition guarantees that, during the
first iteration, a column corresponding to the true support will be
selected. In a similar way, this is also true for all subsequent
iterations. In the second step, another, different from the previously
selected column (as it has already been stated), will be chosen. At
step $k_0$, the last remaining active column, corresponding to the
true support, is selected and this necessarily results to zero
error. To this end, it suffices to set $\epsilon_0$
equal to zero.

\subsubsection{The LARS Algorithm}\index{LARS}

The Least Angle Regression (LARS) algorithm, \cite{Efro-10}, shares
the first two steps with OMP. It selects $j_i$ to be an index outside
the currently available  active set so that to maximize the
correlation with the residual vector. However, instead of performing
an LS fit to compute the nonzero components of $\bm{\theta}^{(i)}$,
these are computed so that the residual to be equicorrelated with all
the columns in the active set, i.e.,
\[
|\mathbf{x}_{j}^T (\bm{y}-X\bm{\theta}^{(i)})| = \text{constant},
\quad\forall j\in S^{(i)},
\]
where we have assumed that the columns of $X$ are normalized, as it is
common in practice (recall, also, the Remarks \ref{rem:normalize.X}). In
other words, in contrast to the OMP, where the error vector is forced to be
orthogonal to the active columns, LARS demands this error to form equal
angles with each one of them. Likewise OMP, it can be shown that, provided
the target vector is sufficiently sparse and under incoherence of the
columns of $X$, LARS can exactly recover the sparsest solution,
\cite{Tsai-10}.

A further small modification leads to the so-called LARS-LASSO
algorithm\index{LARS-LASSO}. According to this version, a previously
selected index in the active set can be removed at a later stage. This
gives the algorithm the potential to ``recover'' from a previously bad
decision. Hence, this modification departs from the strict rationale that
defines the greedy algorithms. It turns out that this version solves the
LASSO optimization task. This algorithm is the same as the one suggested in
\cite{Osbo-10} and it is known as {\it homotopy} algorithm. Homotopy
methods\index{Homotopy methods} are based on a continuous transformation
from one optimization task to another. The solutions to this sequence of
tasks lie along a continuous parameterized path. The idea is that, while
the optimization tasks may be difficult to solve by themselves, one can
trace this path of solutions by slowly varying the parameters. For the
LASSO task, it is the $\lambda$ parameter which is varying, see, e.g.,
\cite{Plum-10, Maliou-10, Asif-10}. Take as an example the
  LASSO task in its regularized version in \eqref{ch10lassos1}.  For $\lambda=0$, the task
  minimizes the $\ell_2$ norm and for $\lambda\rightarrow \infty$ the
  task minimizes the $\ell_1$ norm, and for this case the solution tends to
  zero.  It turns out that the solution path, as $\lambda$ changes from
  large to small values, is polygonal. Vertices on this solution path
  correspond to vectors having nonzero elements only on a subset of
  entries. This subset remains unchanged, till $\lambda$ reaches the next
  critical value, which corresponds to a new vertex of the polygonal path
  and to a new subset of potential nonzero values.  Thus, the solution is
  obtained via this sequence of steps along this polygonal path.

\subsubsection{Compressed Sensing Matching Pursuit (CSMP) Algorithms}
\label{ch10-cosmp}

Strictly speaking, these algorithms are not greedy, yet, as it is stated in
\cite{Needl-10}, they are at heart greedy algorithms. Instead of performing
a single term optimization per iteration step, in order to increase the
support by one, as it is the case with OMP, these algorithms attempt to
obtain first an estimate of the support and then use this information to
compute a least squares estimate of the target vector, constrained on the
respective active columns.  The quintessence of the method lies in the
near-orthogonal nature of the sensing matrix, assuming that this obeys the
RIP.

Assume that $X$ obeys the RIP for some small enough value $\delta_k$ and
sparsity level, $k$, of the unknown vector.  Let, also, that the
measurements are exact, i.e., $\bm{y}=X\bm{\theta}$.  Then,
$X^T\bm{y}=X^TX\bm{\theta}\approx \bm{\theta}$. Therefore, intuition
indicates that it is not unreasonable to select, in the first iteration
step, the $t$ (a user-defined parameter) largest in magnitude components of
$X^T\bm{y}$ as indicative of the nonzero positions of the sparse target
vector. This reasoning carries on for all subsequent steps, where, at the
$i$th iteration, the place of $\bm{y}$ is taken by the residual
$\bm{e}^{(i-1)} \coloneqq \bm{y}-X\bm{\theta}^{(i-1)}$, where
$\bm{\theta}^{(i-1)}$ indicates the estimate of the target vector at the
$(i-1)$th iteration. Basically, this could be considered as a
generalization of the OMP. However, as we will soon see, the difference
between the two mechanisms is more substantial.

\begin{algo}[\textbf{The CSMP Scheme}]\mbox{}

\begin{enumerate}
\item\label{1st.t} Select the value of $t$.
\item Initialize the algorithm: $\bm{\theta}^{(0)} = \bm{0}$,
  $\bm{e}^{(0)} = \bm{y}$.
\item For $i=1,2,\ldots$, execute the following.
\begin{enumerate}
\item\label{2nd.t} Obtain the current support:
\begin{equation*}
S^{(i)} \coloneqq \supp\left(\bm{\theta}^{(i-1)}\right)\cup
\left\{\text{indices of the}\ t\ \text{largest in magnitude} \atop
\text{components of}\ X^T\bm{e}^{(i-1)} \right\}.
\end{equation*}
\item Select the active columns: Construct $X^{(i)}$ to comprise the
  active columns of $X$ in accordance to $S^{(i)}$.
  Obviously, $X^{(i)}$ is a $N\times r$
  matrix, where $r$ denotes the cardinality of the support set $S^{(i)}$.
\item\label{csmp.ls.task} Update the estimate of the parameter vector:
  solve the LS task
\[
\tilde{\bm{\theta}} \coloneqq \argmax_{\bm{z}\in \Real^r}
\norm{\bm{y}-X^{(i)}\bm{z}}_2^2.
\]
Obtain $\hat{\bm{\theta}}^{(i)}\in\Real^l$ having the $r$ elements
of $\tilde{\bm{\theta}}$ in the respective locations, as indicated by
the support, and the rest of the elements being zero.
\item $\bm{\theta}^{(i)} \coloneqq
  H_k\left(\hat{\bm{\theta}}^{(i)}\right)$. The mapping $H_k$ denotes the
  \textit{hard thresholding} operator; that is, it returns a vector with
  the $k$ largest in magnitude components of the argument, and the rest are
  forced to zero.
\item Update the error vector: $\bm{e}^{(i)} =
  \bm{y}-X\bm{\theta}^{(i)}$.
\end{enumerate}
\end{enumerate}
\end{algo}

The algorithm requires as input the sparsity level $k$. Iterations carry on
until a halting criterion is met. The value of $t$, that determines the
largest in magnitude values in Steps \ref{1st.t} and \ref{2nd.t}, depends
on the specific algorithm. In CoSaMP (Compressive Sampling Matching
Pursuit, \cite{Needl-10}), $t=2k$ and in the SP (Subspace
Pursuit, \cite{Dai-10}), $t=k$.

Having stated the general scheme, a major difference with OMP becomes
readily apparent. In OMP, only one column is selected per iteration
step. Moreover, this remains in the active set for all subsequent
steps. If, for some reason, this was not a good choice, the scheme
cannot recover from such a bad decision. In contrast, the support and
hence the active columns of $X$ are continuously updated in CSMP and
the algorithm has the ability to correct a previously bad decision, as
more information is accumulated and iterations progress. In
\cite{Dai-10}, it is shown that if the measurements are exact
($\bm{y}=X\bm{\theta}$) then SP can recover the $k$-sparse true
vector in a finite number of iteration steps, provided that $X$
satisfies the RIP with $\delta_{3k}<0.205$. If the measurements are
noisy, performance bounds have been derived, which hold true for
$\delta_{3k}<0.083$. For the CoSaMP, performance bounds have been
derived for $\delta_{4k}<0.1$.

\subsection{Iterative Shrinkage Algorithms (IST)}\index{Iterative
  shrinkage algorithms}\label{itershr-10}

This family of algorithms have also a long history, see, e.g.,
\cite{Jans-10, Donojon-10, Hoch-10, Kins-10}. However, in the
``early'' days, the developed algorithms had some sense of heuristic
flavor, without establishing a clear bridge with optimizing a cost
function. Later attempts were substantiated by sound theoretical
arguments concerning issues such as convergence and convergence rate,
e.g., \cite{Figu-10, Daub1-10, Eladalg-10, Comb-10}.

The general form of this algorithmic family has a striking resemblance with
the classical linear algebra iterative schemes for approximating the
solution of large linear systems of equations, known as {\it stationary
  iterative} or {\it iterative relaxation} methods.  The classical
Gauss-Seidel and Jacobi algorithms, e.g., \cite{Hagel-10}, in numerical
analysis can be considered as members of this family. Given a linear system
of $l$ equations with $l$ unknowns, $\bm{z}=A\bm{x}$, the basic iteration
at step $i$ has the following form
\beqan
\bm{x}^{(i)}& = &\left
(I-QA\right )\bm{x}^{(i-1)}+Q\bm{z}\\ &=&\bm{x}^{(i-1)}+Q\bm{e}^{(i-1)},
\qquad\bm{e}^{(i-1)} \coloneqq \bm{z}-A\bm{x}^{(i-1)},
\eeqan
which does not come as a surprise. It is of the same form as most of the
iterative schemes for numerical solutions! The matrix $Q$ is chosen so that
to guarantee convergence and different choices lead to different algorithms
with their pros and cons. It turns out that this algorithmic form can also
be applied to underdetermined systems of equations, $\bm{y}=X\bm{\theta}$,
with a ``minor'' modification, which is imposed by the sparsity constraint
of the target vector. This leads to the following general form of iterative
computation
\[
\bm{\theta}^{(i)} = T_i\left ( \bm{\theta}^{(i-1)}+
Q\bm{e}^{(i-1)}\right), \qquad \bm{e}^{(i-1)} =
\bm{y}-X\bm{\theta}^{(i-1)},
\]
starting from an initial guess of $\bm{\theta}^{(0)}$ (usually
$\bm{\theta}^{(0)} = \bm{0}$, $\bm{e}^{(0)} = \bm{y}$). In
certain cases, $Q$ can be made to be iteration-dependent. The operator
$T_i(\cdot)$ is a nonlinear thresholding operator, that is applied
\textit{entrywise}, i.e., \textit{component-wise}. Depending on the specific
scheme, this can be either the hard thresholding operator, denoted as
$H_k$, or the soft thresholding operator, denoted as $S_{\alpha}$. Hard
thresholding, as we already know, keeps the $k$ largest components of a
vector unaltered and sets the rest equal to zero. Soft thresholding was
introduced in Section \ref{ch10:lasso}. All components with magnitude less
than $\alpha$ are forced to zero and the rest are reduced in magnitude by
$\alpha$; that is, the $j$th component of a vector, $\bm{\theta}$, after
soft thresholding becomes
\[
(S_{\alpha}(\bm{\theta}))_j=\sign(\theta_j)(|\theta_j|-\alpha)_+.
\]
 Depending on a) the choice of $T_i$, b) the
specific value of the parameter $k$ or $\alpha$ and c) the matrix $Q$,
different instances occur. A most common choice for $Q$ is $\mu X^T$ and
the generic form of the main iteration becomes
\beq \bm{\theta}^{(i)} =
T_i\left( \bm{\theta}^{(i-1)}+ \mu
X^T\bm{e}^{(i-1)}\right),\label{iteralg-10}
\eeq
where $\mu$ is a relaxation (user-defined) parameter, which can also be
left to vary with each iteration step. The choice of $X^T$ is intuitively
justified, once more, by the near-orthogonal nature of $X$. For the first
iteration step and for a linear system of the form $\bm{y}=X\bm{\theta}$,
starting from a zero initial guess, we have
$X^T\bm{y}=X^TX\bm{\theta}\approx \bm{\theta}$ and we are close to the
solution.

Although intuition is most important in scientific research, it is not
enough, by itself, to justify decisions and actions. The generic
scheme in \eqref{iteralg-10} has been reached from different paths,
following different perspectives that lead to different choices of the
respective parameters. Let us spend some more time on that, with the
aim to make the reader more familiar with techniques that address
optimization tasks of non-differentiable loss functions. The term in
the parenthesis in (\ref{iteralg-10}) coincides with the gradient
descent iteration step if the cost function were the unregularized LS
loss, i.e.,
\[
J(\bm{\theta})=\frac{1}{2}\norm{\bm{y}-X\bm{\theta}}_2^2.
\]
In this case, the gradient descent rationale leads to
\begin{align*}
\bm{\theta}^{(i-1)}-\mu\frac{\partial
  J\left(\bm{\theta}^{(i-1)}\right)}{\partial \bm{\theta}} & =
\bm{\theta}^{(i-1)}-\mu X^T(X\bm{\theta}^{(i-1)}-\bm{y})\\
&=\bm{\theta}^{(i-1)}+\mu
X^T\bm{e}^{(i-1)}.
\end{align*}
It is well known and it can easily be shown that the gradient descent
can alternatively be viewed as the result of minimizing a regularized
version of the linearized loss function, i.e.,
\begin{align}
\bm{\theta}^{(i)} = \argmin_{\bm{\theta}\in \Real^l}
\left\{J\left(\bm{\theta}^{(i-1)}\right) \right. & +
\left(\bm{\theta} - \bm{\theta}^{(i-1)}\right)^T \frac{\partial
  J\left(\bm{\theta}^{(i-1)}\right)}{\partial\bm{\theta}} \nonumber \\
& + \left. \frac{1}{2\mu}\norm{\bm{\theta} -
  \bm{\theta}^{(i-1)}}^2_2 \right\}. \label{ch10:linearized}
\end{align}
One can adopt this view of the gradient descent philosophy as a
kick-off point to minimize iteratively the following LASSO task, i.e.,
\begin{equation*}
\min_{\bm{\theta}\in\Real^l} \left\{L(\bm{\theta},\lambda) =
\frac{1}{2}\norm{\bm{y}-X\bm{\theta}}_2^2 + \lambda\norm{\bm{\theta}}_1
\nonumber = J(\bm{\theta})+\lambda\norm{\bm{\theta}}_1
\right\}.\label{ch10-lassosrev}
\end{equation*}
The difference now is that the loss function comprises two terms. One
which is  smooth (differentiable)  and a non-smooth one. Let  the
current estimate be $\bm{\theta}^{(i-1)}$.  The updated estimate is
obtained by
\begin{align*}
\bm{\theta}^{(i)} = \argmin_{\bm{\theta}\in\Real^l}
\left\{J\left(\bm{\theta}^{(i-1)}\right) \right. & +
\left(\bm{\theta}-\bm{\theta}^{(i-1)}\right)^T\frac{\partial
  J(\bm{\theta}^{(i-1)})}{\partial \bm{\theta}} \\
& + \left
. \frac{1}{2\mu}\norm{\bm{\theta}-\bm{\theta}^{(i-1)}}^2_2 +
\lambda\norm{\bm{\theta}}_1 \right\},
\end{align*}
which, after ignoring constants, becomes
\beq
\bm{\theta}^{(i)}=\argmin_{\bm{\theta}\in\Real^l} \left
\{\frac{1}{2}\norm{\bm{\theta}-\tilde{\bm{\theta}}}^2_2 +
\lambda\mu\norm{\bm{\theta}}_1\right\}
\eeq
where
\beq
\tilde{\bm{\theta}}:= \bm{\theta}^{(i-1)}-\mu\frac{\partial
  J(\bm{\theta}^{(i-1)})}{\partial \bm{\theta}}.\label{ch10-proximal2}
\eeq
 Following exactly the same steps as those that led to the derivation of
\eqref{ch10-l1min2} from \eqref{ch10lassos1} (after replacing
$\hat{\bm{\theta}}_{LS}$ with $\tilde{\bm{\theta}}$) we obtain
\beqa
\bm{\theta}^{(i)} & = & S_{\lambda\mu} (\tilde{\bm{\theta}}) = S_{\lambda\mu}\left
(\bm{\theta}^{(i-1)}-\mu\frac{\partial
  J(\bm{\theta}^{(i-1)})}{\partial \bm{\theta}}\right )  \label{ch10-proximal4}
\\
&=& S_{\lambda\mu}\left (\bm{\theta}^{(i-1)}+\mu
X^T\bm{e}^{(i-1)}\right ). \label{ch10-eqithsc}
\eeqa
This is very interesting and practically useful. The only effect of the
presence of the non-smooth $\ell_1$ norm in the loss function is an extra
simple thresholding operation, which as we know is an operation performed
\textit{individually} on each component. It can be shown, e.g.,
\cite{Beck-10}, that this algorithm converges to a minimizer
$\bm{\theta}_*$ of the LASSO \eqref{ch10lassos1}, provided that $\mu\in
(0,1/\lambda_{\max}(X^TX))$, where $\lambda_{\max}(\cdot)$ denotes the maximum eigenvalue of $X^TX$. The convergence rate is
dictated by the rule
\[
L(\bm{\theta}^{(i)},\lambda)- L(\bm{\theta}_*,\lambda)\approx O(1/i),
\]
which is known as {\it sublinear}\index{Sublinear convergence rate}
global rate of convergence. Moreover, it can be shown that
\[
L(\bm{\theta}^{(i)},\lambda)- L(\bm{\theta}_*,\lambda)\le
\frac{C\norm{\bm{\theta}^{(0)}-\bm{\theta}_*}^2_2}{2i}.
\]
The latter result indicates that if one wants to achieve an accuracy
of $\epsilon$, then this can be obtained by at most $\left\lfloor
\frac{C\norm{\bm{\theta}^{(0)}-\bm{\theta}_*}^2_2}{2\epsilon}\right\rfloor$
iterations, where $\lfloor \cdot \rfloor$ denotes the floor operator.

In \cite{Daub1-10}, \eqref{iteralg-10} was obtained from a nearby corner,
building upon arguments from the classical \textit{proximal-point} methods
in optimization theory, e.g., \cite{Rocka-10}. The original LASSO
regularized cost function is modified to the so called {\it surrogate
  objective},
\[
J(\bm{\theta},\tilde{\bm{\theta}}) = \frac{1}{2}\norm{\bm{y}-X\bm{\theta}}_2^2
+ \lambda\norm{\bm{\theta}}_1 + \frac{1}{2}d(\bm{\theta},\tilde{\bm{\theta}}),
\]
where
\[
d(\bm{\theta},\tilde{\bm{\theta}})\coloneqq
c\norm{\bm{\theta}-\tilde{\bm{\theta}}}_2^2 -
\norm{X\bm{\theta}-X\tilde{\bm{\theta}}}^2_2.
\]
If $c$ is appropriately chosen (larger than the largest eigenvalue of
$X^TX$), the surrogate objective is guaranteed to be strictly convex. Then
it can be shown that the minimizer of the surrogate
objective is given by
\beq
\hat{\bm{\theta}}=S_{\lambda/c}\left
(\tilde{\bm{\theta}} + \frac{1}{c}X^T(\bm{y}-X\tilde{\bm{\theta}})\right).
\eeq
In the iterative formulation, $\tilde{\bm{\theta}}$ is
selected to be the previously obtained estimate; in this way, one
tries to keep the new estimate close to the previous one. The
procedure readily results to our generic scheme in \eqref{iteralg-10},
using soft thresholding with parameter $\lambda/c$. It can be shown
that such a strategy converges to a minimizer of the original LASSO
problem. The same algorithm was reached in \cite{Figu-10}, using {\it
  majorization-minimization} techniques from optimization theory. So,
from this perspective, the IST family has strong ties with algorithms
that belong to the convex optimization category.

In \cite{Wright-10}, the \textit{Sparse Reconstruction by Separable
  Approximation} (SpaRSA)\index{SpaRSA} algorithm is proposed, which is a
modification of the standard IST scheme. The starting point is
(\ref{ch10:linearized}); however, the multiplying factor, $\frac{1}{2\mu}$,
instead of being constant is now allowed to change from iteration to
iteration according to a rule. This results in a speed up in the
convergence of the algorithm. Moreover, inspired by the homotopy family of
algorithms, where $\lambda$ is allowed to vary, SpaRSA can be extended to
solve a sequence of problems which are associated with a corresponding
sequence of values of $\lambda$.  Once a solution has been obtained for a
particular value of $\lambda$, it can be used as a ``warm-start'' for a
nearby value. Solutions can therefore be computed for a range of values, at
a small extra computational cost, compared to solving for a single value
from a ``cold start''. This technique abides with to the so-called {\it
  continuation strategy}, which has been used in the context of other
algorithms as well, e.g., \cite{Halle-10}.  Continuation has been shown to
be a very successful tool to increase the speed of convergence.

An interesting modification of the basic IST scheme has been
proposed in \cite{Beck-10}, which improves the convergence rate to
$O(1/i^2)$, by only a simple modification with almost no extra
computational burden. The scheme is known as {\it Fast Iterative
  Shrinkage-Thresholding Algorithm} (FISTA)\index{FISTA}. This scheme
is an evolution of  \cite{Nest-10}, which introduced the basic idea
for the case of differentiable costs, and consists of the following
steps:
\beqan
\bm{\theta}^{(i)} & = & S_{\lambda\mu}\left (\bm{z}^{(i)}+\mu
X^T\left(\bm{y}-X\bm{z}^{(i)}\right)\right ), \\
\bm{z}^{(i+1)} & \coloneqq &
\bm{\theta}^{(i)}+\frac{t_i-1}{t_{i+1}}\left(\bm{\theta}^{(i)}-\bm{\theta}^{(i-1)}
\right),
\eeqan
where
\[
t_{i+1} \coloneqq \frac{1+\sqrt{1+4t^2_i}}{2},
\]
with initial points $t_1=1$ and $\bm{z}^{(1)}=\bm{\theta}^{(0)}$. In
words, in the thresholding operation, $\bm{\theta}^{(i-1)}$ is
replaced by  $\bm{z}^{(i)}$, which is a specific linear combination of
two successive updates of $\bm{\theta}$. Hence, at a marginal increase
of the computational cost, a substantial increase in convergence speed
is achieved.

In \cite{Blumen-10} the hard thresholding version has been used, with
$\mu=1$ and the thresholding operator $H_k$ uses the sparsity level
$k$ of the target solution, that is assumed to be known. In a later
version, \cite{Blumen1-10}, the relaxation parameter is left to change
so that, at each iterations step, the error is maximally reduced. It
has been shown that the algorithm converges to a local minimum of the
cost function $\norm{\bm{y}-X\bm{\theta}}_2$, under the constraint that
$\bm{\theta}$ is a $k$-sparse vector. Moreover, the latter version is
a stable one and it results to a near optimal solution if a form of
RIP is fulfilled.

A modified version of the generic scheme given in (\ref{iteralg-10}), that
evolves along the lines of \cite{Luo-10}, obtains the updates
component-wise, one vector component at a time. Thus, a ``full'' iteration
consists of $l$ steps. The algorithm is known as \textit{coordinate
  descent}\index{Coordinate descent algorithm} and its basic iteration has
the form,
\beq
\theta_j^{(i)} =
S_{\lambda/\norm{\mathbf{x}_j}_2^2} \left (\theta^{(i-1)}_j +
\frac{{\mathbf{x}_j^T\bm{e}^{(i-1)}}}{\norm{\mathbf{x}_j}^2_2} \right),
\quad j=1,2,\ldots, l.
\eeq
This algorithm
replaces the constant $c$, in the previously reported soft thresholding
algorithm, with the norm of the respective column of $X$, if the columns of
$X$ are not normalized to unit norm. It has been shown that the parallel
coordinate descent algorithm also converges to a LASSO minimizer of \eqref{ch10lassos1},
\cite{Eladalg-10}. Improvements of the algorithm, using line search
techniques to determine the most descent direction for each iteration, have
also been proposed, see, \cite{Zibul-10}.

The main contribution to the complexity for the iterative shrinkage
algorithmic family comes from the two matrix-vector products, which
amounts to $\mathcal{O}(Nl)$, unless $X$ has a
special structure, e.g., DFT,  that can be exploited to reduce the
load.

In \cite{Malek-10}, the so-called Two Stage Thresholding
(TST)\index{Two stage thresholding scheme} scheme is presented, which
brings together arguments from the iterative shrinkage family and the
OMP. This algorithmic scheme involves two stages of thresholding. The
first step is exactly the same as in \eqref{iteralg-10}. However, this
is now used only for determining ``significant'' nonzero locations,
just as in Compressed Sensing Matching Pursuit (CSMP) algorithms,
presented in the previous subsection. Then, a LS problem is solved to
provide the updated estimate, under the constraint of the available
support. This is followed by a second step of thresholding. The
thresholding operations in the two stages can be different. If hard
thresholding, $H_k$, is used in both steps, this results to the
algorithm proposed in \cite{Fouc-10}.  For this latter scheme,
convergence and performance bounds are derived if the RIP holds for
$\delta_{3k}<0.58$. In other words, the basic difference between the
TST and CSMP approaches is that, in the latter case, the most
significant non-zero coefficients are obtained by looking at the
correlation term $X^T\bm{e}^{(i-1)}$ and in the TST family at
$\bm{\theta}^{(i-1)}+ \mu X^T\bm{e}^{(i-1)}$. The differences among
different approaches can be minor and the crossing lines among the
different algorithmic categories are not necessarily crispy
clear. However, from a practical point of view, sometimes small
differences may lead to substantially improved performance.

\begin{remarks}\label{rem:iterative.thresholding}\mbox{}
\begin{itemize}

\item The minimization in (\ref{ch10-proximal4}) bridges the IST
  algorithmic family with another powerful tool in convex optimization,
  which builds around the notion of \textit{proximal mapping}\index{Proximal
    mapping} or \textit{Moreau envelopes}\index{Moreau envelopes}, see, e.g.,
  \cite{Rocka-10, Combettes.Pesquet.proximal.2010}. Given a convex function
  $h:\mathcal{R}^l\rightarrow \mathcal{R}$, and a $\mu>0$, the proximal
  mapping, $\prox_{\mu h}: \Real^l \rightarrow \Real^l$, with respect to
  $h$, and of index $\mu$, is defined as the (unique) minimizer
\beq
\prox_{\mu h}(\bm{x}) \coloneqq \argmin_{\bm{u}\in \Real^l}\left
\{h(\bm{u})+\frac{1}{2\mu}\norm{\bm{x}-\bm{u}}_2^2\right\}, \quad \forall
\bm{x}\in \Real^l.
\eeq
Let us now assume that we want to minimize a convex function, which is
given as the sum
\[
f(\bm{\theta})=J(\bm{\theta})+h(\bm{\theta}),
\]
where $J(\cdot)$ is  convex and differentiable, and $h(\cdot)$ is also a
convex, but not necessarily a smooth one.  Then it can be shown that the
following iterations converge to a minimizer of $f(\cdot)$,
\beq
\bm{\theta}^{(i)}=\prox_{\mu h}\left
(\bm{\theta}^{(i-1)}-\mu\frac{\partial J(\bm{\theta}^{(i-1)})}{\partial
  \bm{\theta}}\right ),\label{ch10-proximal3}
\eeq
where $\mu>0$ and it can also be made iteration dependent, i.e.,
$\mu_i>0$. If we now use this scheme to minimize our familiar cost, i.e.,
\[
J(\bm{\theta})+\lambda\norm{\bm{\theta}}_1,
\]
we obtain (\ref{ch10-proximal4}); this is so, because the proximal operator
of $h(\bm{\theta}):=\lambda\norm{\bm{\theta}}_1$ is shown
(\cite{Combettes.Pesquet.proximal.2010, Comb-10}) to be identical to the soft thresholding
operator, i.e.,
\[
\prox_{h}(\bm{\theta})=S_{\lambda}(\bm{\theta}).
\]
In order to feel more comfortable with this operator, note that if
$h(\bm{x}):=0$, its proximal operator is equal to $\bm{x}$, and in this case
(\ref{ch10-proximal3}) becomes our familiar gradient descent algorithm.

\item All the non-greedy algorithms, which have been discussed so far, have
  been developed to solve the task defined in the formulation
  (\ref{ch10lassos1}). This is mainly because this is an easier task to
  solve; once $\lambda$ has been fixed, it is an unconstrained optimization
  task. However, there are algorithms which have been developed to solve
  the alternative formulations.

The NESTA\index{NESTA} algorithm has been proposed in \cite{Becker-10} and
solves the task in its (\ref{ch10lassos3}) formulation. Adopting this path
can have an advantage since $\epsilon$ may be given as an estimate of the
uncertainty associated with the noise, which can readily be obtained in a
number of practical applications. In contrast, selecting a-priori the value
for $\lambda$ is more intricate. In \cite{Chen-10}, the value
$\lambda=\sigma_\eta\sqrt{2\ln l}$, where $\sigma_\eta$ is the noise
standard deviation, is argued to have certain optimality properties;
however this argument hinges on the assumption of the orthogonality of $X$.
NESTA relies heavily on Nesterov's generic scheme (\cite{Nest-10}), hence
its name. The original Nesterov's algorithm performs a constrained
minimization of a smooth convex function $f(\bm{\theta})$, i.e.,
\[
\min_{\bm{\theta}\in Q}f(\bm{\theta})
\]
where $Q$ is a convex set, and in our case this is associated with the
quadratic constraint in (\ref{ch10lassos3}). The algorithm consists of
three basic steps. The first one is similar with the step in
(\ref{ch10:linearized}), i.e
\beq
\bm{w}^{(i)} = \argmin_{\bm{\theta}\in Q}
\left\{ \left(\bm{\theta} - \bm{\theta}^{(i-1)}\right)^T \frac{\partial
  J\left(\bm{\theta}^{(i-1)}\right)}{\partial\bm{\theta}}
+\frac{L}{2}\norm{\bm{\theta} - \bm{\theta}^{(i-1)}}^2_2
\right\},\label{ch10:nesta1}
\eeq
where $L$ is an upper bound on the Lipschitz coefficient, which the gradient
of $f(\cdot)$ has to satisfy. The difference with (\ref{ch10:linearized})
is that the minimization is now a constrained one. However, Nesterov has
also added a second step involving another auxiliary variable,
$\bm{z}^{(i)}$, which is computed in a similar way as $\bm{w}^{(i)}$ but
the linearised term is now replaced by a weighted cumulative gradient,
\[
\sum_{k=0}^{i-1}\alpha_k \left(\bm{\theta} - \bm{\theta}^{(k)}\right)^T \frac{\partial
  J\left(\bm{\theta}^{(k)}\right)}{\partial\bm{\theta}}.
  \]
 The effect of this term is to smooth out the ``zig-zagging'' of the path
 towards the solution, whose effect is to  increase significantly  the
 convergence speed. The final step of the scheme involves an averaging of
 the previously obtained variables,
 \[
 \bm{\theta}^{(i)}=t_i\bm{z}^{(i)}+(1-t_i)\bm{w}^{(i)}.
 \]
 The values of the parameters $\alpha_k$, $k=0,\ldots,i-1$, and $t_i$ result
 from the theory so that convergence is guaranteed. As it was the case with
 its close relative FISTA, the algorithm enjoys an $O(1/i^2)$ convergence
 rate. In our case, where the function to be minimized,
 $\norm{\bm{\theta}}_1$, is not smooth, NESTA uses a smoothed prox-function
 of it. Moreover, it turns out that close-form updates are obtained for
 $\bm{z}^{(i)}$ and $\bm{w}^{(i)}$. If $X$ is chosen so that to have
 orthonormal rows, the complexity per iteration is $O(l)$ plus the
 computations needed for performing the product $X^TX$, which is the most
 computational thirsty part. However, this complexity can substantially be
 reduced if the sensing matrix is chosen to be a submatrix of a unitary
 transform, which admits fast matrix-vector product computations, e.g., a
 subsampled DFT matrix. For example, for the case of a subsampled DFT
 matrix, the complexity amounts to $O(l)$ plus the load to perform the two Fast Fourier
 Transforms (FFT). Moreover, the continuation strategy can also be employed
 to accelerate convergence. In \cite{Becker-10}, it is demonstrated that
 NESTA exhibits good accuracy results, while retaining a complexity that is
 competitive with algorithms developed around the (\ref{ch10lassos3})
 formulation and scales in an affordable way for large size
 problems. Furthermore, NESTA, and in general Nesterov's scheme, enjoy a
 generality that allows their use to other optimization tasks as well.

 \item The task in (\ref{ch10lassos2}) has been considered in \cite{van-10}
   and \cite{Osbo-10}. In the former, the algorithm comprises a projection
   on the $l_1$ ball $\norm{\bm{\theta}}_1\le \rho$ (see also, Section
   \ref{cha10-dikomasmegale}) per iteration step. The most computationally
   dominant part of the algorithm consists of matrix-vector products. In
   \cite{Osbo-10}, a homotopy algorithm is derived for the same task, where
   now the bound $\rho$ becomes the homotopy parameter which is left to
   vary. This algorithm is also referred as the LARS-LASSO, as it has
   already been reported before.
\end{itemize}
\end{remarks}

\subsection{Which Algorithm Then: Some Practical Hints}
\label{ch-10-whichalg}

We have already discussed a number of algorithmic alternatives to obtain
solutions to the $\ell_0/\ell_1$ norm minimization task. Our
focus was on schemes whose computational demands are rather low and they
scale well to very large problem sizes. We have not touched more expensive
methods such as interior point methods for solving the $\ell_1$ convex
optimization task. A review of such methods is provided in
\cite{Kim-10}. Interior point methods
evolve along the Newton-type recursion and their complexity per iteration
step is at least of the order $\mathcal{O}(l^3)$. As it is most often
the case, there is a trade off. Schemes of higher complexity tend to result
in enhanced performance. However, such schemes become impractical in
problems of large size. Some examples of other algorithms, that were not
discussed, can be found in \cite{Yin-10, van-10, Daub2-10, Wright-10}.
Talking about complexity, it has to be pointed out that what really
matters at the end is not so much the complexity per iteration step but the
overall required resources in computer time/memory for the algorithm to
converge to a solution within a specified accuracy. For example, an
algorithm may be of low complexity per iteration step but it may need an
excessive number of iterations to converge.

Computational load is only one among a number of indices that characterize
the performance of an algorithm. Other performance measures, refer to convergence
rate, tracking speed (for the adaptive algorithms), and stability with
respect to the presence of noise and/or finite word length computations. No
doubt, all these performance measures are also of interest here,
too. However, there is an additional aspect that is of particular
importance when quantifying performance of sparsity-promoting algorithms.
This is related to the so called \textit{undersampling-sparsity tradeoff}
or the \textit{phase transition curve}\index{Phase transition curve}.

One of the major issues, on which we focused in this paper, was to
derive and present the conditions that guarantee uniqueness of the
$\ell_0$ minimization and its equivalence with the $\ell_1$
minimization task, under an underdetermined set of measurements,
$\bm{y}=X\bm{\theta}$, for the recovery of sparse enough
signals/vectors. While discussing the various algorithms in this
section, we reported a number of different RIP-related conditions that
some of the algorithms have to satisfy in order to recover the target
sparse vector. As a matter of fact, it has to be admitted that this
was quite confusing, since each algorithm had to satisfy its own
conditions. In addition, in practice, these conditions are not easy to
be verified. Although such results are, no doubt, important to
establish convergence and make us more confident and also understand
better why and how an algorithm works, one needs further experimental
evidence in order to establish good performance bounds for an
algorithm. Moreover, all the conditions that we have dealt with,
including coherence and RIP, are sufficient conditions. In practice,
it turns out that sparse signal recovery is possible with sparsity
levels much higher than those predicted by the theory, for given $N$
and $l$. Hence, proposing a new algorithm or selecting an algorithm
from an available palette, one has to demonstrate experimentally the
range of sparsity levels that can be recovered by the algorithm, as a
percentage of the number of measurements and the
dimensionality.  Thus, in order to select an algorithm, one should
  cast his/her vote for the algorithm which, for given $l$ and $N$, has
  the potential to recover $k$-sparse vectors with $k$ being as high
  as possible, for most of the cases, that is, with {\it high
    probability}.

\begin{figure}[!tbp]
\centering
\includegraphics[scale=1]{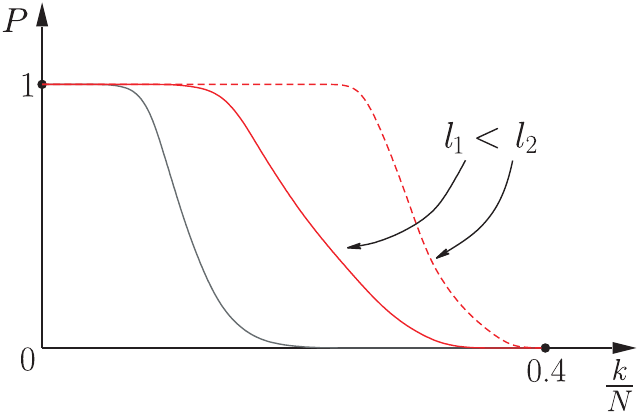}
\caption{For any algorithm, the transition between the regions of 100\% success and of a complete failure is very sharp. For the algorithm corresponding to the red curve, this transition occurs at higher sparsity values and, from this point of view, it is a better algorithm than the one associated with the black curve. Also, given a algorithm, the higher the dimensionality the higher the sparsity level where this transition occurs, as indicated by the two red curves.}\label{fig:prob.vs.sparsity.level}
\end{figure}

Fig.~\ref{fig:prob.vs.sparsity.level} illustrates the type of curve that is
expected to result in practice. The vertical axis is the probability of
exact recovery of a target $k$-sparse vector and the horizontal axis shows
the ratio $k/N$, for a given number of measurements, $N$, and the
dimensionality of the ambient space, $l$. Three curves are shown. The red
ones correspond to the same algorithm, for two different values of the
dimensionality, $l$, and the gray one corresponds to another
algorithm. Curves of this shape are expected to result from experiments of
the following set up. Assume that we are given a sparse vector,
$\bm{\theta}_0$, with $k$ nonzero components in the $l$-dimensional
space. Using a sensing matrix $X$, we generate $N$ measurements
$\bm{y}=X\bm{\theta}_0$. The experiment is repeated a number of, say, $M$
times, each time using a different realization of the sensing matrix and a
different $k$-sparse vector. For each instance, the algorithm is run to
recover the target sparse vector. This is not always possible. We count the
number, $m$, of successful recoveries, and compute the corresponding
percentage of successful recovery (probability), $m/M$, which is plotted on
the vertical axis of Fig.~\ref{fig:prob.vs.sparsity.level}. The procedure
is repeated for a different value of $k$, $1\le k\le N$. A number of issues
now jump into the scene: a) How one selects the ensemble of sensing
matrices and b) how one selects the ensemble of sparse vectors. There are
different scenarios and some typical examples are described next.

\begin{enumerate}
\item The $N\times l$ sensing matrices $X$ are formed by:
\begin{enumerate}
\item   Different i.i.d.\ realizations  with elements drawn from a Gaussian
  $\mathcal{N}(0,1/N)$.
\item   Different i.i.d.\ realizations from the uniform distribution on
  the unit sphere in $\Real^N$, which is also known as the uniform
  spherical ensemble.
\item   Different i.i.d.\ realizations with elements drawn from
  Bernoulli type distributions.	
\item   Different i.i.d.\ realizations of partial Fourier matrices, each
  time using a different set of $N$ rows.
\end{enumerate}
\item The $k$-sparse target vector $\bm{\theta}_0$ is formed by selecting
  the locations of at most $k$ nonzero elements randomly, by ``tossing a
  coin'' with probability $p=k/l$, and fill the values of the nonzero
  elements according to a statistical distribution, e.g., Gaussian,
  uniform, double exponential, Cauchy.
\end{enumerate}

Other scenarios are also possible. Some authors set all nonzero values
to one, \cite{Blum-10}, or to $\pm 1$, with the randomness imposed on
the choice of the sign.  It must be stressed out that the performance
of an algorithm may vary significantly under different experimental
scenarios, and this may be indicative of the stability of an
algorithm. In practice, a user may be interested in a specific
scenario, which is more representative of the available data.

Looking at Fig.~\ref{fig:prob.vs.sparsity.level}, the following conclusions
are in order. In all curves, there is a sharp transition between two
levels. From the $100\%$ success to the $0\%$ success. Moreover, the higher
the dimensionality, the sharper the transition is. This has also been shown
theoretically in \cite{Dono05-10}. For the algorithm corresponding to the
red curves, this transition occurs at higher values of $k$, compared to the
algorithm that generates the curve drawn in gray. Provided that the
computational complexity of the ``red'' algorithm can be accommodated by
the resources, which are available for a specific application, this seems
to be the more sensible choice between the two algorithms. However, if the
resources are limited, concessions are unavoidable.

Another way to ``interrogate'' and demonstrate the performance of an
algorithm, with respect to its robustness to the range of values of
sparsity levels that can be successfully recovered, is via the
so-called {\it phase transition curve}\index{Phase Transition
  curve}. To this end define:
\begin{itemize}
\item $\alpha \coloneqq \frac{N}{l}$, which is a normalized measure of the
  problem indeterminacy.
\item $\beta \coloneqq \frac{k}{N}$, which is a normalized measure of
  sparsity.
\end{itemize}
In the sequel, plot a graph having $\alpha\in [0,1]$ in the horizontal axis
and $\beta\in [0,1]$ in the vertical one. For each point, $(\alpha,\beta)$,
in the $[0,1]\times [0,1]$ region, compute the probability of the algorithm
to recover a $k$-sparse target vector. In order to compute the probability,
one has to adopt one of the previously stated scenarios. In practice, one
has to form a grid of points that cover densely enough the region
$[0,1]\times [0,1]$ in the graph. Use a varying intensity level scale to color the
corresponding $(\alpha, \beta)$ point. Black corresponds to probability one
and red to probability zero. Fig.~\ref{fig:alpha_vs_beta}, illustrates
the type of graph that is expected to be recovered in practice, for large
values of $l$. That is, the transition from the region (phase) of ``success''
(black) to that of ``fail'' (red) is very sharp. As a matter of fact, there
is a curve that separates the two regions. The theoretical aspects of this curve have been studied in the context of combinatorial geometry in
\cite{Dono05-10} for the asymptotic case, $l\rightarrow \infty$, and in \cite{DonoTanPha-10} for finite values of $l$.
Observe that
the larger the value of $\alpha$ (larger percentage of measurements) the
larger the value of $\beta$ at which the transition occurs. This is in line
with what we have said so far in this paper, and the problem gets
increasingly harder as one moves up and to the left in the graph. In
practice, for smaller values of $l$, the transition region from white to
black is smoother, and it gets narrower as $l$ increases. In practice, one
can draw an approximate curve that separates the ``success'' and ``fail''
regions, using regression techniques, see, e.g., \cite{Malek-10}.

\begin{figure}[!tbp]
\centering
\includegraphics[scale=1]{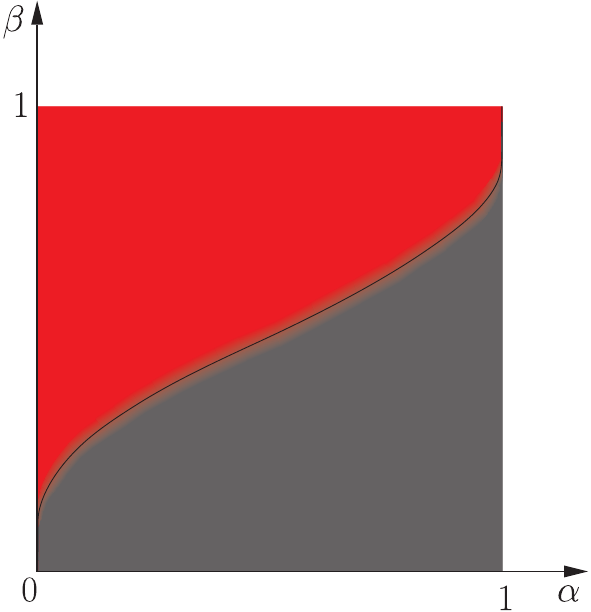}
\caption{ Typical phase transition behavior of a sparsity promoting algorithm. Black corresponds to 100\% success of recovering the sparsest solution and red to 0\%. For high dimensional spaces, the transition is very sharp, as it is the case in the figure. For lower dimensionality spaces, the transition from black to red is smoother and involves a region of varying color intensity.}\label{fig:alpha_vs_beta}
\end{figure}

The reader may already be aware of the fact that, so far, we have avoided
to talk about the performance of individual algorithms. We have just
discussed some ``typical'' behavior that algorithms tend to exhibit in
practice. What the reader might have expected is to discuss comparative
performance tests and draw related conclusions. We have not done it since
we feel that it is early in time to have ``definite'' performance
conclusions, and this field is still in an early stage. Most authors
compare their newly suggested algorithm with a few other algorithms,
usually within a certain algorithmic family and, more important, under some
specific scenarios, where the advantages of the newly suggested algorithm
are documented. However, the performance of an algorithm can change
significantly by changing the experimental scenario, under which the tests
are carried out. The most comprehensive comparative performance study, so
far, has been carried out in \cite{Malek-10}. However, even in this one,
the scenario of exact measurements has been considered and there are no
experiments concerning the robustness of individual algorithms to the
presence of noise. It is important to say that this study involved a huge
effort of computation. We will comment on some of the findings from this
study, which will also reveal to the reader that different experimental
scenarios can significantly affect the performance of an algorithm.


\begin{figure}[!tbp]
\centering
    \includegraphics[scale=0.8]{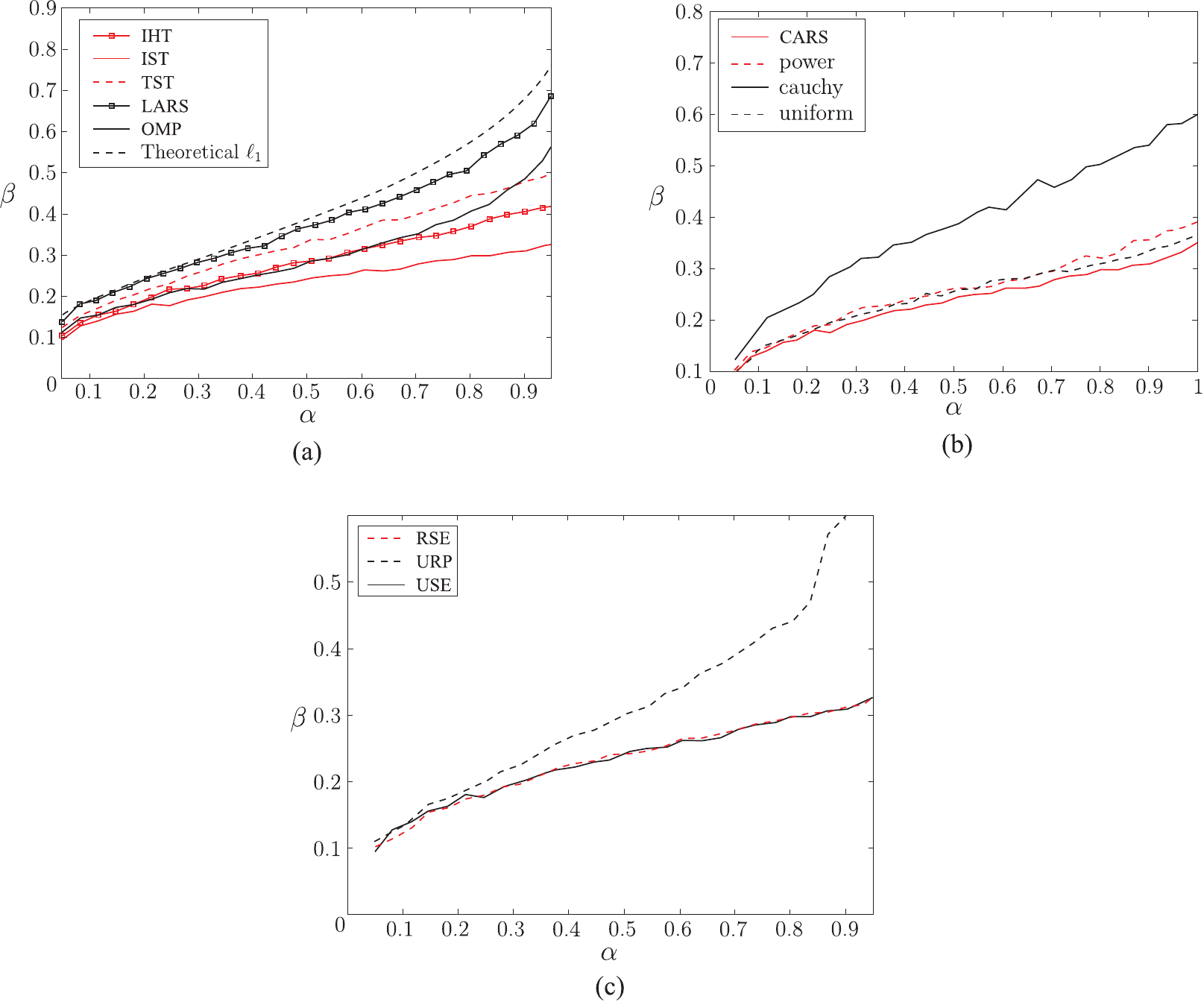}
        \caption{(a) The obtained phase transition curves for different algorithms under the same experimental scenario, together with the theoretical one. (b) Phase transition curve for the IST algorithm under different experimental scenarios for generating the target sparse vector. (c) The phase transition for the IST algorithms under different experimental scenarios for generating the sensing matrix $X$.}\label{fig:maleki}
\end{figure}

Fig.~\ref{fig:maleki}a shows the obtained phase transition curves for a) the
iterative hard thresholding (IHT), b) the iterative soft thresholding
scheme of \eqref{iteralg-10} (IST), c) the Two-Stage-Thresholding scheme
(TST), as discussed earlier on, d) the LARS and e) the OMP algorithms,
together with the theoretically obtained one using $\ell_1$
minimization. All algorithms were tuned with the optimal values, with
respect to the required user-defined parameters, after extensive
experimentation. The results in the Figure correspond to the uniform
spherical scenario, for the generation of the sensing matrices. Sparse
vectors were generated according to the $\pm 1$ scenario, for the nonzero
coefficients. The interesting observation is that, although the curves
deviate from each other as we move to larger values of $\beta$, for smaller
values, the differences in their performance become less and less. This is
also true for computationally simple schemes, such as the IHT one. The
performance of LARS is close to the optimal one. However, this comes at the
cost of computational increase. The required computational time for
achieving the same accuracy, as reported in \cite{Malek-10}, favor the TST
algorithm. In some cases, LARS required excessively longer time to reach
the same accuracy, in particular when the sensing matrix was the partial
Fourier one and fast schemes to perform matrix vector products can be
exploited.  For such matrices, the thresholding schemes (IHT, IST, TST)
exhibited a performance that scales very well to large size problems.

Fig.~\ref{fig:maleki}b indicates the phase transition curve for one of the
algorithms (IST) as we change the scenarios for generating the sparse
(target) vectors, using different distributions; a) $\pm 1$, with equiprobable
selection of signs (Constant Amplitude Random Selection (CARS)), b) double
exponential (power), c) Cauchy and d) uniform in $[-1,1]$.  This is
indicative and typical for other algorithms as well, with some of them
being more sensitive than others. Finally, Fig.~\ref{fig:maleki}c shows
the transition curves for the IST algorithm by changing the sensing matrix
generation scenario. Three curves are shown corresponding to a) uniform
spherical ensemble (USE), b) random sign ensemble (RSE), where the elements
are $\pm 1$ with signs uniformly distributed and c) the uniform random
projection (URP) ensemble. Once more, one can observe the possible
variations that are expected due to the use of different matrix
ensembles. Moreover, changing ensembles affects each algorithm in a
different way.

Concluding this section it must be emphasized that the field of
algorithmic development is still an ongoing research field and it is
early to come with definite and concrete comparative performance
conclusions. Moreover, besides the algorithmic front, existing theories often
fall short to predict what is observed in practice, with respect to their phase
transition performance. For a related discussion, see, e.g., \cite{DonoTan-10}.

\begin{example}\label{ch10toy}

We are given a set of $N=20$ measurements stacked in the $\bm{y}\in
\mathcal{R}^N$ vector. These were taken by applying a sensing matrix $X$ on
an ``unknown''  vector in $\mathcal{R}^{50}$, which is known to be sparse
with $k=5$ nonzero components; the location of these nonzero components in
the unknown vector are not known. The sensing matrix was a random matrix
with elements drawn from a normal distribution $\mathcal{N}(0,1)$ and then
the columns were normalized to unit norm. There are two scenarios for the
measurements. In the first one, we are given the exact measurements while
in the second one white Gaussian noise of variance $\sigma^2=0.025$ was
added.

In order to recover the unknown sparse vector, the CoSaMP algorithm was
used for both scenarios.


\begin{figure}[!tbp]
\centering
    \includegraphics[scale=1]{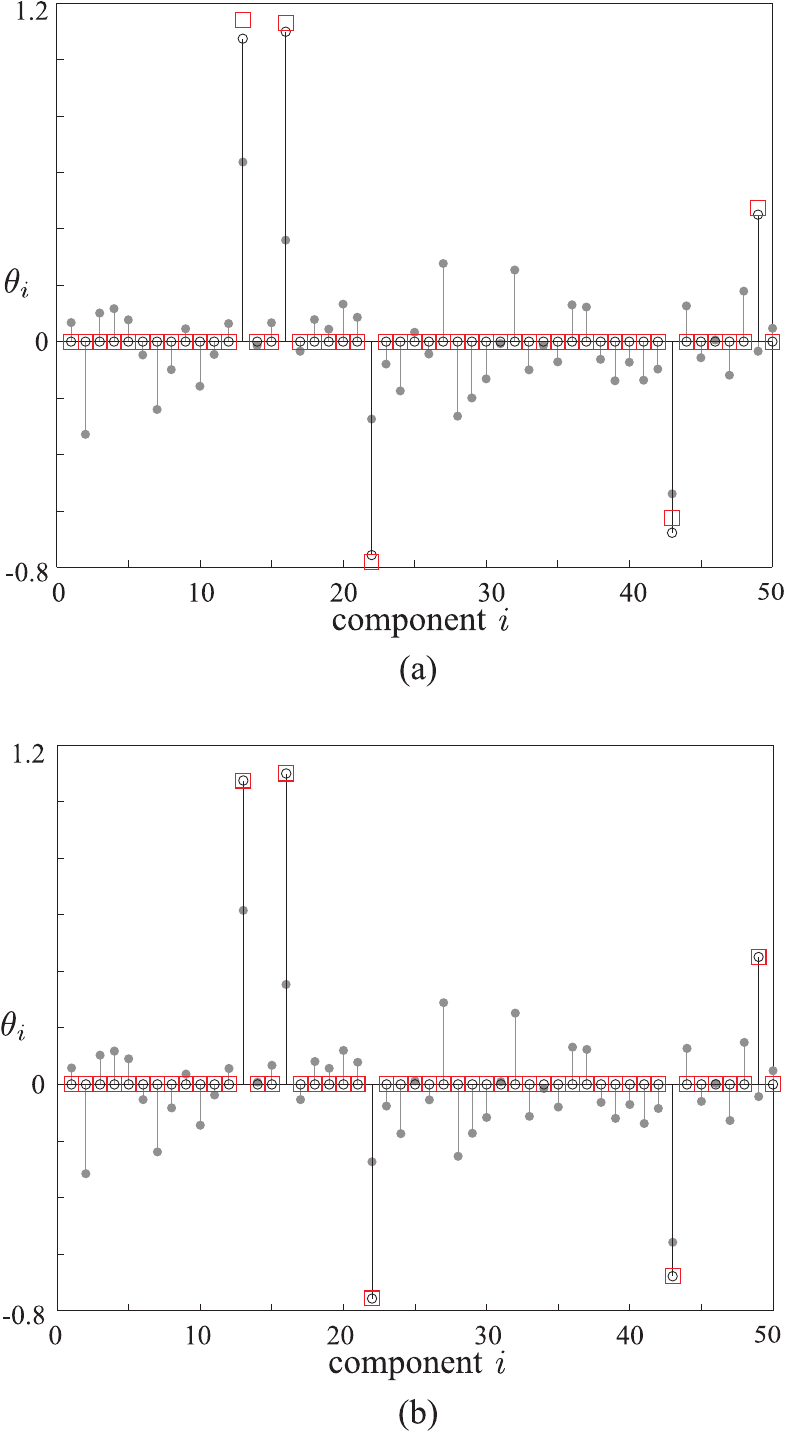}
        \caption{(a) Noiseless case. The values of the true vector, which generated the data for
  the Example~\ref{ch10toy}, are shown with stems topped with open
  circles. The recovered points, using the CoSaMP, are shown with
  squares. An exact recovery of the signal has been obtained. The stems
  topped with gray circles correspond to the minimum Euclidean norm LS solution. (b)
  This figure corresponds to the noisy counterpart of that in (a)
  In the presence of noise, exact
  recovery is not possible and the more the power of the noise is the less
  accurate the results are.}\label{ch10fig:toy}
\end{figure}

The results are shown in Figs.~\ref{ch10fig:toy}a and
\ref{ch10fig:toy}b for the noiseless and noisy scenarios,
respectively.  The values of the true unknown vector $\bm{\theta}$ are
represented with black stems topped with open circles. Note that all but
five of them are zero. In Fig.~\ref{ch10fig:toy}a exact recovery of the
unknown values is succeeded; the estimated values of $\theta_i,
i=1,2\ldots,50$, are indicated with squares in red color. In the noisy case
of Fig.~\ref{ch10fig:toy}b, the resulted estimates, which are
denoted with squares, deviate from the correct values. Note that estimated
values very close to zero ($|\theta|\le 0.01$) have been omitted from the
figure in order to facilitate visualizing. In both figures, the stemmed
gray filled circles correspond to the  minimum $\ell_2$ norm LS
solution. The advantages of adopting a sparsity-promoting approach to
recover the solution are obvious.  The CoSaMP algorithm was provided with the exact
number of sparsity. The reader is advised to play with this example by
experimenting with different values of the parameters and see how results
are affected.
\end{example}

\section{Variations on the Sparsity-Aware Theme}
\label{ch10-sparsvar}

In our tour, so far, we have touched a number of aspects of the
sparsity-aware learning that come from  the main stream of the theoretical
developments. However, more and more variants appear, which are developed
with the goal to address problems of a more special structure and/or to
propose alternatives, which can be beneficial in boosting the performance
in practice, by serving the needs of specific applications. These variants
focus either on the regularization term in \eqref{ch10lassos1} or on the misfit-measuring term
or on both.
Once more, research
activity in this direction is dense and our purpose is to simply highlight
possible alternatives and make the reader alert of the various
possibilities that spring from the basic theory.

In a number of tasks, it is a-priori known that the nonzero coefficients in
the target signal/vector occur in groups and they are not randomly spread
in all possible positions. Such a typical example is the echo path in
internet telephony, where the nonzero coefficients of the impulse response
tend to cluster together, see Fig.~\ref{fig:echo.path}. Other examples of
``structured'' sparsity can be traced in DNA microarrays, MIMO channel
equalization, source localization in sensor networks,
magnetoencephalography or in neuroscience problems, e.g., \cite{Baramb-10,
  Parva-10, Garri-10, Bara1-10}. As it is always the case in Machine
Learning, being able to incorporate a-priori information in the
optimization can only be of benefit for improving performance, since the
estimation task is externally assisted in its effort to search for the
target solution.

The \textit{group LASSO}\index{Group LASSO}, \cite{Bakin-10, Yuan-10,
  Oboz-10, Fried-10} addresses the task where it is a-priori known
that the nonzero components occur in groups. The unknown vector
$\bm{\theta}$ is divided into, say, $L$ groups, i.e.,
\[
\bm{\theta}^T=[\bm{\theta}^T_1,\ldots,\bm{\theta}^T_L]^T,
\]
each of them of a predetermined size, $s_i,~i=1,2,\ldots,L$, with
$\sum_{i=1}^L s_i=l$. The regression model can then be written as
\[
\bm{y}=X\bm{\theta}+\bm{\eta}=\sum_{i=1}^L X_i\bm{\theta}_i+\bm{\eta},
\]
where each $X_i$ is a submatrix of $X$ comprising the corresponding
$s_i$ columns. The solution of the group LASSO is given by the
following LS regularized task
\beq
\hat{\bm{\theta}} = \argmin_{\bm{\theta}\in {\cal R}^l} \left
(\norm{\bm{y}-\sum_{i=1}^LX_i\bm{\theta}_i}^2_2+\lambda
\sum_{i=1}^L\sqrt{s_i}\norm{\bm{\theta}_i}_2 \right),
\eeq
where $\norm{\bm{\theta}_i}_2$ is the Euclidean norm (not the squared
one) of $\bm{\theta}_i$, i.e.,
\[
\norm{\bm{\theta}_i}_2=\sqrt{\sum _{j=1}^{s_i}|\theta_{i,j}|^2}.
\]
In other words, the individual components of $\bm{\theta}$ that contribute
to the formation of the $\ell_1$ norm, in the standard LASSO formulation
are now replaced by the square root of
the energy of each individual block. In this setting, it is not the individual
components but blocks of them which are forced to zero, when their contribution
to the LS misfit measuring term is not significant. Sometimes, this type of
regularization is coined as the $\ell_1/\ell_2$ regularization. It is
straightforward to see that if $L=l$, then the group LASSO becomes the
standard LASSO method. An alternative formulation of the group sparse model
using greedy algorithms is considered in \cite{Eldargoup-10}.  Theoretical
results that extend the RIP to the so-called block RIP have been developed
and reported, see, e.g., \cite{Blumstr-10, Lu-10}.

In \cite{Cevh-10, Baramb-10}, the so-called \textit{model based} Compressed
Sensing is addressed. The $(k,C)$ model allows the significant coefficients
of a $k$-sparse signal to appear in at most $C$ clusters, whose size is
unknown. This is a major difference with the group LASSO, that was reported
before. In Section \ref{ch10compressive}, it was commented that searching
for a $k$-sparse solution takes place in a union of subspaces, each one of
dimensionality $k$. Imposing a certain structure on the target solution
restricts the searching in a subset of these subspaces and leaves a number
of these out of the game. This obviously facilitates the optimization
task. In \cite{Cevh-10}, a dynamic programming technique is adopted to
obtain the solution. In \cite{CevhMag-10}, structured sparsity is
considered in terms of graphical models. An even more advanced block
sparsity model is the C-HiLasso, which allows each block to have a sparse
structure itself, \cite{Sprech-10}.
%

\begin{figure}[!tbp]
    \centering
    \includegraphics[scale=1]{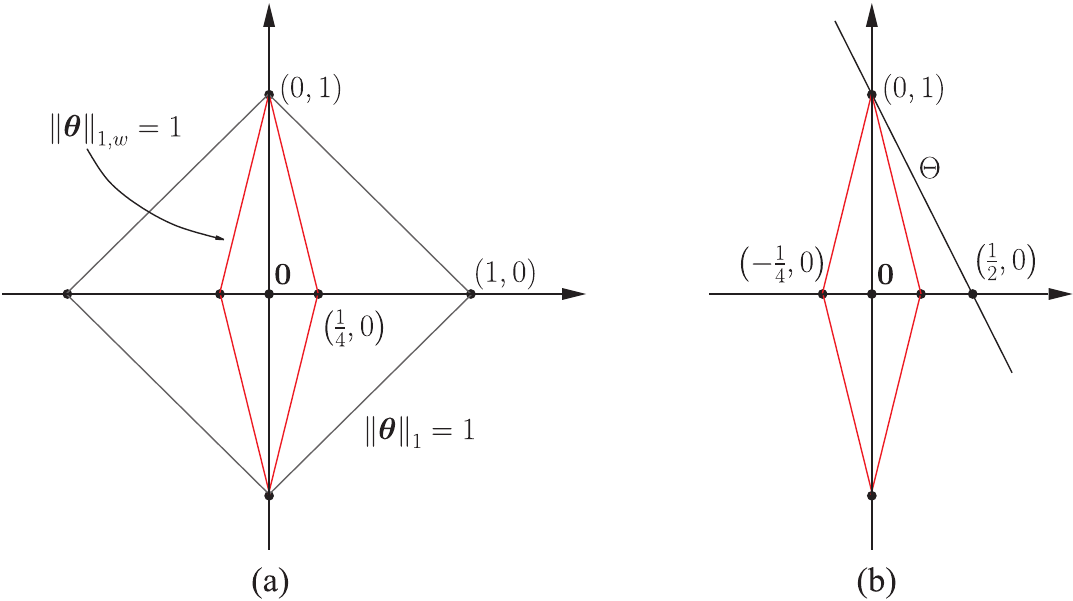}
    \caption{(a) The isovalue curves for the $\ell_1$ and the weighted $\ell_1$ norms for the same value. The weighted $\ell_1$ is sharply pinched around one of the axis, depending on the weights. (b) Adopting to minimize the weighted $\ell_1$ norm, for the setup of Figure \ref{fig:example.l1.min} the correct sparse solution is obtained. }\label{fig:l1.weighted}
\end{figure}

In \cite{Cand4-10}, it is suggested to replace the $\ell_1$ norm by a
weighted version of it. To justify such a choice, let us recall
Example \ref{example:l1.min} and the case where the ``unknown'' system
was sensed using $\bm{x}=[2,1]^T$. We have seen that by ``blowing'' up
the $\ell_1$ ball, the wrong sparse solution was obtained. Let us now
replace the $\ell_1$ norm in \eqref{ch10:liminim} with its weighted
version
\[
\norm{\bm{\theta}}_{1,w} \coloneqq w_1|\theta_1|+w_2|\theta_2|, \quad w_1,
w_2>0,
\]
and set $w_1=4$ and $w_1=1$. Fig.~\ref{fig:l1.weighted}a shows the
isovalue curve $\norm{\bm{\theta}}_{1,w}=1$, together with that resulting
from the standard $\ell_1$ norm. The weighted one is sharply ``pinched''
around the vertical axis, and the larger the value of $w_1$ is, compared to
that of $w_2$, the sharper the corresponding ball will
be. Fig.~\ref{fig:l1.weighted}b shows what happens when ``blowing''
the weighted $\ell_1$ ball. It will first touch the point $(0,1)$, which is
the true solution. Basically, what we have done is to ``squeeze'' the
$\ell_1$ ball to be aligned more to the axis that contains the (sparse)
solution. For the case of our example, any weight $w_1>2$ would do the job.

Considering now the general case of a weighted norm
\[
\norm{\bm{\theta}}_{1,w} \coloneqq \sum_{j=1}^lw_j|\theta_j|, \quad w_j>0.
\]
The ideal choice of the weights would be
\[
w_j=\begin{cases}
\frac{1}{|\theta_{0,j}|}, & \theta_{0,j}\neq 0, \\
\infty, & \theta_{0,j} = 0,
\end{cases}
\]
where $\bm{\theta}_0$ is the target true vector, and where we have
silently assumed that $0\cdot \infty=0$. In other words, the
smaller a coefficient is the larger the respective weight becomes. This is
justified, since large weighting will force respective coefficients
towards zero during the minimization process. Of course, in practice
the values of the true vector are not known, so it is suggested to use
their estimates during each iteration of the minimization
procedure. The resulting scheme is of the following form.

\begin{algo}\label{algo:l1.weighted}\mbox{}
\begin{enumerate}
\item Initialize weights to unity, $w^{(0)}_j=1$, $j=1,2,\ldots,l$.
\item Minimize the weighted $\ell_1$ norm,
\beqan
\bm{\theta}^{(i)}&=&\argmin_{\bm{\theta}\in \Real^l} \norm{\bm{\theta}}_{1,w}\\
&& \text{s.t.}\ \bm{y}=X\bm{\theta}.
\eeqan
\item Update the weights
\[
w_j^{(i+1)}=\frac{1}{\left|\theta_j^{(i)}\right|+\epsilon}, \quad j=1,2,\ldots,l.
\]
\item Terminate when a stopping criterion is met, otherwise return to step 2.
\end{enumerate}
\end{algo}

The constant $\epsilon$ is a small user-defined parameter to guarantee
stability when the estimates of the coefficients take very small values.
Note that if the weights have constant preselected values, the task retains
its convex nature; this is no longer true when the weights are changing.
It is interesting to point out that this intuitively motivated weighting
scheme can result if the $\ell_1$ norm is replaced by
$\sum_{j=1}^l\ln\left(|\theta_j|+\epsilon\right)$ as the regularizing
term of \eqref{ch10lassos1}. Fig.~\ref{fig:l1.and.ln} shows the respective graph, in the
one-dimensional space together with that of the $\ell_1$ norm. The graph of
the logarithmic function reminds us of the $\ell_p$, $p<0<1$, ``norms'' and
the comments made in Section \ref{ch10:norm}. This is no more a convex
function and the iterative scheme, given before, is the result of a
majorization-minimization procedure in order to solve the resulting
non-convex task, \cite{Cand4-10}.

\begin{figure}[!tbp]
\centering
\includegraphics[scale=1]{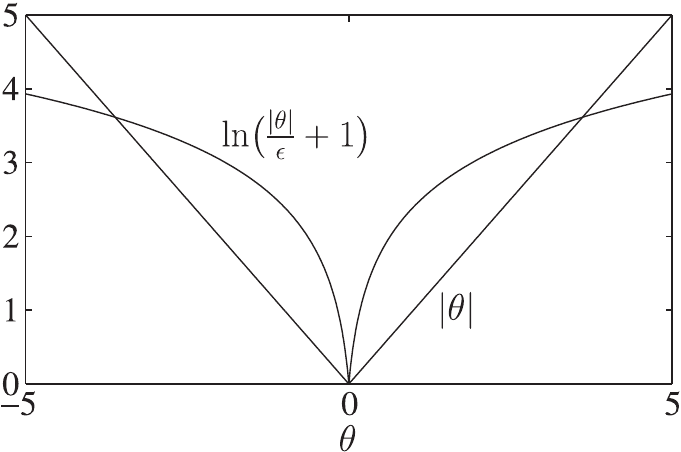}
\caption{One-dimensional graphs of the $\ell_1$ norm and the logarithmic
  regularizer $\ln\bigl(\frac{|\theta|}{\epsilon}+1 \bigr)=\ln\bigl(|\theta|+\epsilon\bigr)-\ln\epsilon$, with $\epsilon =
  0.1$. The term $\ln\epsilon$ was subtracted for illustration purposes only and does not affect the optimization. Notice the nonconvex nature of the logarithmic regularizer.}\label{fig:l1.and.ln}
\end{figure}

The concept of the iterative weighting, as used before, has also been
applied in the context of the {\it iterative reweighted least squares
  algorithm}. Observe that the $\ell_1$ norm can be written as
\[
\norm{\bm{\theta}}_1 =
\sum_{j=1}^l|\theta_j|=\bm{\theta}^T\mathcal{W}_{\theta}\bm{\theta},
\]
where
\[
\mathcal{W}_{\theta} = \begin{bmatrix}
\frac{1}{|\theta_1|} & 0 & \cdots & 0\\
0 & \frac{1}{|\theta_2|} & \cdots & 0\\
\vdots & \vdots & \ddots & \vdots\\
0 & 0 & \cdots & \frac{1}{|\theta_l|}
\end{bmatrix},
\]
and where in the case of $\theta_i=0$, for some $i\in\{1,2,\ldots,l\}$, the
respective coefficient of $\mathcal{W}_{\theta}$ is defined to be $1$. If
$\mathcal{W}_{\theta}$ were a constant weighting matrix, i.e.,
$\mathcal{W}_{\theta} \coloneqq \mathcal{W}_{\tilde{\theta}}$, for some
fixed $\tilde{\bm{\theta}}$, then obtaining the minimum
\[
\hat{\bm{\theta}} = \argmin_{\bm{\theta}\in \Real^l}
\norm{\bm{y}-X\bm{\theta}}^2_2 + \lambda
\bm{\theta}^T\mathcal{W}_{\tilde{\theta}}\bm{\theta},
\]
is straightforward and similar to the ridge regression. In the iterative
reweighted scheme, $\mathcal{W}_{\theta}$ is replaced by
$\mathcal{W}_{{\theta}^{(i)}}$, formed by using the respected estimates of
the coefficients which have been obtained from the previous iteration,
i.e., $\tilde{\bm{\theta}} \coloneqq \bm{\theta}^{(i)}$, as we did
before. In the sequel, each iteration solves a weighted ridge regression
task. Variants of this basic iteratively weighting scheme have also been
proposed, see, e.g., \cite{Daub2-10} and the references therein.

In \cite{candestao-10}, the LASSO task is modified by replacing the
square error term with one involving  correlations and the
minimization task becomes
\beqan
\hat{\bm{\theta}}: \min_{\bm{\theta}\in \Real^l} & &  \norm{\bm{\theta}}_1 \nonumber \\
\text{s.t.} & &  \norm{X^T(\bm{y}-X\bm{\theta})}_\infty \le \epsilon,
\eeqan
where $\epsilon$ is related to $l$ and the noise variance. This task is
known as the \textit{Dantzig selector}\index{Dantzig selector}. That
is, instead of constraining the energy of the error, the constraint,
now, imposes an upper limit to the correlation of the error vector
with any of the columns of $X$. In \cite{Bick-10, Asifa-10}, it is shown that
under certain conditions  the LASSO
estimator and the Dantzig selector become identical.

\textit{Total Variation (TV)}\index{Total Variation} \cite{rudin10} is a
closely related to $\ell_1$ sparsity promotion notion and it has been
widely used in image processing.  Most of the grayscale image arrays, $I\in
\mathcal{R}^{l\times l}$, consist of slowly varying pixel intensities
except at the edges.  As a consequence, the discrete gradient of an image
array will be approximately sparse (compressible).  The discrete
directional derivatives of an image array are defined pixel-wise as
\beqa
\nabla_{x}(I)(i,j)& \coloneqq &I(i+1,j)-I(i,j), \quad \forall i\in\{1,2,
\ldots, l-1\},\\
\nabla_y(I)(i,j)& \coloneqq &I(i,j+1)-I(i,j), \quad \forall j\in\{1,2,
\ldots, l-1\},
\eeqa
and
\beq
\nabla_x(I)(l,j) \coloneqq \nabla_y(I)(i,l) \coloneqq 0, \quad \forall
i,j\in\{1,2,\ldots, l-1\}.
\eeq
The discrete gradient transform
\[
\nabla:\mathcal{R}^{l\times l} \rightarrow \mathcal{R}^{l\times 2l},
\]
is defined in terms of a matrix form as
\beq
\nabla(I)(i,j) \coloneqq [\nabla_x(i,j), \nabla_y(i,j)], \quad \forall
i,j\in \{1,2,\ldots l\}.
\eeq
The total variation of the image array is defined as the $\ell_1$ norm of
the \textit{magnitudes} of the elements of the discrete gradient transform, i.e.,
\beq
\norm{I}_{\text{TV}} \coloneqq \sum_{i=1}^l\sum_{j=1}^l
\norm{\nabla(I)(i,j)}_2 =
\sum_{i=1}^l\sum_{j=1}^l\sqrt{\nabla_x(I)^2(i,j)+\nabla_y(I)^2(i,j)}.
\eeq
Note that this is a mixture of $\ell_2$ and $\ell_1$ norms. The sparsity
promoting optimization around the total variation is defined as
\beqa
I_*&\in&\mbox{arg}\min_I \norm{I}_{TV} \nonumber \\
&& \text{s.t.}\ \norm{\bm{y}-\mathcal{F}(I)}_2\le \epsilon,\label{ch10TV}
\eeqa
where $\bm{y}\in \mathcal{R}^N$ is the measurements vector and
$\mathcal{F}(I)$ denotes the result in vectorized form of the application
of a linear operator on $I$. For example, this could be the result of the
action of a partial two-dimensional DFT on the image. Subsampling of the
DFT matrix as a means to form sensing matrices has already been discussed
in Section \ref{ch10isom}. The task in (\ref{ch10TV}) retains its convex
nature and it basically expresses our desire to reconstruct an image which
is as smooth as possible given the available measurements. The NESTA
algorithm can be used for solving the total variation minimization task;
besides it, other efficient algorithms for this task can be found in, e.g.,
\cite{goldstein10, yang10}.

It has been shown in \cite{Cand1b}, for the exact measurements case
$(\epsilon=0)$, and in \cite{Needel1210}, for the erroneous measurements
case, that conditions and bounds which guarantee recovery of an image array
from the task in (\ref{ch10TV}) can be derived and are very similar with
those that we have discussed for the case of the $\ell_1$ norm.

\begin{example}[\textit{Magnetic Resonance Imaging (MRI)}]\label{ch10:MRI}

In contrast to ordinary imaging systems, which directly acquire pixel
samples, MRI scanners sense the image in an encoded form. Specifically, MRI
scanners sample components in the spatial frequency domain, known as
``$k$-space'' in the MRI nomenclature. If all the components in this
transform domain were available, one could apply the inverse 2D-DFT to
recover the exact MR image in the pixel domain. Sampling in the $k$-space
is realized along particular trajectories in a number of successive
acquisitions. This process is time consuming, merely due to physical
constraints. As a result, techniques for efficient image recovery from a
{\it limited number of measurements} is of high importance, since they can
reduce the required acquisition time for performing the measurements.  Long
acquisition times are not only inconvenient but even impossible, since the
patients have to stay still for long time intervals. Thus, MRI was among
the very first applications where compressed sensing found its way to offer
its elegant solutions.

Fig.~\ref{ch10fig:loganmri}a shows the ``famous'' Shepp-Logan phantom,
and the goal is to recover it via a limited number of (measurements)
samples in its frequency domain. The MRI measurements are taken across 17
radial lines in the spatial frequency domain, as shown in
Fig. \ref{ch10fig:loganmri}(b). A ``naive'' approach to recover the image
from this limited number of measuring samples would be to adopt a
zero-filling rationale for the missing components. The recovered image
according to this technique is shown in
Fig.~\ref{ch10fig:loganmri}(c). Fig.~\ref{ch10fig:loganmri}(d) shows the
recovered image using the approach of minimizing the total variation, as
explained before. Observe that the results for this case are astonishingly
good.  The original image is almost perfectly recovered.  The constrained
minimization was performed via the NESTA algorithm. Note that if the
minimization of the $\ell_1$ norm of the image array were used in place of
the total variation, the results would not be as good; the phantom image is
sparse in the discrete gradient domain, since it contains large sections
which share constant intensities.

\begin{figure}[!tbp]
\centering
\includegraphics[scale=1]{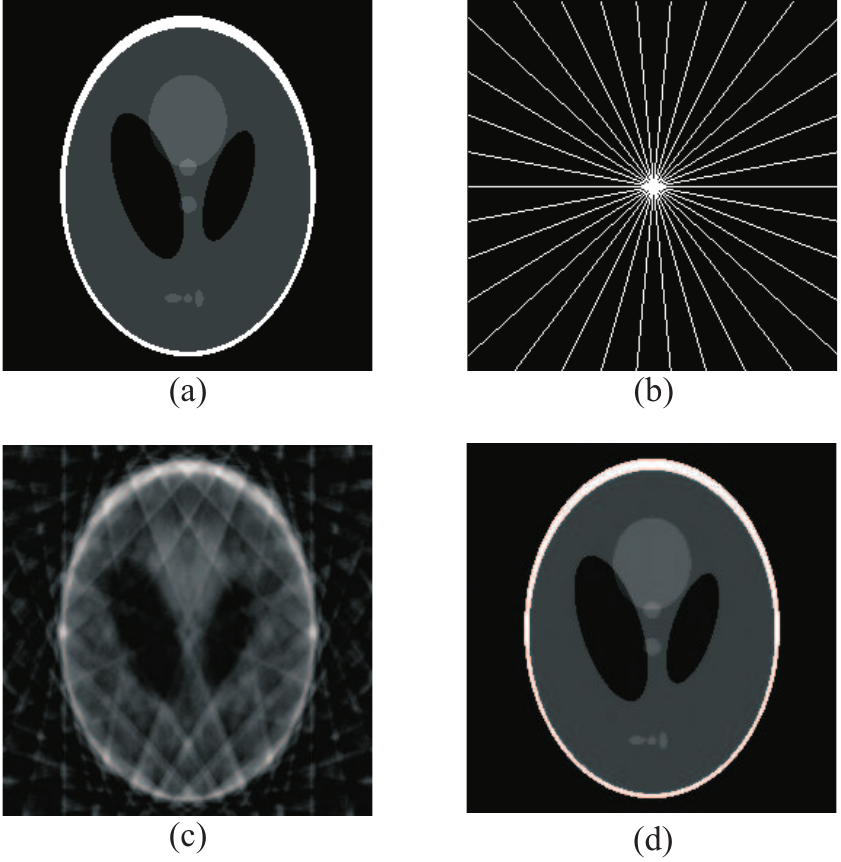}
\caption{a) The original Shepp-Logan image phantom. b) The white lines
  indicate the directions across which the sampling in the spatial Fourier
  transform were obtained. c) The recovered image after applying the
  inverse DFT having first filled with zeros the missing values in the DFT
  transform. d) The recovered image using the total variation minimization
  approach.}
\label{ch10fig:loganmri}
\end{figure}
\end{example}

\section{Online Time-Adaptive Sparsity-Promoting Algorithms}
\label{ch10-adspaawa}

In this section, online (time-recursive) schemes for sparsity-aware
learning are presented. There is a number of reasons that one has to resort
to such schemes in
various signal processing tasks, for example, when the data arrive sequentially. Under such a
scenario, using batch processing techniques to obtain an estimate of an
unknown target parameter vector would be highly inefficient, since the
number of training points keeps increasing. Such an approach is prohibited
for real time applications. Moreover, time-recursive schemes can easily
incorporate the notion of adaptivity, when the learning environment is not
stationary but it undergoes changes as time evolves.  Besides signal
processing applications, there is an increasing number of machine learning
applications where online processing is of paramount importance, such as
bioinformatics, hyperspectral imaging, and data mining. In such
applications, the number of training points easily amounts to a few
thousand up to hundred of thousand points. Concerning the dimensionality of
the ambient (feature) space, one can claim numbers that lie in similar
ranges. For example, in \cite{Lang-10}, the task is to search for sparse
solutions in feature spaces with dimensionality as high as $10^9$ having
access to data sets as large as $10^7$ points. Using batch techniques, in a
single computer, is out of question with today's technology.

Let us assume that there is
an unknown parameter vector that generates data according the standard
regression model
\[
y_n=\bm{x}_n^T\bm{\theta}+\eta_n, \quad \forall n,
\]
and the training samples are received sequentially $(y_n,\bm{x}_n)$,
$n=1,2,\ldots$. In the case of a stationary environment, we would expect
our algorithm to converge, asymptotically as $n\rightarrow \infty$, to or
``near to'' the true parameter vector that gives birth to the measurements,
$y_n$, when it is sensed by $\bm{x}_n$. For time varying environments, the
algorithms should be able to track the underlying changes as time goes
by. Before we proceed, a comment is important. Since the time index, $n$,
is left to grow, all we have said in the previous sections with respect to
underdetermined systems of equations, looses its meaning. Sooner or later
we are going to have more measurements than the dimensionality of the
space. Our major concern here becomes the issue of asymptotic convergence,
for the case of stationary environments.  The obvious question, that is now
raised, is why not using a standard algorithm, e.g., LMS, RLS or APSM \cite{SayedBook, Haykin, TheodoridisSlavYamReview},
since we know that these algorithms
converge to, or near enough in some sense, to the solution; that is, the algorithm will
identify the zeros asymptotically. The answer is that if such algorithms
are modified to be aware for the underlying sparsity, convergence is
significantly speeded up; in real life applications, one has not the
``luxury'' to wait long time for the solution. In practice, a good
algorithm should be able to provide a good enough solution, and in the case
of sparse solutions to \textit{obtain the support}, after a reasonably
small number of iteration steps. In this section, the
powerful theory around the $\ell_1$ norm regularization will be used to obtain
sparsity-aware time adaptive schemes.

\subsection{LASSO: Asymptotic Performance}
\label{ch10-asperf}

The notions of bias, variance and consistency, are major indices for assessing
the performance of an estimator.  In a number of cases,
such performance measures are derived asymptotically. For example, it is well known that the maximum likelihood estimator is asymptotically unbiased and
consistent \cite{Theod-10}. Also the LS estimator is asymptotically consistent. Moreover, under the assumption that the noise
samples are i.i.d., the LS estimate, $\hat{\bm{\theta}}_N$, that is obtained
using $N$ measurement (training) samples, is itself a random vector, that
satisfies the $\sqrt{N}$-estimation consistency\index{Root square
  estimation consistence}, e.g., \cite{Kay}, i.e.,
\[
\sqrt{N} \left(\hat{\bm{\theta}}_N-\bm{\theta}_0 \right)
\xrightarrow{d} \mathcal{N} \left(\bm{0},\sigma^2\Sigma^{-1}\right),
\]
where $\bm{\theta}_0$ is the true vector that generates the
measurements, $\sigma^2$ denotes the variance of the noise source and
$\Sigma$ is the covariance matrix $\expect[\bm{x}\bm{x}^T]$ of the
input sequence, which has been assumed to be zero mean and the limit
denotes convergence in distribution.

The LASSO in \eqref{ch10lassos1} is the task of minimizing the $\ell_1$ norm regularized version
of the LS cost. However, nothing has been said, so far, about the
statistical properties of this estimator. The only performance measure that
we referred to was the error norm bound given in
\eqref{ch10-lassperfrip}. However, this bound, although important in the
context it was proposed for, does not provide much statistical information.
Since the introduction of the LASSO estimator, a number of papers have
addressed problems related to its statistical performance, see, e.g.,
\cite{DonoJo-10, Knight-10, Fan-10, Zou-10}.

When dealing with sparsity-promoting estimators, such as the LASSO, two
crucial issues emerge: a) whether the estimator, even asymptotically,
can obtain the support, if the true vector parameter is a sparse one
and b) quantify the performance of the estimator with respect to the
estimates of the nonzero coefficients, i.e., those whose index belongs
to the support. Especially for LASSO, the latter issue becomes to
study whether LASSO behaves as well as the unregularized LS with
respect to these nonzero components. This task was addressed, for a
first time and in a more general setting, in \cite{Fan-10}. Let the
support of the true, yet unknown, $k$-sparse parameter vector
$\bm{\theta}_0$ be denoted as $S$. Let also $\Sigma_{|S}$ be the
$k\times k$ covariance matrix $\expect[\bm{x}_{|S}\bm{x}_{|S}^T]$, where
$\bm{x}_{|S}\in\Real^k$ is the vector that contains only the $k$
components of $\bm{x}$, with indices in the support $S$. Then, we say
that an estimator satisfies asymptotically the {\it oracle
  properties}\index{Oracle properties} if:
\begin{itemize}
\item $\lim_{N\rightarrow \infty}\prob
  \left\{S_{\hat{\bm{\theta}}_N} = S \right\}=1 $. This is known as {\it
    support consistency}\index{support consistency}.
\item
  $\sqrt{N}\left(\hat{\bm{\theta}}_{N|S}-\bm{\theta}_{0|S}
  \right)\xrightarrow{d}
  \mathcal{N}\left(\bm{0},\sigma^2\Sigma^{-1}_{|S}\right)$. This is the
  $\sqrt{N}$-{\it estimation consistency}\index{Square root estimation
    consistency}.
\end{itemize}
We denote as $\bm{\theta}_{0|S}$ and $\bm{\theta}_{N|S}$ the
$k$-dimensional vectors which result from
$\bm{\theta}_0,~\hat{\bm{\theta}}_N$, respectively, if we keep the
components whose indices lie in the support $S$.  In other words,
according to the oracle properties, a good sparsity-promoting estimator
should be able a) to predict, asymptotically, the true support and b)
its performance with respect to the nonzero components should be as
good as that of a genie-aided LS estimator, which is informed, in advance,
of the positions of the nonzero coefficients.

Unfortunately, the LASSO estimator \textit{cannot} satisfy simultaneously
both conditions. It has been shown, \cite{Knight-10, Fan-10, Zou-10}
that:
\begin{itemize}
\item For support consistency, the regularization parameter
  $\lambda\coloneqq \lambda_N$ should be time varying such as
\[
\lim_{N\rightarrow\infty}\frac{\lambda_N}{\sqrt{N}}= \infty, \quad
\lim_{N\rightarrow\infty}\frac{\lambda_N}{N}=0.
\]
That is, $\lambda_N$ must grow faster than $\sqrt{N}$, but slower than
$N$.
\item For $\sqrt{N}$-consistency, $\lambda_N$ must grow as
\[
\lim_{N\rightarrow\infty}\frac{\lambda_N}{\sqrt{N}}= 0,
\]
i.e., it grows slower than $\sqrt{N}$.
\end{itemize}

The previous two conditions are conflicting and the LASSO estimator
cannot comply with the two oracle conditions simultaneously. The
proofs of the previous two points are somewhat technical and are not
given here. The interested reader can obtain them from the previously
given references. However, before we proceed, it is instructive to see
why the regularization parameter has to grow slower than $N$, in any
case. Without being too rigorous mathematically, recall that the LASSO
solution comes from equation \eqref{ch10lassos1}. This can be written
as
\beq
\bm{0} \in -\frac{2}{N}\sum_{n=1}^N\bm{x}_ny_n +\frac{2}{N}\left
(\sum_{n=1}^N\bm{x}_n\bm{x}_n^T \right)
\bm{\theta} +
\frac{\lambda_N}{N} \partial\norm{\bm{\theta}}_1, \label{ch10-lassadeq}
\eeq
where we have divided by $N$ both sides. Taking the limit as
$N\rightarrow \infty$, if $\lambda_N/N\rightarrow 0$, then we
are left with the first two terms; this is exactly what we would have
if the unregularized LS had been chosen as the cost function. In this case, the solution asymptotically
converges\footnote{Recall that this convergence is with probability
  $1$.} (under some general assumptions, which are assumed to hold
true, here) to the true parameter vector; that is, we have strong
consistency, e.g., \cite{Kay}.

\subsection{The Adaptive Norm-Weighted LASSO}\index{Adaptive LASSO}
\label{ch10-sadlasso}

There are two ways to get out of the previously stated conflict. One is to
replace the $\ell_1$ norm with a nonconvex function and this can lead to an
estimator that satisfies the oracle properties simultaneously
\cite{Fan-10}. The other is to modify the $\ell_1$ norm by replacing it
with a weighted version. Recall that the weighted $\ell_1$ norm was
discussed in Section \ref{ch10-sparsvar}, as a means to assist the
optimization procedure to unveil the sparse solution. Here the notion of
weighted $\ell_1$ norm comes as a necessity imposed by our willingness to
satisfy the oracle properties. This gives rise to the \textit{adaptive
  norm-weighted LASSO} cost estimator defined as\footnote{To emphasize
  that the number of training points is now increasing, we have used $n$ in
  place of $N$. Capital $N$ was previously used to denote a fixed
  number of points.}
\beq
\hat{\bm{\theta}}= \argmin_{\bm{\theta}\in \Real^l}\left \{
\sum_{j=1}^{n}\beta^{n-j}\left( y_j-\bm{x}_j^T\bm{\theta}\right)^2 +
\lambda_n\sum_{i=1}^l w_{i}(n)|\theta_i|\right\}, \label{ch10-timadlasso}
\eeq
where $\beta\le 1$ is used as the forgetting factor to allow for
tracking slow variations. The time varying weighting sequences is
denoted as $w_i(n)$. There are different options. In \cite{Zou-10}
and under a stationary environment with $\beta=1$, it is shown that if
\[
w_i(n)=\frac{1}{|\theta_i^{\text{est}}|^\gamma},
\]
where $\theta_i^{\text{est}}$ is the estimate of the $i$th component obtained
by \textit{any} $\sqrt{n}$-consistent estimator, such as the
unregularized LS, then for specific choices of $\lambda_n$ and
$\gamma$ the estimator satisfies the oracle properties
simultaneously. The main reasoning behind the weighted norms is that
as time goes by, and the $\sqrt{n}$-consistent estimator provides
better and better estimates, then the weights corresponding to indices
outside the true support (zero values) are inflated and those
corresponding to the true support converge to a finite value. This
helps the algorithm, simultaneously, to locate the support and obtain
unbiased (asymptotically) estimates of the large coefficients.

Another choice for the weighing sequence is related to the so called
\textit{Smoothly Clipped Absolute Deviation (SCAD)} \cite{Fan-10,
  Zou1-10}. This is defined as
\[
w_i(n)=\chi_{(0, \mu_n)}(|\theta^{\text{est}}_i|) + \frac{\left
  (\alpha\mu_n-|\theta_i^{\text{est}}|\right)_+}{(\alpha-1)\mu_n}
\chi_{(\mu_n,\infty)}(|\theta^{\text{est}}_i|),
\]
where $\chi(\cdot)$ stands for the characteristic function,
$\mu_n=\lambda_n/n$, and $\alpha>2$. Basically, this corresponds to a
quadratic spline function. It turns out, \cite{Zou1-10}, that if
$\lambda_n$ is chosen to grow faster that $\sqrt{n}$ and slower that
$n$, the adaptive LASSO, with $\beta=1$ satisfies both oracle
conditions, simultaneously.

A time adaptive scheme for solving the time adaptive norm-weighted (TNWL)
LASSO was presented in \cite{Agge-10}. The cost function of the
adaptive LASSO in \eqref{ch10-timadlasso} can be written as
\[
J(\bm{\theta})= \bm{\theta}^TR_n\bm{\theta} -
\bm{r}_n^T\bm{\theta}+\lambda_n\norm{\bm{\theta}}_{1,w(n)},
\]
where
\[R_n \coloneqq \sum_{j=1}^n\beta^{n-j}\bm{x}_j\bm{x}_j^T,
\quad\bm{r}_n \coloneqq \sum_{j=1}^n\beta^{n-j}y_j\bm{x}_j,
\]
and $\norm{\bm{\theta}}_{1,w(n)}$ is the weighted $\ell_1$ norm. It is straightforward to see, that
\[
R_n= \beta R_{n-1}+\bm{x}_n\bm{x}_n^T,
\quad\bm{r}_n=\beta\bm{r}_{n-1}+y_n\bm{x}_n.
\]
The complexity for both of the previous updates, for matrices of a
general structure, amounts to $\mathcal{O}(l^2)$ multiply/add
operations. One alternative is to update $R_n$ and $\bm{r}_n$ and then
solve a convex optimization task for each time instant, $n$, using any
standard algorithm. However, this is not appropriate for real time
applications, due to its excessive computational cost. In
\cite{Agge-10}, a time recursive version of a coordinate descent algorithm has
been developed. As we have seen in Section \ref{itershr-10},
coordinate descent algorithms update one component at each recursive
step. In \cite{Agge-10}, recursive steps are associated with time
updates, as it is always the case with the time-recursive
algorithms. As each new training pair $(y_n,\bm{x}_n)$ is received, a
single component of the unknown vector is updated. Hence, at each time
instant, a scalar optimization task has to be solved and its solution
is given in closed form, which results in a simple soft thresholding
operation (OCCD-TWL). If the weighted norm is to be used in place of the
$\ell_1$, a RLS is run in parallel to provide the necessary
weights. One of the drawbacks of the coordinate techniques is that
each coefficient is updated every $l$ time instants, which, for large
values of $l$, can slow down convergence. Variants of the basic scheme
that cope with this drawback are also addressed in \cite{Agge-10},  referred \
to as online cyclic coordinate descent (OCCD-TNWL). The
complexity of the scheme is of the order of
$\mathcal{O}(l^2)$. Computational savings are possible, if the input
sequence is a time series and fast schemes for the updates of $R_n$
and the RLS can then be exploited. However, if an RLS-type algorithm is used in parallel,
the convergence of the overall scheme may be slowed down, since the RLS-type algorithm has to
converge first, in order to provide reliable estimates for the weights, as pointed out before.

\subsection{Adaptive CoSaMP Algorithm (AdCoSaMP)}
\label{adcosmpach10}

In \cite{Mile-10}, an adaptive version of the CoSaMP algorithm, which was
presented in Section \ref{ch10-cosmp}, was proposed. Iteration steps, $i$,
now coincide with time updates, $n$, and the LS solver in Step
\ref{csmp.ls.task} of the general CSMP scheme is replaced by an LMS one.

Let us focus first on the quantity $X^T\bm{e}^{(i-1)}$ in Step \ref{2nd.t}
of the CSMP scheme, which is used to compute the support at iteration
$i$. In the adaptive setting and at (iteration) time $n$, this quantity is
now ``rephrased'' as
\[
X^T\bm{e}(n-1)=\sum_{j=1}^{n-1}\bm{x}_je(j).
\]
In order to make the algorithm  flexible to adapt to variations of the
environment, as the time index, $n$, increases, the previous correlation
sum is modified to
\[
\bm{p}(n):=
\sum_{j=1}^{n-1}\beta^{n-1-j}\bm{x}_je(j)=\beta\bm{p}(n-1)+\bm{x}_{n-1}e(n-1).
\]
The LS task, constrained on the active columns that correspond to the
indices in the support $S$ in Step \ref{csmp.ls.task}, is performed in an
adaptive rationale by involving the basic LMS recursions, i.e.,
\beqan
\tilde{e}(n)& \coloneqq &y_n-\bm{x}_{n|S}^T\tilde{\bm{\theta}}_{|S}(n-1)
\\ \tilde{\bm{\theta}}_{|S}(n)& \coloneqq &\tilde{\bm{\theta}}_{|S}(n-1) +
\mu\bm{x}_{n|S}\tilde{e}(n),
\eeqan
where $\tilde{\bm{\theta}}_{|S}(\cdot)$ and $\bm{x}_{n|S}$ denote the
respective subvectors corresponding to the indices in the support $S$. The
resulting algorithm is given as follows.

\begin{algo}[The AdCoSaMP Scheme]\mbox{}
\begin{enumerate}
\item Select the value of $t=2k$.
\item Initialize the algorithm: $\bm{\theta}(1) = \bm{0}$,
  $\tilde{\bm{\theta}}(1) = \bm{0}$, $\bm{p}(1) = \bm{0}$,
  $e(1)= y_1$.
\item Choose $\mu$ and $\beta$.
\item For $n=2,3,\ldots$, execute the following steps.
\begin{enumerate}
\item $\bm{p}(n)=\beta\bm{p}(n-1)+\bm{x}_{n-1}e(n-1)$.
\item Obtain the current support:
\begin{equation*}
S =  \supp\{\bm{\theta}(n-1)\}\cup \left\{\text{indices of
  the}\ t\ \text{largest}\atop \text{in magnitude components
  of}\ \bm{p}(n)\right\}.
\end{equation*}
\item Perform the LMS update:
\beqan
\tilde{e}(n) & =  & y_n-\bm{x}_{n|S}^T\tilde{\bm{\theta}}_{|S}(n-1), \\
 \tilde{\bm{\theta}}_{|S}(n)& = &\tilde{\bm{\theta}}_{|S}(n-1) +
 \mu\bm{x}_{n|S}\tilde{e}(n).
 \eeqan
 \item Obtain the set $S_k$ of the indices of the $k$ largest
   components of $\tilde{\bm{\theta}}_{|S}(n)$.
\item\label{adcosamp.hard.thresholding} Obtain $\bm{\theta}(n)$ such that:
\[
\bm{\theta}_{|S_k}(n) = \tilde{\bm{\theta}}_{|S_k},
\quad\text{and}\ \bm{\theta}_{|S^c_k}(n) = \bm{0},
\]
where $S_k^c$ is the complement set of $S_k$.
\item Update the error:
  $e(n)=y_n-\bm{x}_n^T\bm{\theta}(n)$.
\end{enumerate}
\end{enumerate}
In place of the standard LMS, its normalized version can alternatively be
adopted. Note that Step \ref{adcosamp.hard.thresholding} is directly
related to the hard thresholding operation.
\end{algo}

In \cite{Mile-10}, it is shown that if the sensing matrix, which is
now time dependent and keeps increasing in size, satisfies a
condition similar to RIP, for each time instant, called {\it Exponentially
Weighted Isometry Property} (ERIP)\index{Exponentially weighted isometry property}, which depends on $\beta$, then the
algorithm asymptotically satisfies an error bound, which is similar to the one
that has been derived for CoSaMP in \cite{Needl-10}, plus an extra
term that is due to the excess Mean Square Error,
which is the price paid by replacing the LS solver by the LMS.

\subsection{Sparse Adaptive Parallel Projection onto Convex Sets Method (SpAPSM)}\label{cha10-dikomasmegale}

The APSM family of algorithms is one
among the most powerful techniques for adaptive learning \cite{TheodoridisSlavYamReview}. A major advantage of this algorithmic family is that one can readily
incorporate convex constraints. The rationale behind APSM is that since our data are
known to be generated by a regression model, then the unknown vector could
be estimated by finding a point in the intersection of a sequence of
hyperslabs, that are defined by the data points, i.e., $S_n[\epsilon]
\coloneqq \bigl\{\bm{\theta}\in \Real^l: \left|y_n-\bm{x}^T_n\bm{\theta}
\right|\le \epsilon \bigr\}$. Such a model
is most natural when the noise is bounded, (which, after all, it is the case in any practical application). In case the noise is assumed unbounded, a choice of $\epsilon$ of the order say, $\sigma$, can guarantee, with high probability, that the unknown solution lies inside these hyperslabs.

 The APSM family builds upon the elegant philosophy that runs across the classical projections onto convex sets (POCS) theory. Recall that the basic rationale behind POCS is that starting from an arbitrary point in the space and sequentially projecting onto a \textit{finite number} of convex sets then the sequence of projections converges, in some sense, into the intersection of all these sets, assuming this is not empty. The theory was extended to embrace the online processing setting in \cite{YamadaOguraHybridQuasine, OguraYamadaNonStrictly, KostasAPSMatNFAO}. In contrast to the classical POCS theory, here the number of the involved convex sets is \textit{infinite}. It turns out that, under certain general conditions, a sequence of projections over all these sets also converges to a point in their intersection.

 To fit the theory into our needs, the place of the aforementioned convex sets is taken by the hyperslabs, which are formed by the received training data, as mentioned before. Thus, the resulting algorithms involves (metric) projections onto these hyperslabs (see Appendix). However, when dealing with sparse vectors, there is an extra projetion associated with the convex set formed by the $\ell_1$ ball; that is, $\norm{\bm{\theta}}_1\le \rho$ (see, also, the LASSO
formulation \eqref{ch10lassos2}). Hence, this task fits nicely in the APSM
rationale and the basic recursion can be readily written, without much
thought or derivation, as follows; for any arbitrarily chosen initial point
$\bm{\theta}(0)$, define $\forall n$,
\[
\bm{\theta}(n) \coloneqq P_{B_{\ell_1}[\delta]}\Biggl (
\bm{\theta}(n-1)+\mu_n\biggl( \frac{1}{q} \sum_{i=n-q+1}^n
P_{S_{i}[\epsilon]}\bigl(\bm{\theta}(n-1)\bigr)-\bm{\theta}(n-1)\biggr) \Biggr)
\]
where $P_{S_{i}[\epsilon]}$ is the metric projection onto the hyperslab $S_{i}[\epsilon]$ (see Appendix).  Note, that in the previous recursion we have used $q$, instead of one, hyperslabs whose metric projections are averaged out at time $n$. It turns out that such an averaging improves convergence significantly. Parameter $\mu_n$ is an extrapolation parameter, which takes values in the interval
$(0,2\mathcal{M}_n)$, where
\begin{equation}
\mathcal{M}_n \coloneqq \begin{cases}
\frac{\sum_{i=n-q+1}^n\omega^{(n)}_i
  \norm{P_{S_{i}[\epsilon]}(\bm{\theta}(n-1))-\bm{\theta}(n-1)}^2}
     {\norm{\sum_{i=n-q+1}^n\omega^{(n)}_i
P_{S_{i}[\epsilon]}(\bm{\theta}(n-1))-\bm{\theta}(n-1)}^2}, & \\
& \hspace{-150pt} \text{if}\ \norm{\sum_{i=n-q+1}^n\omega^{(n)}_i
P_{S_{i}[\epsilon]}(\bm{\theta}(n-1))-\bm{\theta}(n-1)}\neq 0,\\
1, & \hspace{-150pt} \text{otherwise},
\end{cases} \label{Mu.n}
\end{equation}
and $P_{{B_{\ell_1}[\rho]}}(\cdot)$ is the projection operator onto the
$\ell_1$ ball $B_{\ell_1}[\rho] \coloneqq \bigl\{\bm{\theta}\in \Real^l:
\norm{\bm{\theta}}_1\le \rho\bigr\}$, since the solution is constrained
to live within this ball. Note, that the previous recursion is analogous to the
iterative soft thresholding shrinkage algorithm in the batch processing
case, \eqref{ch10-eqithsc}. There, we saw that the only difference that the
sparsity imposes on an iteration, with respect to its unconstrained
counterpart, is an extra soft thresholding.  This is exactly the case
here. The term in the parenthesis is the iteration for the unconstrained
task. Moreover, as it has been shown in \cite{Duch-10}, projection on the
$\ell_1$ ball is equivalent to a soft thresholding
operation. It can be shown that the previous iteration converges arbitrarily close to a point in the
intersection
\[
B_{\ell_1}[\delta] \cap \bigcap_{n\ge n_0}S_n[\epsilon],
\]
for some finite value of $n_0$ \cite{YamadaOguraHybridQuasine, OguraYamadaNonStrictly, KostasAPSMatNFAO, TheodoridisSlavYamReview}. In \cite{Kops-10, Kops1-10} the
weighted $\ell_1$ ball has been used to improve convergence as well as
the tracking speed of the algorithm, when the environment is
time varying. The weights were adopted in accordance to what was discussed
in Section \ref{ch10-sparsvar}, i.e.,
\[
w_i(n) \coloneqq \frac{1}{|\theta_i(n-1)|+\acute{\epsilon}_n}, \quad
\forall i\in \{1,2,\ldots, l\},
\]
where $(\acute{\epsilon}_n)_{n\geq 0}$ is a sequence (can be also constant) of
small numbers to avoid division by zero. The basic time iteration
becomes as follows; for any arbitrarily chosen initial point
$\bm{\theta}(0)$, define $\forall n$,
\beq
\bm{\theta}(n) \coloneqq P_{B_{\ell_1}[\bm{w}(n),\rho]}\Biggl(
\bm{\theta}(n-1)+\mu_n\biggl( \sum_{i=n-q+1}^n\omega^{(n)}_i
P_{S_{i}[\epsilon]}\bigl(\bm{\theta}(n-1)\bigr)-\bm{\theta}(n-1)\biggr)
\Biggr),
\eeq
where $\mu_n\in (0,2\mathcal{M}_n)$ and $\mathcal{M}_n$ is given in
\eqref{Mu.n}. Fig.~\ref{fig:APSM.l1} illustrates the associated geometry of
the basic iteration in $\Real^2$ and for the case of $q=2$. It
comprises two parallel projections on the hyperslabs followed by one
projection onto the weighted $\ell_1$ ball. In \cite{Kops-10}, it is shown that a good bound for the weighted $\ell_1$ norm is
the sparsity level $k$ of the target vector, which is assumed to be known
and it is a user-defined parameter. In \cite{Kops-10}, it is shown that
asymptotically, and under some general assumptions, this algorithmic scheme
converges arbitrarily close to the intersection of the hyperslabs with the
weighted $\ell_1$ balls, i.e.,
\[
\bigcap_{n\ge n_0}\left(P_{B_{\ell_1}[\bm{w}(n),\rho]} \cap
S_j[\epsilon]\right),
\]
for some non-negative integer $n_0$. It has to be pointed out that, in the
case of weighted $\ell_1$ norms, the constraint is \textit{time varying}
and the convergence analysis is not covered by the standard analysis used
for APSM, and had to be extended to this more
general case. The complexity of the algorithm amounts to
$\mathcal{O}(ql)$. The larger the $q$ the faster the convergence rate, at
the expense of higher complexity. In \cite{Kops1-10}, in order to reduce
the dependence of the complexity on $q$, the notion of the
\textit{sub-dimensional} projection is introduced, where projections onto
 the $q$ hyperslabs could be restricted along the directions of the most
significant coefficients, of the currently available estimates. The dependence on $q$
now becomes $\mathcal{O}(qk_n)$ where $k_n$ is the sparsity level of the
currently available estimate, which, after a few steps of the algorithm,
gets much lower than $l$. The total complexity amounts to $\mathcal{O}(l)+
\mathcal{O}(qk_n)$, per iteration step. This allows the use of
large values of $q$, which drives the algorithm to a performance close to
that of the adaptive weighted LASSO, at only a small extra computational
cost.

\begin{figure}[!tbp]
\centering
\includegraphics[scale=0.8]{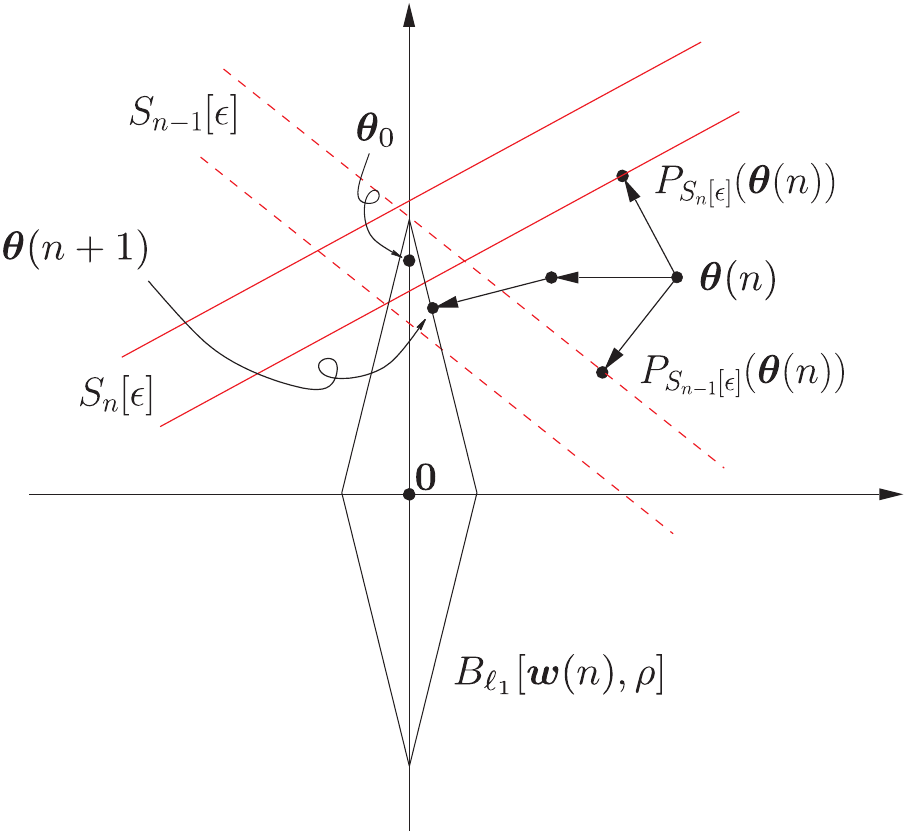}
\caption{Geometric illustration of the update steps involved in the SpAPSM algorithm, for the case of $q=2$. The update at time $n+1$ is obtained by first convexly combining the projections onto the current and previously formed hyperslabs, $S_{n}[\epsilon],~S_{n-1}[\epsilon]$ and then projecting onto the weighted $\ell_1$ ball. This brings the update closer to the target solution $\bm{\theta}_*$.}\label{fig:APSM.l1}
\end{figure}

\subsubsection{Projection onto the Weighted $\ell_1$ Ball}

Projecting onto an $\ell_1$ ball is equivalent to a soft thresholding
operation. Projection onto the weighted $\ell_1$ norm results to a slight
variation of the soft thresholding, with different threshold values per
component. In the sequel, we give the iteration steps for the more general
case of the weighted $\ell_1$ ball. The proof is a bit technical and
lengthy and it will not be given here. It was derived, for the first time,
via purely geometric arguments, and without the use of the classical
Lagrange multipliers, in \cite{Kops-10}. Lagrange multipliers have been
used instead in \cite{Duch-10}, for the case of the $\ell_1$ ball.

Given a point outside the ball, $\bm{\theta}\in\Real^l
\setminus B_{{\ell_1}}[\bm{w},\rho]$, then its projection onto the
weighted $\ell_1$ ball is the point
$P_{B_{\ell_1}[\bm{w},\rho]}(\bm{\theta})\in
B_{{\ell_1}}[\bm{w},\rho] \coloneqq \{\bm{z}\in \Real^l:
\sum_{i=1}^lw_i|z_i|\le \rho\}$, that lies closest to $\bm{\theta}$
in the Euclidean sense. If $\bm{\theta}$ lies within the ball then it
coincides with its projection. Given the weights and the value of
$\rho$, the following iterations provide the projection.

\begin{algo}[Projection onto the weighted $\ell_1$ ball
    {$B_{\ell_1}[\bm{w}, \rho]$}]\mbox{}

\begin{enumerate}
\item Form the vector
  $[|\theta_1|/w_1,\ldots,|\theta_l|/w_l]^T\in\Real^l$.

\item Sort the previous vector in a non-ascending order, so that
  $|\theta_{\tau(1)}|/w_{\tau(1)}\ge \ldots \ge
  |\theta_{\tau(l)}|/w_{\tau(l)}$. The notation $\tau$ stands for the
  permutation, which is implicitly defined by the sorting
  operation. Keep in memory the inverse $\tau^{-1}$, which is the
  index of the  position of the element in the original vector.

\item $r_1 \coloneqq l$.

\item Let $m=1$. While $m\le l$, do:

\begin{enumerate}
\item\label{start.inner.cycle} $m_*\coloneqq m$.

\item Find the maximum $j_*$ among those $j\in \{1,2,\ldots,r_m\}$
  such that
  $\frac{|\theta_{\tau(j)}|}{w_{\tau(j)}} >
  \frac{\sum_{i=1}^{r_m}w_{\tau(i)}|\theta_{\tau(i)}|-\rho}{\sum_{i=1}^{r_m}w^2_{\tau(i)}}$.

\item If $j_*=r_m$ then break the loop.

\item Otherwise set $r_{m+1} \coloneqq j_*$.

\item Increase $m$ by $1$ and go back to Step \ref{start.inner.cycle}.

\end{enumerate}

\item Form the vector $\hat{\bm{p}}\in {\cal R}^{r_{m_*}}$ whose
  $j$-th component, $j=1,...,r_{m_*}$, is given by
\[
\hat{p}_j\coloneqq |\theta_{\tau(j)}|-
\frac{\sum_{i=1}^{r_{m_*}}w_{\tau(i)}|
  \theta_{\tau(i)}|-\rho}{\sum_{i=1}^{r_{m_*}}w^2_{\tau(i)}}w_{\tau(j)}.
\]
\item Use the inverse mapping $\tau^{-1}$ to insert the element
  $\hat{p}_j$ into the $\tau^{-1}(j)$ position of the $l$-dimensional
  vector $\bm{p}$, $\forall j\in\{1,2,\ldots r_{m_*}\}$, and fill in
  the rest with zeros.

\item The desired projection is
  $P_{B_{\ell_1}[\bm{w},\rho]}(\bm{\theta}) = [\sign(\theta_1)p_1,\ldots,
  \sign(\theta_l)p_l]^T$.
\end{enumerate}
\end{algo}

\begin{remarks}\label{rem12}
Projections onto both $\ell_1$ and weighted $\ell_1$ balls impose convex sparsity inducing constraints via properly performed soft thresholding operations. More recent advances within the SpAPSM framework \cite{Kops_icassp12}, allow the substitution of $P_{B_{\ell_1}[\rho]}$ and $P_{B_{\ell_1}[\bm{w},\rho]}$ with a \textit{generalized thresholding}, built around the notions of SCAD, nonegative garrote, as well as a number of thresholding functions corresponding to the non-convex, $\ell_p$, $p<1$ penalties. Moreover, it is shown shown that such generalized thresholding operators (GT) are  nonlinear mappings with their fixed point set being a union of subspaces, i.e., the non-convex object which lies at the heart of any sparsity-promoting technique. Such schemes are very useful for low values of $q$, where one can improve upon the performance obtained by the LMS-based  AdCoSAMP, at comparable complexity levels.
\end{remarks}

\begin{example}\label{exampleonline}(\textit{Time varying signal})
In this example, the performance curves of the most typical online algorithms, mentioned before, are studied in the context of a time varying environment. A typical simulation setup, which is commonly adopted by the adaptive filtering community, in order to study the tracking agility of an algorithm, is that of an unknown vector which undergoes an abrupt change after a number of observations. Here, we consider a signal, $\bm{s}$, with a sparse wavelet representation, i.e., $\bm{s}=\Phi \bm{\theta}$, where $\Phi$ is the corresponding transformation matrix. In particular, we set $l=1024$ with $100$ nonzero wavelet coefficients. After $1500$ measurements (observations), ten arbitrarily picked wavelet coefficients change their values to new ones selected uniformly at random from the interval $[-1~1]$. Note that this may affect the sparsity level of the signal, and we can now end with up to 110 nonzero coefficients. A total of $N=3000$ sensing vectors are used, which result from the wavelet transform of the input vectors $\bm{x}_n \in \Real^l,~n=1,2,\ldots,3000$, having elements drawn from $\mathcal{N}(0,1)$. In this way, the adaptive algorithms do not estimate the signal itself, but its sparse wavelet representation, $\bm{\theta}$. The observations are corrupted by additive white Gaussian noise of variance $\sigma_n^2=0.1$. Regarding SpAPSM, the extrapolation parameter $\mu_n$ is set equal to $1.8 \times
\mathcal{M}_n$, the hyperslabs parameter $\epsilon$ was set equal to $1.3\sigma_n$ and $q=390$. The parameters for all algorithms were selected in order to optimize their performance. Since the sparsity level of the signal may change (from $k=100$ up to $k=110$) and since in practice it is not possible to know in advance the exact value of $k$, we feed the algorithms with an overestimate, $k$, of the true sparsity value and  in particular we used $\hat{k}=150$ (i.e., $50\%$ overestimation up to the 1500-th iteration).

\begin{figure}[!tbp]
\centering
\includegraphics[scale=0.8]{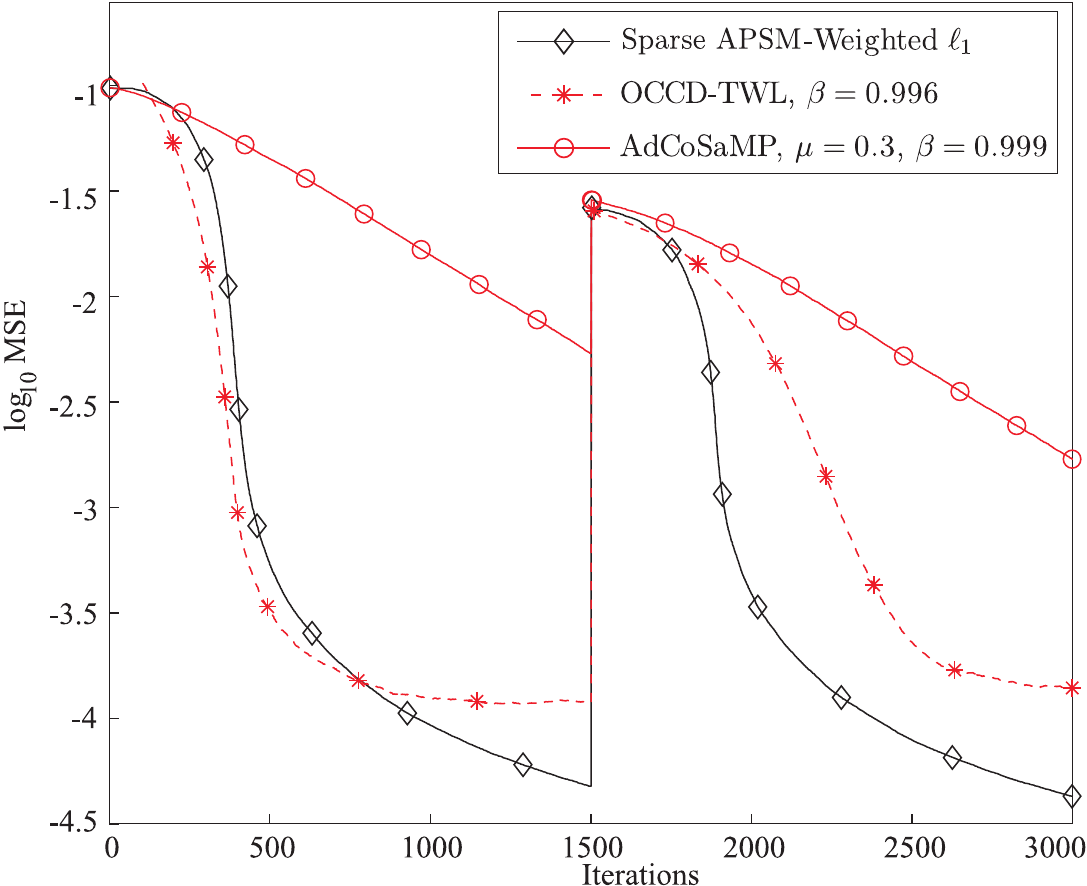}
\caption{MSE learning curves for AdCoSAMP, SpAPSM and OCCD-TWL for the simulation example discussed in \ref{exampleonline}. The vertical axis shows the $\log_{10}$ of the Mean Squares Error, i.e., $\log_{10}\left(\frac{1}{2}\norm{\bm{s}-\Phi \bm{\theta}(n)}^2_2\right)$ and the horizontal shows the time index. At time $n=1500$, the system undergoes a sudden change.}\label{fig:timevarfig_a}.
\end{figure}

The results are shown in Fig.~\ref{fig:timevarfig_a}. Note the enhanced performance obtained via the SpAPSM algorithm. However, it has to be pointed out that the complexity of the AdCoSAMP is much lower compared to the other two algorithms, for the choice of $q=390$ for the SpAPSM. The interesting observation is that SpAPSM achieves a better performance compared to OCCD-TWL, albeit at significantly lower complexity. If on the other hand complexity is of major concern, as it has already been pointed out in \ref{rem12}, use of SpAPSM offers the flexibility to use GT operators, which lead to improved performance for small values of $q$ at complexity comparable to that of LMS-based sparsity promoting algorithms \cite{Kops_iscass_submitted}.
\end{example}

\section{Learning Sparse Analysis Models}
\label{ch10:analysis}

All our discussion, so far, has been exhausted in the terrain of signals
which are either sparse themselves or they can be sparsely represented in
terms of the atoms of a dictionary in a synthesis model, as introduced in
(\ref{ch-10signrepr}), i.e.,
\[
\bm{s}=\sum_{i\in \mathcal{I}}\theta_i\bm{\psi}_i.
\]
As a matter of fact, most of the research activity over the last decade or
so has been focused on the synthesis model. This may be partly due to the
fact that the synthesis modeling path may provide a more intuitively
appealing structure to describe the generation of the signal in terms
of the elements (atoms) of a dictionary. Recall from Section
\ref{ch10compressive} that the sparsity assumption was imposed on $\bm{\theta}$
in the synthesis model and the corresponding optimization task was
formulated in (\ref{ch10sens-1a}) and (\ref{ch10sens-1}) for the exact and
noisy cases, respectively.

However, this is not the only way to attack the task of sparsity
modeling. Very early in this paper, in Section~\ref{ch10:sparse}, we
referred to the analysis model,
\[
\bm{S}=\Phi^H\bm{s}
\]
and pointed out that in a number of real life applications the resulting
transform $\bm{S}$ is sparse. To be fair, for such an experimental
evidence, the most orthodox way to deal with the underlying model sparsity
would be to consider $\norm{\Phi^H \bm{s}}_0$. Thus, if one wants to estimate
$\bm{s}$, a very natural way would be to cast the related optimization task
as
\beqa
\min_{\bm{s}} & & \norm{\Phi^H\bm{s}}_0,\nonumber \\
\mbox{s.t.} & & \bm{y}=X\bm{s},\ \text{or}\ \norm{\bm{y}-X\bm{s}}_2^2\le
\epsilon, \label{ch10:analysis1}
\eeqa
depending on whether the measurements via a sensing matrix, $X$, are exact
or noisy. Strictly speaking, the total variation minimization approach,
which was used in Example \ref{ch10:MRI}, falls under this analysis model
formulation umbrella, since what is minimized is the $\ell_1$ norm of the
gradient transform of the image.

The optimization tasks in either of the two formulations given in
(\ref{ch10:analysis1}) build around the assumption that the signal of
interest has {\it sparse analysis representation}\index{Sparse analysis
  representation}. The obvious question that is now raised is whether the
optimization tasks in (\ref{ch10:analysis1}) and their counterparts in
(\ref{ch10sens-1a}) or (\ref{ch10sens-1}) are any different. One of the
first efforts to shed light in this problem was in \cite{Eladsa-10}. There,
it is pointed out that the two tasks, although related, yet they are in
general different. Moreover, their comparative performance depends on the
specific problem at hand. However, it is fair to say that this is a new field of research
and more definite conclusions are currently being shaped.  An easy answer
can be obtained for the case where the involved dictionary corresponds to
an orthonormal transformation matrix, e.g., DFT. In this case, we already know
that the analysis and synthesis matrices are related as
\[
\Phi=\Psi=\Psi^{-H},
\]
which leads to an equivalence between the two previously stated
formulations. Indeed, for such a transform we have
\[
\underbrace{\bm{S}=\Phi^H\bm{s}}_{\text{Analysis}}\quad \Leftrightarrow
\quad \underbrace{\bm{s}=\Phi\bm{S}}_{\text{Synthesis}}.
\]
Using the last formula into the (\ref{ch10:analysis1}), the tasks in
(\ref{ch10sens-1a}) or (\ref{ch10sens-1}) are readily obtained by replacing $\bm{\theta}$ by $\bm{s}$. However,
this reasoning cannot be carried out to the case of overcomplete
dictionaries; it is for these cases, where the two optimization tasks may
lead to different solutions.

The previous discussion, concerning the comparative performance between the
synthesis or analysis-based sparse representations, is not only of a
``philosophical'' value. It turns out that, often in practice, the nature of
certain overcomplete dictionaries does not permit the use of the synthesis
based formulation. These are the cases where the columns of the
overcomplete dictionary exhibit high degree of dependence; that is, the
coherence of the matrix, as defined in section \ref{ch10mutcoh}, has large
values. Typical examples of such overcomplete dictionaries are the Gabor
frames, the curvelet frames and the oversampled DFT. The use of such
dictionaries lead to enhanced performance in a number of applications,
e.g., \cite{starck10-1, starck10-2}. Take as an example the case of our
familiar DFT transform. This transform provides a representation of our
signal samples in terms of sampled exponential sinusoids, whose frequencies
are multiples of $\frac{2\pi}{lT}$, where $T$ is the sampling frequency and
$lT$ is the length of our signal segment $\bm{s}$; that is, \beq
\bm{s}:=\left [ \begin{array}{c} s(0)\\ s(T)\\ \vdots
    \\s((l-1)T) \end{array}\right
]=\sum_{i=0}^{l-1}S_i\bm{\psi}_i,\label{ch10dft1} \eeq where $S_i$ are the
DFT coefficients and $\bm{\psi}_i$ is the sampled sinusoid with frequency
equal to $\frac{2\pi}{lT}i$, i.e., \beq \bm{\psi}_i=\left [\begin{array}{c}
    1\\ \exp \left (-j\frac{2\pi}{lT}iT\right ) \\ \vdots \\ \exp \left
    (-j\frac{2\pi}{lT}i(l-1)T\right )\end{array}\right ].\label{ch10dft2}
\eeq However, this is not necessarily the most efficient representation.
For example, it is highly unlikely that a signal comprises only frequencies
which are multiples of the basic one; only such signals can result in a
sparse representation using the DFT basis. Most probably, in general, there
will be frequencies lying in between the frequency samples of the DFT
basis, which result in non-sparse representations. This can be remedied by
increasing the number $l$ of the atoms, and form a dictionary that involves
sinusoids with frequencies taken at smaller frequency
intervals. However, in such a dictionary the atoms are no more linearly
independent and the coherence of the respective (dictionary) matrix increases.

Once a dictionary exhibits high coherence, then there is no way of finding
a sensing matrix, $X$, so that $X\Psi$ to obey the RIP. Recall that at the
heart of the sparsity-aware learning lies the concept of stable embedding, that
allows the recovery of a vector/signal after projecting it in a lower
dimensional space; this is what all the available conditions, e.g., RIP,
guarantee. However, no stable embedding is possible with highly coherent
dictionaries. Take as an extreme example the case where the first and
second atoms are identical. Then no sensing matrix $X$ can achieve a signal
recovery that distinguishes the vector $[1,0,\ldots,0]^T$ from
$[0,1,0,\ldots,0]^T$. Can then one conclude that for highly coherent
overcomplete dictionaries compressed sensing techniques are not possible?
Fortunately, the answer to this is negative. After all, our goal in compressed
sensing has always been the recovery of the signal $\bm{s}=\Psi\bm{\theta}$
and not the identification of the sparse vector $\bm{\theta}$ in the
synthesis model representation. The latter was just a means to an
end. While the unique recovery of $\bm{\theta}$ cannot be guaranteed for
highly coherent dictionaries, this does not necessarily cause any problems
for the recovery of $\bm{s}$, using a small set of measurement samples. The
escape route will come by considering  the analysis model
formulation. However, prior to this treatment, it will be of no harm to
refresh our basics concerning the theory of \textit{frames}\index{Frames} and
recall some key definitions.

\subsection{Some Hints from the Theory of Frames}
\label{ch10:frames}

In order to remain in the same framework as the one already adopted for this paper and comply with the notation previously used,
we will adhere to the real data case,
although everything we are going to say
is readily extended to the complex case, by replacing transposition with its Hermitian counterpart.

A frame in a vector space\footnote{We constrain
  our discussion in this section to finite dimensional Euclidean
  spaces. The theory of frames has been developed for general Hilbert
  spaces.}  $V\subseteq \mathcal{R}^l$ is a generalization of the notion of a basis. Recall form
our linear algebra basics that a {\it basis} is a set of vectors
$\bm{\psi}_i$, $i\in \mathcal{I}$, with the following two properties: a)
$V=\linspan \{\bm{\psi}_i: i\in \mathcal{I}\}$, where the cardinality
$\card(\mathcal{I})=l$ and b) $\bm{\psi}_i$, $i\in \mathcal{I}$, are
mutually independent. If, in addition,
$\innprod{\bm{\psi}_i}{\bm{\psi}_j} =\delta_{i,j}$ then the basis is
known as orthonormal. If we now relax the second condition and allow
$l<\card(\mathcal{I}) \coloneqq p$, we introduce redundancy in the signal
representations, which, as it has already been mentioned, can offer a
number of advantages in a wide range of applications. However, once
redundancy is introduced we lose uniqueness in the signal representation
\beq
\bm{s}=\sum_{i\in \mathcal{I}}\theta_i\bm{\psi}_i, \label{ch10:frame0}
\eeq
due to the dependency among the vectors $\bm{\psi}_i$. The question
that is now raised is whether there is a simple and systematic way to
compute the coefficients $\theta_i$ in the previous expansion.

\begin{definition}
The set $\bm{\psi}_i,~ i\in \mathcal{I}$, which spans a vector space, $V$,
is called a {\it frame}\index{Frames} if there exist positive real numbers,
$A$ and $B$, such that for any non-zero $\bm{s}\in V$,
\beq
0<A\norm{\bm{s}}_2^2\le\sum_{i\in
  \mathcal{I}}|\innprod{\bm{\psi}_i}{\bm{s}}|^2 \le
B\norm{\bm{s}}^2_2, \label{ch10:frame1}
\eeq
where $A$ and $B$ are known as the bounds of the frame.
\end{definition}

Note that if $\bm{\psi}_i$, $i\in \mathcal{I}$, comprise an orthonormal
basis, then $A=B=1$ and (\ref{ch10:frame1}) is the celebrated Parseval's
theorem. Thus, (\ref{ch10:frame1}) can be considered as a generalization of
Parseval's theorem. Looking at it more carefully, we notice that this is a
stability condition that closely resembles our familiar RIP condition in
(\ref{ch10iso-1}). Indeed, the upper bound guarantees that the expansion
never diverges (this applies to infinite dimensional spaces) and the lower
bound guarantees that no non-zero vector, $\norm{\bm{s}}\ne 0$, will ever
become zero after projecting it along the atoms of the frame. To look at it
from a slightly different perspective, form the dictionary matrix
\[
\Psi=[\bm{\psi}_1,\bm{\psi}_2,\ldots,\bm{\psi}_p],
\]
where we used $p$ to denote the cardinality of $\mathcal{I}$. Then, the
lower bound in (\ref{ch10:frame1}) guarantees that $\bm{s}$ can be
reconstructed from its transform samples $\Psi^T\bm{s}$; note that in such a case, if
$\bm{s}_1\ne \bm{s}_2$, then their respective transform values will be
different.

It can be shown that if condition (\ref{ch10:frame1}) is valid,
then there exists another set of vectors,
$\tilde{\bm{\psi}}_i$, $i\in\mathcal{I}$, known as the \textit{dual
  frame}\index{Dual frames}, with the following elegant property
\beq
\bm{s}=\sum_{i\in\mathcal{I}}\innprod{\tilde{\bm{\psi}}_i}
{\bm{s}}\bm{\psi}_i=\sum_{i\in\mathcal{I}}\innprod{\bm{\psi}_i}
{\bm{s}}\tilde{\bm{\psi}}_i, \quad \forall \bm{s}\in
V. \label{cha10:frame2}
\eeq
Once a dual frame is available, the coefficients in the expansion of a
vector in terms of the atoms of a frame are easily obtained. If we form the
matrix $\tilde{\Psi}$ of the dual frame vectors, then it is easily checked
out that since condition (\ref{cha10:frame2}) is true for any $\bm{s}$, it
implies that
\beq
\tilde{\Psi}\Psi^T=\Psi\tilde{\Psi}^T=I, \label{ch10:frame3}
\eeq
where $I$ is the $l\times l$ identity matrix.  Note that all of us have
used the property in (\ref{cha10:frame2}), possibly in a disguised form,
many times in our professional life. Indeed, consider the simple case of
two independent vectors in the two-dimensional space (in order to make
things simple). Then, (\ref{ch10:frame0}) becomes
\[
\bm{s}=\theta_1\bm{\psi}_1+\theta_2\bm{\psi}_2=\Psi\bm{\theta}.
\]
Solving for the unknown $\bm{\theta}$ is nothing but the solution of a
linear set of equations; note that the involved matrix $\Psi$ is
invertible. Let us rephrase a bit our familiar solution
\beq
\bm{\theta}=\Psi^{-1}\bm{s}:=\tilde{\Psi}\bm{s}:=\left
   [\begin{array}{c}\tilde{\bm{\psi}}_1^T
       \\ \tilde{\bm{\psi}}_2^T \end{array}\right ]\bm{s},
\eeq
where $\tilde{\bm{\psi}}_i^T$, $i=1,2$, are the rows of the inverse
matrix. Using now the previous notation, it is readily seen that
\[
\bm{s}=\innprod{\tilde{\bm{\psi}}_1}{\bm{s}}
\bm{\psi}_1+\innprod{\tilde{\bm{\psi}}_2}
{\bm{s}}\bm{\psi}_2.
\]
Moreover, note that in this special case of independent vectors, the
respective definitions imply
\[
 \left [\begin{array}{c}\tilde{\bm{\psi}}_1^T \\
         \tilde{\bm{\psi}}_2^T\end{array}\right]
[\bm{\psi}_1,\bm{\psi}_2]=I,\]
and the dual frame is not only unique but it also fulfils the
\textit{biorthogonality}\index{Biorthogonality condition} condition, i.e.,
\beq
\innprod{\tilde{\bm{\psi}}_i}{\bm{\psi}_j}=\delta_{i,j}.
\eeq
In the case of a general frame, the dual frames are neither
biorthogonal nor uniquely defined. The latter can also be verified by the
condition (\ref{ch10:frame3}) that defines the respective
matrices. $\Psi^T$ is a rectangular tall matrix and its left inverse is not
unique. There is, however, a uniquely defined dual frame, known as the
\textit{canonical} dual frame\index{Canonical dual frame}, given as
\beq
\tilde{\bm{\psi}}_i \coloneqq (\Psi\Psi^T)^{-1}\bm{\psi}_i,\ \text{or}\ \tilde{\Psi}
:= (\Psi\Psi^T)^{-1}\Psi. \label{Cha0-canonical}
\eeq

Another family of frames of special type are the so-called \textit{tight
  frames}\index{Tight Frames}. For tight frames, the two bounds in
(\ref{ch10:frame1}) are equal, i.e., $A=B$. Thus, once a tight frame is
available, we can normalize each vector in the frame as
\[
\bm{\psi}_i\mapsto\frac{1}{\sqrt{A}}\bm{\psi}_i,
\]
which then results to the so-called \textit{Parseval tight
  frame}\index{Parseval tight frame}; the condition \eqref{ch10:frame1}
now becomes similar in appearance with our familiar Parseval's theorem for
orthonormal bases
\beq
\sum_{i\in \mathcal{I}}|\innprod{\bm{\psi}_i}{\bm{s}}|^2=
\norm{\bm{s}}^2_2. \label{ch10:frame4}
\eeq
Moreover, it can be shown that a Parseval tight frame
coincides with its canonical dual frame (that is, it is self dual) and we
can write
\[
\bm{s}=\sum_{i\in\mathcal{I}}\innprod{\bm{\psi}_i}{\bm{s}}\bm{\psi}_i,
\]
or in matrix form
\beq
\tilde{\Psi}=\Psi,
\eeq
which is similar with what we know for orthonormal bases;  however in this
case, orthogonality does not hold in general.

We will conclude this subsection with a simple example of a Parseval
(tight) frame, known as the \textit{Mercedes Benz (MB)},
\[
\Psi=\left [\begin{array}{ccc} 0 & -\frac{1}{\sqrt{2}} & \frac{1}{\sqrt{2}} \\
   \sqrt{\frac{2}{3}} & -\frac{1}{\sqrt{6}} & -\frac{1}{\sqrt{6}} \end{array} \right ].
   \]
One can easily check that all the properties of a Parseval tight frame are
fulfilled.  If constructing a frame, especially in high dimensional spaces,
may sound a bit like magic, the following theorem (due to Naimark, see,
e.g., \cite{Han10}) offers a systematic way for such constructions.

\begin{theorem}
A set $\{\bm{\psi}_i\}_{i\in\mathcal{I}}$ in a Hilbert space $\hilbert_s$ is a
Parseval tight frame, if and only if it can be obtained via
  orthogonal projection, $P_{\hilbert_s}: \hilbert\rightarrow \hilbert_s$,
  of an orthonormal basis $\{\bm{e}_i\}_{i\in\mathcal{I}}$ in a larger
  Hilbert space $\hilbert$, such that $\hilbert_s\subset \hilbert$.
\end{theorem}

To verify the theorem, check that the MB frame is obtained by orthogonally
projecting the three-dimensional orthonormal basis
\[
\bm{e}_1=\left [\begin{array}{r} 0\\ -\frac{1}{\sqrt{2}}
    \\ \frac{1}{\sqrt{2}}\end{array}\right ],\
\bm{e}_2=\left [\begin{array}{r} \sqrt{\frac{2}{3}}\\-\frac{1}{\sqrt{6}}
    \\ -\frac{1}{\sqrt{6}}\end{array}\right ],\
\bm{e}_3=\left [ \begin{array}{r} \frac{1}{\sqrt{3}}\\ \frac{1}{\sqrt{3}}
    \\ \frac{1}{\sqrt{3}}\end{array}\right],
\]
using the projection matrix
\[
P_{\hilbert_s} \coloneqq \left [\begin{array}{rrr} \frac{2}{3} &
    -\frac{1}{3} & -\frac{1}{3} \\
 -\frac{1}{3} & \frac{2}{3} & -\frac{1}{3} \\
  -\frac{1}{3} & -\frac{1}{3} & \frac{2}{3}
  \end{array}\right].
  \]
Observe that the effect of the projection
\[
P_{\hilbert_s}[\bm{e}_1,\bm{e}_2,\bm{e}_3]=\Psi^T,
\]
is the deletion of the last column in the basis matrix.

Frames were introduced by Duffin and Schaeffer in their study on
nonharmonic Fourier series in 1952 \cite{Duffin} and they remained rather
obscured till they were used in the context of wavelet theory, e.g.,
\cite{Daubechieswav10}. The interested reader can obtain the proofs of what
has been said in this section from these references. An introductory review
with a lot of engineering flavor can be found in \cite{Kovacevic10}, where
the major references in the field are given.

\subsection{Compressed Sensing for Signals Sparse in Coherent Dictionaries}
\label{ch10:CoSeCoDi}

Our goal in this subsection is to establish conditions that guarantee
recovery of a signal vector, which accepts a sparse representation in
a redundant and coherent dictionary, using a small number of signal-related
measurements. Let the dictionary at hand be a tight frame, $\Psi$. Then,
our signal vector is written as
\beq
\bm{s}=\Psi\bm{\theta}, \label{ch10:drip1}
\eeq
where $\bm{\theta}$ is assumed to be $k$-sparse. Recalling the properties
of a tight frame, as they were summarized in the previous subsection, the
coefficients in the expansion  (\ref{ch10:drip1}) can be written as
$\innprod{\bm{\psi}_i}{\bm{s}}$, and the respective vector as
\[
\bm{\theta}=\Psi^T\bm{s},
\]
since a tight frame is self dual. Then, the analysis counterpart of the synthesis formulation in \eqref{ch10sens-1} can be cast as
\beqa
\min_{\bm{s}} && \norm{\Psi^T\bm{s}}_1, \nonumber \\
\text{s.t.}&& \norm{\bm{y}-X\bm{s}}_2^2\le\epsilon.\label{ch10:drip2}
\eeqa
The goal now is to investigate the accuracy of the recovered solution to
this convex optimization task. It turns out that similar strong theorems
are also valid for this problem as with the case of the synthesis
formulation, which was studied earlier.

\begin{definition}
Let $\Sigma_k$ be the union of all subspaces spanned by all subsets of $k$
columns of $\Psi$. A sensing matrix, $X$, obeys the restricted isometry
property adapted to $\Psi$, ($\Psi$-RIP) with $\delta_k$, if
\beq
 (1-\delta_k)\norm{\bm{s}}_2^2\le \norm{X\bm{s}}^2_2\le
 (1+\delta_k)\norm{\bm{s}}_2^2,
\eeq
for all $\bm{s}\in \Sigma_k$.
\end{definition}

The union of subspaces, $\Sigma_k$, is the image under $\Psi$ of all
$k$-sparse vectors. This is the difference with the RIP definition given in
Section \ref{ch10isom}. All the random matrices discussed earlier in this
paper can be shown to satisfy this form of RIP, with overwhelming
probability, provided the number of measurements, $N$, is at least of the
order of $k\ln(l/k)$. We are now ready to establish the main theorem
concerning our $l_1$ minimization task.

\begin{theorem}
Let $\Psi$ be an arbitrary tight frame and $X$ a sensing matrix that
satisfies the $\Psi$-RIP with $\delta_{2k}\le 0.08$, for some positive
$k$. Then the solution, $\bm{s}_*$, of the minimization task in
(\ref{ch10:drip2}) satisfies the following property
 \beq
 \norm{\bm{s}-\bm{s}_*}_2\le
 C_0k^{-\frac{1}{2}}\norm{\Psi^T\bm{s}-(\Psi^T\bm{s})_k}_1 +
 C_1 \sqrt{\epsilon}, \label{ch10:drip3}
 \eeq
where $C_0, C_1$ are constants depending on $\delta_{2k}$,
$(\Psi^T\bm{s})_k$ denotes the best $k$-sparse approximation of $\Psi^Ts$;
i.e., it results by setting all but the $k$ largest in magnitude components
of $\Psi^Ts$ equal to zero.
\end{theorem}

The bound in (\ref{ch10:drip3}) is the counterpart of that given in
(\ref{ch10-lassperfrip}). In other words, the previous theorem states that
if $\Psi^T\bm{s}$ decays rapidly, then $\bm{s}$ can be reconstructed from
just a few (compared to the signal length $l$) measurements. The theorem
was first given in \cite{Candesdrip10} and it is the first time that such a
theorem provides results for the sparse analysis model formulation in a
general context.

\subsection{Cosparsity}\label{ch10:cosparse}

In \cite{nam10}, the task of sparse analysis modeling was approached via an
alternative route, employing the tools which were developed in
\cite{luunion10, Blumstr-10} for treating general union-of-subspaces
models. This complementary point of view will also unveil different aspects
of the problem by contributing to its deeper understanding.  We have done
it before, where the notions of spark, coherence and RIP were all mobilized
to shed light from different corners to the sparse synthesis modeling task.

In the sparse synthesis formulation, one searches for a solution in a
union of subspaces, which are formed by all possible combinations of $k$
columns of the dictionary, $\Psi$. Our signal vector lies in one of these
subspaces; the one which is spanned by the columns of $\Psi$ whose indices
lie in the support set (Section \ref{ch10:greedy}). In the sparse analysis
approach things get different. The kick off point is the sparsity of the
transform $\bm{S} \coloneqq \Phi^T\bm{s}$, where $\Phi$ defines the
transformation matrix or analysis operator. Since $\bm{S}$ is assumed to be
sparse, there exists an index set $\mathcal{I}$ such that $\forall i\in
\mathcal{I}$, $S_i = 0$. In other words, $\forall i\in \mathcal{I}$,
$\bm{\phi}^T_i \bm{s}\coloneqq \innprod{\bm{\phi}_i}{\bm{s}}=0$, where
$\bm{\phi}_i$ stands for the $i$th column of $\Phi$. Hence, the subspace in
which $\bm{s}$ lives is the orthogonal complement of the subspace formed by
those columns of $\Phi$, which correspond to a zero in the transform
vector $\bm{S}$. Assume, now, that $\card(\mathcal{I})= C_o$. The
signal, $\bm{s}$, can be identified by searching the \textit{orthogonal
  complements} of the subspaces formed by all possible combinations of
$C_o$ columns of $\Phi$, i.e.,
\[
\innprod{\bm{\phi}_i}{\bm{s}}=0, \quad \forall i\in \mathcal{I}.
\]
The difference between the synthesis and analysis problems is illustrated
in Fig.~\ref{fig:analysis.synthesis}. To facilitate the theoretical treatment of
this new setting, the notion of {\it cosparsity} was introduced in
\cite{nam10}.


\begin{figure}[!tbp]
    \centering
    \includegraphics[scale=1]{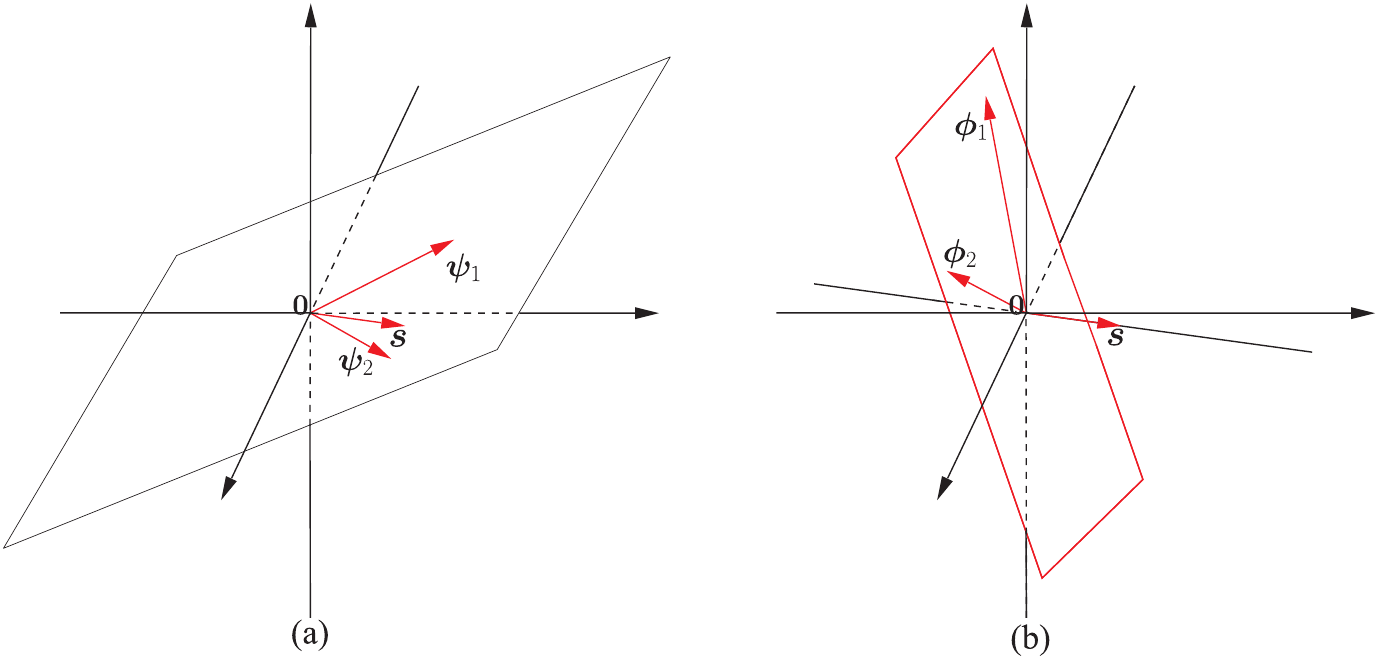}
    \caption{Searching  for a spare vector $\bm{s}$.  (a) In the synthesis model, the sparse vector lies
        in  subspaces formed by  combinations of $k$ (in this case $k=2$) columns of the dictionary $\Psi$. (b) In the analysis model, the sparse vector lies in the orthogonal compliment of the subspace formed by $C_o$ (in this case $C_o=2$) columns of the transformation matrix $\Phi$.}\label{fig:analysis.synthesis}
\end{figure}

\begin{definition} The cosparsity of a signal $\bm{s}\in \mathcal{R}^l$
  with respect to a $p\times l$ matrix $\Phi^T$ is defined as
\beq
C_o:=p-\norm{\Phi^T\bm{s}}_0.
\eeq
\end{definition}

In words, the cosparsity is the number of zeros in the obtained transform
vector $\bm{S}= \Phi^T\bm{s}$; in contrast, the sparsity measures the
number of the nonzero elements of the respective sparse vector. If one assumes that $\Phi$ has
``full spark''\footnote{Recall by Def.~\ref{def:spark} that $\spark(\Phi)$
  is defined for an $l\times p$ matrix $\Phi$ with $p\geq l$ and of full
  rank.}, i.e., $l+1$, then any $l$ of the columns of $\Phi$, and thus
any $l$ rows of $\Phi^T$ are guaranteed to be independent. This indicates
that for such matrices, the maximum value that cosparsity can take is equal
to $C_o=l-1$. Otherwise, the existence of $l$ zeros will necessarily
correspond to a zero signal vector. Higher cosparsity levels are possible,
by relaxing the full spark requirement.

Let now the cosparsity of our signal with respect to a matrix $\Phi^T$ be
$C_o$. Then, in order to dig out the signal from the subspace in which is
hidden, one must form all possible combinations of $C_o$ columns of $\Phi$
and search in their orthogonal complements. In case that $\Phi$ is full
rank, we have seen previously that $C_o<l$, and hence any set of $C_o$
columns of $\Phi$ are linearly independent. In other words, the dimension
of the span of those columns is $C_o$. As a result, the dimensionality of
the orthogonal complement, into which we search for $\bm{s}$, is
$l-C_o$.

We have by now accumulated enough information to elaborate a
  bit more on the statement made before, concerning the different nature of
  the synthesis and analysis tasks. Let us consider a synthesis task using
  an $l\times p$ dictionary and let $k$ be the sparsity level in the
  corresponding expansion of a signal in terms of this dictionary. The
  dimensionality of the subspaces in which the solution is sought is $k$
  ($k$ is assumed to be less than the spark of the respective matrix). Let
  us keep the same dimensionality for the subspaces in which we are going
  to search for a solution in an analysis task. Hence, in this case
  $C_o=l-k$ (assuming a full spark matrix). Also, for the sake of
  comparison assume that the analysis matrix is $p\times l$.  Solving the
  synthesis task, one has to search $\binom{p}{k}$ subspaces, while solving
  the analysis task one has to search for $\binom{p}{C_o=l-k}$
  subspaces. These are two different numbers; assuming that $k\ll l$ and also
  that $l<p/2$, which are natural assumptions for overcomplete
  dictionaries, then the latter of the two numbers is much larger than the former one
  (use your computer to play with some typical values). In other words,
  there are much more analysis than synthesis low-dimensional subspaces to
  be searched for. The large number of low-dimensional subspaces makes the
  algorithmic recovery of a solution from the analysis model a tougher
  task, \cite{nam10}.

Another interesting aspect that highlights the difference
between the two approaches is the following. Assume that the synthesis and
analysis matrices are related as $\Phi=\Psi$, as it was the case for tight frames.
Under this assumption, $\Phi^T\bm{s}$ provides a set of coefficients for the synthesis expansion in terms of the atoms of $\Phi=\Psi$. Moreover, if $\norm{\Phi^T\bm{s}}_0=k$, then the  $\Phi^T\bm{s}$ is a possible
$k$-sparse solution for synthesis model. However, there is no guarantee that this is the
sparsest one.

It is now the time to investigate whether conditions that guarantee
uniqueness of the solution for the sparse analysis formulation can be
derived. The answer is affirmative and it has been established in
\cite{nam10}, for the case of exact measurements.

\begin{lemma}
Let $\Phi$ be a transformation matrix of full spark. Then, for almost all
$N\times l$ sensing matrices and for $N>2(l-C_o)$, the equation
\[
\bm{y}=X\bm{s},
\]
has at most one solution with cosparsity at least $C_o$.
\end{lemma}

The above lemma guarantees the uniqueness of the solution, if one exists,
of the following optimization
\beqa
\min_{\bm{s}} &&\norm{\Phi^T\bm{s}}_0\nonumber \\
\text{s.t.} && \bm{y}=X\bm{s}. \label{chcosp2}
\eeqa
However, solving the previous $l_0$ minimization task is a difficult one
and we know that its synthesis counterpart has been shown to be NP-hard, in
general. Its relaxed convex relative is the $l_1$ minimization
\beqa
\min_{\bm{s}}&& \norm{\Phi^T\bm{s}}_1 \nonumber\\
\text{s.t.} & &\bm{y}=X\bm{s}. \label{chcosp3}
\eeqa
In \cite{nam10}, conditions are derived that guarantee the equivalence of
the $l_0$ and $l_1$ tasks, in (\ref{chcosp2}) and (\ref{chcosp3}),
respectively; this is done in a way similar to that for the sparse
synthesis modeling. Also, in \cite{nam10}, a greedy algorithm inspired by
the Orthogonal Matching Pursuit, discussed in Section \ref{ch10:greedy},
has been derived. Other algorithms that solve the $l_1$ optimization in the
analysis modeling framework can be found in, e.g., \cite{Caij10,
  Eladanal-10, Selean-10}. NESTA can also be used for the analysis
formulation.

\section{A Case Study: Time-Frequency Analysis}
\label{ch10:timeFreq}

The goal of this section is to demonstrate how all the previously stated
theoretical findings can be exploited in the context of a real
application. Sparse modelling has been applied to almost everything. So,
picking up a typical application would not be easy. We preferred to focus
on a less ``publicised'' application; that of analysing echolocation
signals emitted by bats. However, the analysis will take place within the
framework of time-frequency\index{Time-frequency analysis} representation,
which is one of the research areas that significantly inspired the
evolution of compressed sensing theory. Time-Frequency analysis of signals
has been the field of intense research for a number of decades, and it is
one of the most powerful signal processing tools. Typical applications
include speech processing, sonar sounding, communications, biological
signals, EEG processing, to name but a few, see, e.g.,
\cite{Boamash10,flandrin10}.

\subsubsection{Gabor Transform and Frames}
\label{ch10gabor}

It is not our intention to present the theory behind the Gabor
transform. Our goal is to outline some basic related notions and use it as
a vehicle for the less familiar reader so that a) to better understand
how redundant dictionaries are used and b) get more acquainted with their
potential performance benefits.

The Gabor transform was introduced in the middle 1940s by Dennis Gabor
(1900--1979), who was a Hungarian-British engineer. His most notable
scientific achievement was the invention of holography, for which he won
the Nobel prize for Physics in 1971. The discrete version of the Gabor
transform can be seen as a special case of the Short Time Fourier Transform
(STFT), e.g., \cite{Mallatbook10,flandrin10}. In the standard DFT
transform, the full length of a time sequence, comprising $l$ samples, is
used all in ``one-go'' in order to compute the corresponding frequency
content. However, the latter can be time varying, so the DFT will provide
an average information, which cannot be of much use. The Gabor transform
(and the STFT in general) introduces time localization via the use of a
window function, which slides along the signal segment in time, and at each
time instant focusses on a different part of the signal; this is a way that
allows one to follow the slow time variations, which take place in the
frequency domain. The time localization in the context of the Gabor
transform is achieved via a Gaussian window function, i.e.,
\beq g(n) \coloneqq
\frac{1}{\sqrt{2\pi\sigma^2}}\exp\left(-\frac{n^2}{2\sigma^2}\right).
\eeq
Fig.~\ref{cha10fig:atoms}a shows the Gaussian window, $g(n-m)$, centered at
time instant $m$. The choice of the window spreading factor, $\sigma$, will
be discussed later on.

\begin{figure}[!tbp]
\centering
\includegraphics[scale=1]{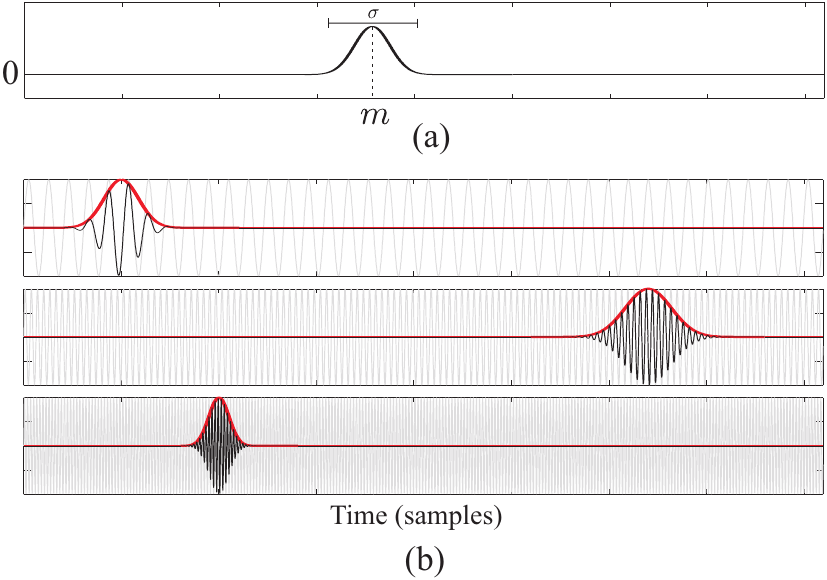}
\caption{(a) The Gaussian window with spreading factor $\sigma$ centered
  at time instant $m$. (b) Pulses obtained by windowing three different
  sinusoids with Gaussian windows of different spread and applied at
  different time instants.}
\label{cha10fig:atoms}
\end{figure}

Let us now construct the atoms of the Gabor dictionary. Recall that in the
case of the signal representation in terms of the DFT in (\ref{ch10dft1}),
each frequency is represented only once, by the corresponding sampled
sinusoid, (\ref{ch10dft2}). In the Gabor transform, each frequency appears
$l$ times; the corresponding sampled sinusoid is multiplied by the Gaussian
window sequence, each time shifted by one sample. Thus, at the $i$th
frequency bin, we have $l$ atoms, $\bm{g}^{(m,i)},m=0,1,\ldots,l-1$, with
elements given by
\beq
g^{(m,i)}(n)=g(n-m)\psi_i(n), \quad n, m, i=0,1,\ldots,l-1,
\eeq
where $\psi_i(n)$ is the $n$th element of the vector $\bm{\psi}_i$ in (\ref{ch10dft2}). This results to
an overcomplete dictionary comprising $l^2$ atoms in the $l$-dimensional
space.  Fig.~\ref{cha10fig:atoms}b illustrates the effect of multiplying
different sinusoids with Gaussian pulses of different spread and at
different time delays. Fig.~\ref{ch10fig:tf_grid} is a graphical
interpretation of the atoms involved in the Gabor dictionary. Each node,
$(m,i)$, in this time-frequency plot, corresponds to an atom of frequency
equal to $\frac{2\pi}{lT}i$ and delay equal to $m$.

\begin{figure}[!tbp]
\centering
\includegraphics[scale=1]{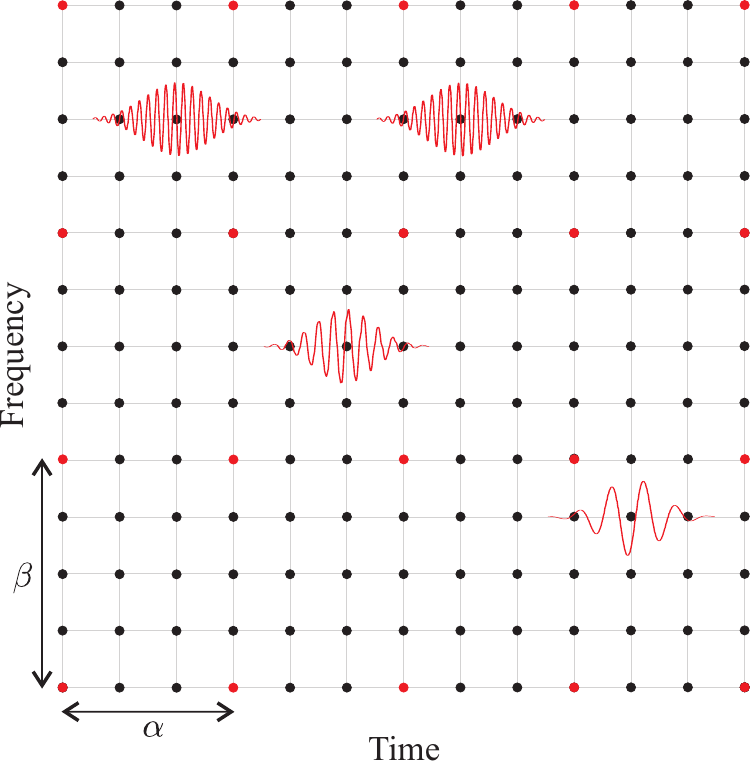}
\caption{Each atom of the Gabor dictionary corresponds to a node in the
  time-frequency grid. That is, it is a sampled windowed sinusoid whose
  frequency and location in time are given by the coordinates of the
  respective node. In practice, this grid may be subsampled by factors $\alpha$ and $\beta$ for the two axes respectively, in order to reduce the number of the involved atoms. }
\label{ch10fig:tf_grid}
\end{figure}

Note that the windowing of a signal of finite duration inevitably
introduces boundary effects, especially when the delay $m$ gets close to
the time segment edges, $0$ and $l-1$. A solution to it, that facilitates
the theoretical analysis, is to use a modulo $l$ arithmetic to wrap around
at the edge points (this is equivalent with extending the signal
periodically), see, e.g., \cite{Strohmer-10}.

Once the atoms have been defined, they can be stacked  one next to the
other to form the columns of the $l\times l^2$ Gabor dictionary, $G$. It
can be shown that the Gabor dictionary is a  tight frame,
\cite{zibuleski10}.

\subsubsection{Time-Frequency Resolution}

By the definition of the Gabor dictionary, it is readily understood that
the choice of the window spread, as measured by $\sigma$, must be a
critical factor, since it controls the localization in time. As it is known
from our Fourier transform basics, when the pulse becomes short, in order
to increase the time resolution, its corresponding frequency content
spreads out, and vice versa. From Heisenberg's principle, we know that we
can never achieve high time and frequency resolution, simultaneously; one
is gained at the expense of the other. It is here where the Gaussian shape
in the Gabor transform is justified.  It can be shown that the Gaussian
window gives the optimal trade-off between time and frequency resolution,
\cite{Mallatbook10, flandrin10}. The time-frequency resolution trade-off is
demonstrated in Fig.~\ref{ch10fig:heisenbergbox}, where three sinusoids
are shown windowed with different pulse durations. The diagram shows the
corresponding spread in the time-frequency plot. The value of $\sigma_t$
indicates the time spread and $\sigma_f$ the spread of the respective
frequency content around the basic frequency of each sinusoid.

\begin{figure}[!tbp]
\centering
\includegraphics[scale=1]{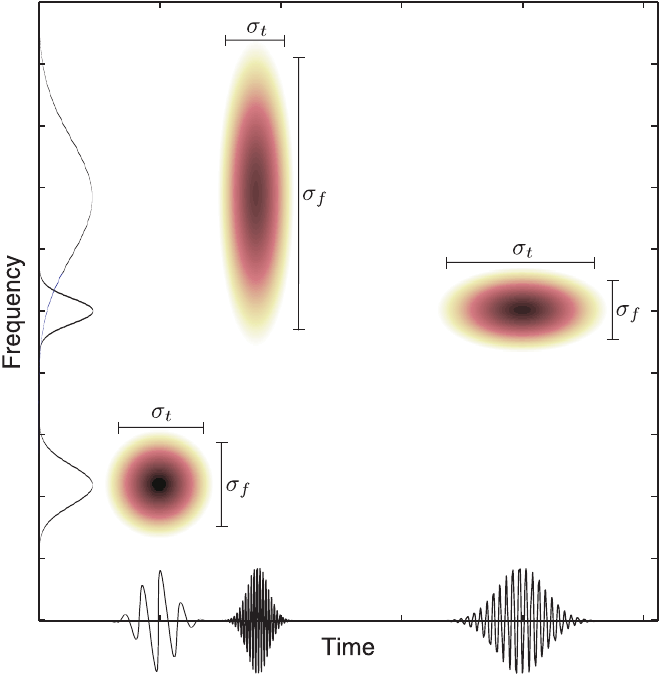}
\caption{The shorter the width of the pulsed (windowed) sinusoid is in time
  the wider the spread of its frequency content around the frequency of the
  sinusoid. The Gaussian-like curves along the frequency axis indicate the
  energy spread in frequency of the respective pulses. The values of
  $\sigma_t$ and $\sigma_f$ indicate the spread in time and frequency,
  respectively.}
\label{ch10fig:heisenbergbox}
\end{figure}

\subsubsection{Gabor Frames}
\label{ch10gaborframes}

In practice, $l^2$ can take large values and it is desirable to see whether
one can reduce the number of the involved atoms, without sacrificing the
frame-related properties.  This can be achieved by an appropriate
subsampling, as this is illustrated in Fig.~\ref{ch10fig:tf_grid}. We
only keep the atoms that correspond to the red nodes. That is, we subsample
by keeping every $\alpha$ nodes in time and every $\beta$ nodes in
frequency in order to form the dictionary, i.e.,
\[
G_{(\alpha,\beta)}=\{\bm{g}^{(m\alpha,i\beta)}\}, \quad
m=0,1,\ldots,\frac{l}{\alpha}-1,\ i=0,1,\ldots,\frac{l}{\beta}-1,
\]
where $\alpha$ and $\beta$ are divisors of $l$. Then, it can be shown, e.g.,
(\cite{flandrin10}), that if $\alpha\beta<l$ the resulting dictionary
retains its frame properties. Once $G_{(\alpha,\beta)}$ is obtained, the
canonical dual frame is readily available via (\ref{Cha0-canonical}) (adjusted for complex data), from which the corresponding set of expansion  coefficients, $\bm{\theta}$, results.

\subsubsection{Time-Frequency Analysis of Echolocation Signals Emitted  by
  Bats}

Bats are using echolocation for navigation (flying around at night), for
prey detection (small insects) and for prey approaching and catching; each
bat adaptively changes the shape and frequency content of its calls in
order to better serve the previous tasks.  Echolocation is used in a
similar way for sonars. Bats emit calls as they fly, and ``listen'' to the
returning echoes in order to build up a sonic map of their surroundings. In
this way, bats can infer on the distance, the size of obstacles as well as
of other flying creatures/insects. Moreover, all bats emit special types of
calls, called social calls, which are used for socializing, flirting, etc.
The fundamental characteristics of the echolocation calls, as for example,
the frequency range and the average time duration, differ from species to
species since, thanks to evolution, bats have adapted their calls in order
to get best suited to the environment in which a species operates.

Time-Frequency analysis of echolocation calls provides information about
the species (species identification) as well as of the specific task and
behaviour of the bats in certain environments. Moreover, the bat-biosonar
system is studied in order humans to learn more about nature and be
inspired for subsequent advances in applications such as sonar navigation
systems, radars, medical ultrasonic devices, etc.

Fig.~\ref{ch10fig:batcall}a shows a case of a recorded echolocation signal
from bats. Zooming at two different parts of the signal, we can observe
that the frequency is changing with time.  In Fig.~\ref{ch10fig:batcall}b,
the DFT of the signal is shown, but there is no much information that can
be drawn from it, except that the signal is compressible in the frequency
domain; most of the activity takes place within a short range of
frequencies.

\begin{figure}[!tbp]
\centering
\includegraphics[scale=1]{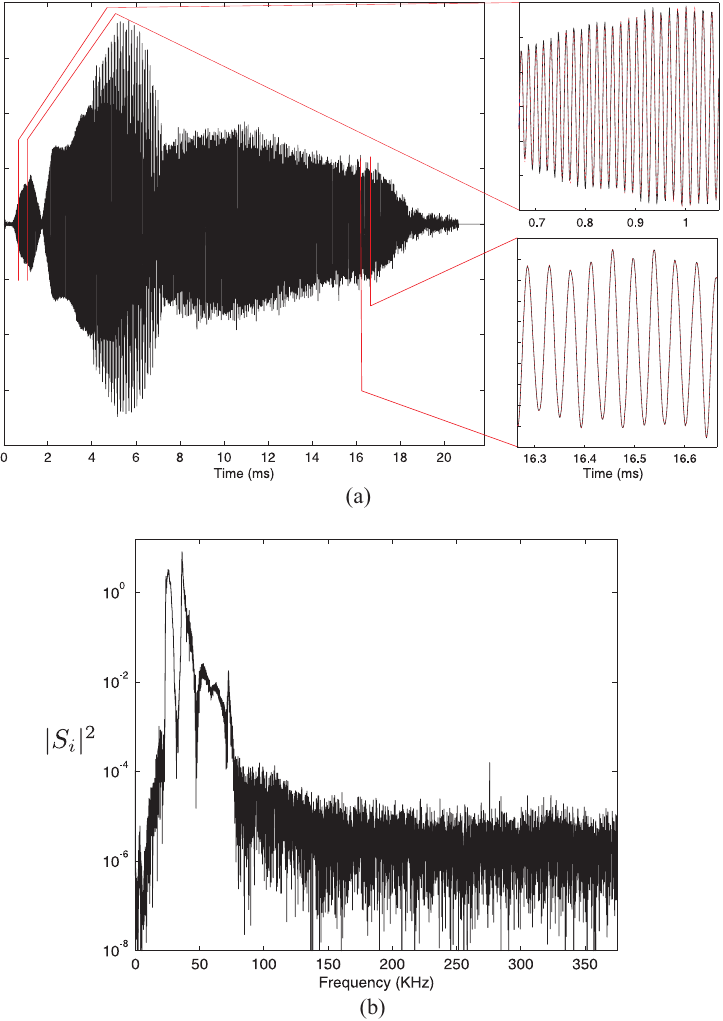}
\caption{(a) The recorded echolocation signal. The frequency of the signal
  is time varying and this is indicated by focussing on two different parts
  of the signal. (b) Plot of the energy of the DFT transform coefficients,
  $S_i$. Observe that most of the frequency activity takes place within a
  short frequency range.}
\label{ch10fig:batcall}
\end{figure}

Our echolocation signal was a recording of total length $T=21.845\text{msecs}$,
\cite{Kopsinisbat10}.  Samples were taken at the sampling frequency
$f_s=750$ KHz, which results in a total of $l=16384$ samples. Although the
signal itself is not sparse in the time domain, we will take advantage of
the fact that it is sparse in a transformed domain. We will assume that the
signal is sparse in its expansion in terms of the Gabor dictionary.

Our goal in this example is to demonstrate that one does not really need
all 16384 samples to perform time-frequency analysis; all the processing
can be carried out by using a reduced number of measurements, by exploiting
the theory of compressed sensing. To form the measurements vector,
$\bm{y}$, the number of measurements was chosen to be $N= 2048$. This
amounts to a reduction of eight times with respect to the number of
available samples. The measurements vector was formed as
\[
\bm{y}=X\bm{s},
\]
where $X$ is a $N\times l$ sensing matrix comprising $\pm 1$ generated in a
random way. This means that once we obtain $\bm{y}$, one does not need to
store the original samples any more, leading to a saving in
memory. Ideally, one could have obtained the reduced number of measurements
by sampling directly the analogue signal at sub-Nyquist rates, as it has
already been discussed at the end of Section
\ref{ch10compressive}. Another goal is to use both the analysis and
synthesis models and demonstrate their difference.

Three different spectrograms were computed. Two of them, shown in
Figs.\ \ref{ch10fig:bat_sparsegabor}b and \ref{ch10fig:bat_sparsegabor}c,
correspond to the reconstructed signals obtained by the analysis, \eqref{chcosp3},  and the
synthesis, \eqref{ch10sens}, formulations, respectively. In both cases, the NESTA algorithm
was used and the $G_{(128,64)}$ frame was employed. Note that the latter
dictionary is redundant by a factor of 2. The spectrograms are the result of plotting the time-frequency grid
and colouring each node $(t,i)$ according to the energy $|\theta|^2$ of the
coefficient associated with the respective atom in the Gabor dictionary. The
full Gabor transform applied on the reconstructed signals to obtain the spectrograms, in order to get a
better coverage of the time-frequency grid.  The scale is logarithmic and
the darker areas correspond to larger values. The spectrogram of the original signal obtained via the full Gabor transform is shown in Fig.~\ref{ch10fig:bat_sparsegabor}d. It is evident, that the
analysis model resulted in a more clear spectrogram, which resembles the original one better. When the frame $G_{(64,32)}$ is employed, which is a highly redundant Gabor dictionary comprising $8 l$ atoms, then the analysis model results in a recovered signal whose spectrogram is visually
indistinguishable from the original one in Fig.~\ref{ch10fig:bat_sparsegabor}d.

Fig.~\ref{ch10fig:bat_sparsegabor}a is the plot of the magnitude of the
corresponding Gabor transform coefficients, sorted in decreasing values. The synthesis model
provides a sparser representation, in the sense that the coefficients
decrease much faster. The third curve is the one that results if we
multiply the dual frame matrix $\tilde{G}_{(128,64)}$ directly with the
vector of the original signal samples and it is shown for comparison
reasons.

To conclude, the curious reader may wonder what do these curves in Fig.~\ref{ch10fig:bat_sparsegabor}d mean after
all. The call denoted by (A)
belongs to a Pipistrellus pipistrellus (!) and the call denoted by (B) is
either a social call or belongs to a different species. The (C) is
the return echo from the signal (A). The large spread in time of (C)
indicates a highly reflective environment, \cite{Kopsinisbat10}.

 \begin{figure}[!tbp]
\centering
\includegraphics[scale=0.95]{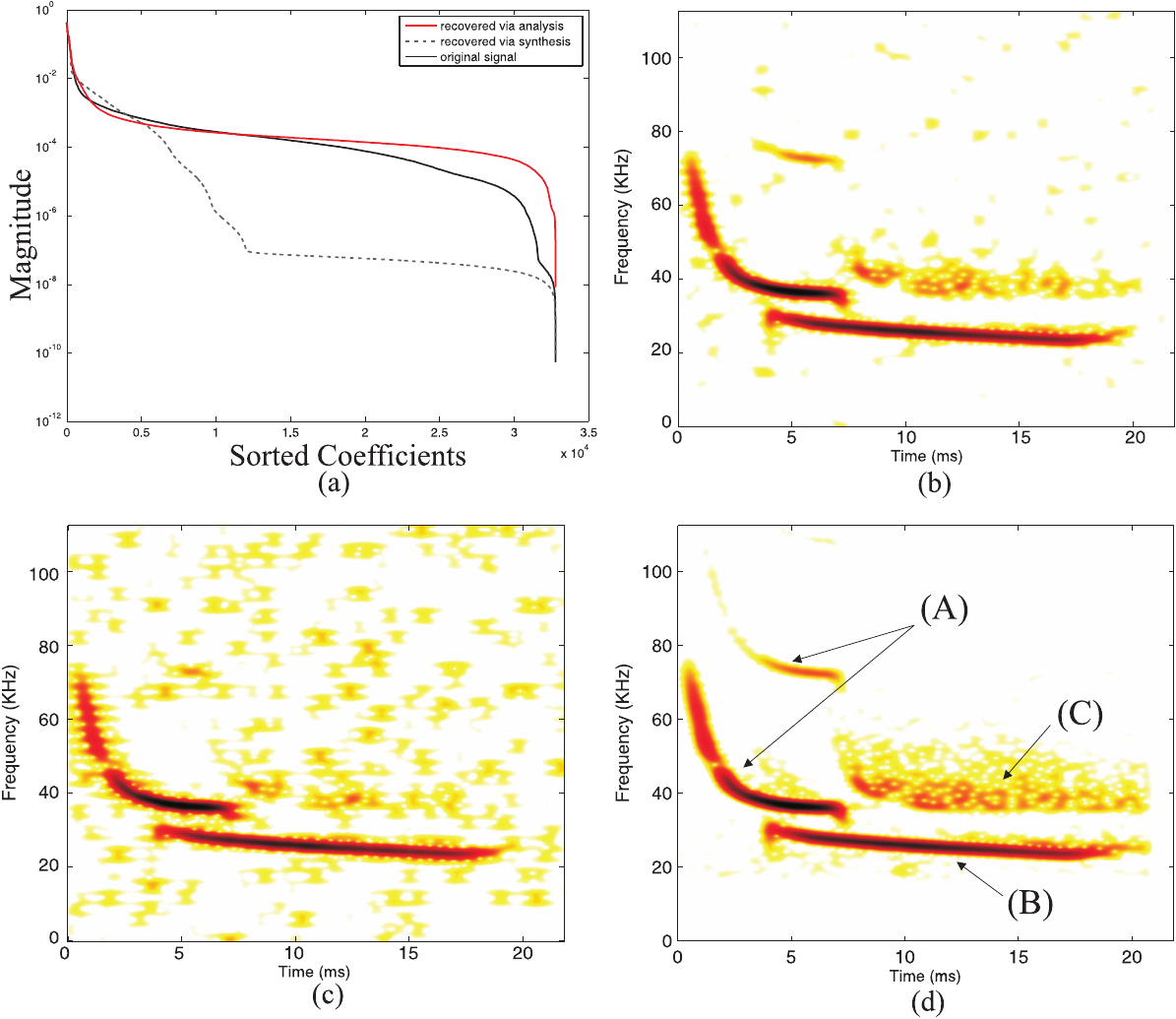}
\caption{(a) Plot of the magnitude of the coefficients, sorted in decreasing
  order, in the expansion in terms of the $G_{(128,64)}$ Gabor frame. The
  results correspond to the analysis and synthesis model formulations. The
  third curve corresponds to the case of analysing the original vector
  signal directly, by projecting it on the dual frame. (b) The spectrogram
  from the analysis and (c) the spectrogram from the synthesis formulations,
  respectively. (d) The spectrogram corresponding to $G_{(64,32)}$ frame
  using the analysis formulation. For all cases, the number of measurements
  used was one eighth of the total number of signal samples. A, B and C
  indicate different parts of the signal, as explained in the text.}
\label{ch10fig:bat_sparsegabor}
\end{figure}

\section{From Sparse Vectors to Low Rank Matrices: A highlight}

In this section, we move beyond sparse vectors and our goal is to investigate if and how notions related to sparsity can be generalized to matrices. We will see that such a generalization builds upon linear algebra tools and notions related to SVD decomposition, low rank approximation and dimensionality reduction. Our goal is to simply highlight the basic concepts and definitions without delving into a deeper treatment. Our aim is to make the reader alert of the problems and their potential for applications.

\subsection{Matrix Completion} \label{ch10:sec_MC}
Consider a signal vector $\bm{s} \in \Real^l$, where only $N$ of its components are observed and the rest are unknown. This is equivalent with sensing $\bm{s}$ via a sensing matrix $X$ having its $N$ rows picked uniformly at random from the standard (canonical) basis $\Phi=I$, where $I$ is the $l\times l$ identity matrix. The question which is now posed is whether it is possible to recover  the missing components of $\bm{s}$ based on these $N$ components. From the theory presented, so far, we know that one can recover all the components of $\bm{s}$, provided that $\bm{s}$ is sparse in some basis or dictionary, $\Psi$, which exhibits low mutual coherence with $\Phi=I$, and $N$ is large enough, as it has been pointed out in Section \ref{ch10compressive}.

Inspired by the theoretical advances in Compressed Sensing, a question similar in flavor and with a prominent impact regarding practical applications was posed in \cite{candes_exact_mc_2009}. Given a $l_1 \times l_2$ matrix $M$, assume that only $N<<l_1 l_2$ among its entries are known. The question now is whether one is able to recover the exact full matrix. This problem is widely known as {\it matrix completion} \cite{candes_exact_mc_2009}. The answer, although it might come as a surprise, is ``yes'' with high probability, provided that a) the matrix is \textit{well structured}, b) it has a \textit{low rank}, $r<<l$, where $l=\min(l_1,l_2)$, and c) that $N$ is large enough. Intuitively, this is plausible because a low rank matrix is fully described in terms of a number of parameters (degrees of freedom), which is much smaller than its total number of entries. These parameters are revealed via its Singular Value Decomposition (SVD)
\beqa
\label{ch10svd}
M =\sum_{i=1}^r \sigma_i \bm{u}_i \bm{v}_i^T = U \begin{bmatrix}
\sigma_1 & & \text{\large{0}} \\ 
& \ddots & & \\
\text{\large{0}} & & \sigma_r \end{bmatrix}  V^T,
\eeqa
where $r$ is the rank of the matrix, $\bm{u}_i \in \Real^{l_1}$ and $\bm{v}_i \in \Real^{l_2}$, $i=1,~2,\ldots,r$, are the left and right orthonormal singular vectors, spanning the column and row spaces of $M$ respectively, $\sigma_i$, $i=1,~2,~,\ldots,~r$, are the corresponding singular values and $U=[\bm{u}_1,\bm{u}_2,\cdots,\bm{u}_r]$, $V=[\bm{v}_1,\bm{v}_2,\cdots,\bm{v}_r]$.

Let $\bm{\sigma}_M$ denote the vector containing all the singular values of $M$, i.e., $\bm{\sigma}_M=[\sigma_1,~\sigma_2,\cdots,\sigma_l]^T$, then $\rank(M) \coloneqq \norm{\bm{\sigma}_{M}}_0$. Counting the parameters associated with the singular values and vectors in \eqref{ch10svd} turns out that the number of degrees of freedom of a rank $r$ matrix is equal to $d_M=r(l_1+l_2)-r^2$ \cite{candes_tao_mc_2010}. When $r$ is small, $d_M$ is much smaller than $l$.

Let us denote with $\Omega$ the set of $N$ pairs of indexes, $(i,j)$, $i=1,~,2,\ldots,l_1$, $j=1,~,2,\ldots,l_2$, of the locations of the known entries of $M$, which have been sampled uniformly at random. Adopting the main reasoning followed so far, one would attempt to recover $M$ based on the following rank minimization problem
\beqa
\min_{\hat{M} \in \Real^{l_1 \times l_2}} & & \norm{\bm{\sigma}_{\hat{M}}}_0 \nonumber \\
\text{s.t.} & &\hat{M}_{i,j}=M_{i,j}, \quad
(i,j) \in \Omega. \label{ch10:rankmin}
\eeqa
It turns out that, assuming that there exist a unique low-rank matrix having the specific known entries, then \eqref{ch10:rankmin} leads to the exact solution \cite{candes_exact_mc_2009}. However, compared to the case of sparse vectors, in the matrix completion problem the uniqueness issue gets much more involved. The following issues play a crucial part concerning the uniqueness of the task in \eqref{ch10:rankmin}.
\begin{enumerate}
  \item If the number of known entries is lower than the degrees of freedom, i.e., $N<d_M$, then there is no way to recover the missing entries whatsoever, since there is an infinite number of low rank matrices consistent with the $N$ observed entries.
  \item Even if $N \ge d_M$, uniqueness is still not guaranteed. It is required that the $N$ elements with indices in $\Omega$ are such that at least one entry per column and one entry per row is observed. Otherwise, even a rank-1 matrix, i.e, $M=\sigma_1 \bm{u}_1 \bm{v}_1^T$, is not possible to be recovered. This becomes clear with a simple example. Assume that $M$ is a rank-1 matrix and that no entry in the first column as well as in the last row is observed. Then, since for this case $M(i,j)=\sigma_1 u_{1i} v_{1j}$, it is clear that no information concerning the first component of $\bm{v}_1$ as well as the last component of $\bm{u}_1$ is available; hence these singular vector components are impossible to be recovered, regardless which method is used. As a consequence, the matrix can not be completed. On the other hand, if the elements of $\Omega$ are picked at random and $N$ is large enough, one can only hope that $\Omega$ is such that to comply with the previous requirement; i.e., at least one entry per row and column is observed, with high probability. It turns out that this problem resembles the famous in probability theory theorem known as the {\it coupon collector's} problem. According to this, at least $N=C_0 l \ln l$ entries are needed, where $C_0$ is a constant \cite{motwani_book1995}. This is the information theoretic limit for exact matrix completion \cite{candes_tao_mc_2010} of any low-rank matrix.
  \item Even if points (1) and (2) before are fulfilled, still uniqueness is not guaranteed. In fact, not every low rank matrix is liable to exact completion, regardless of the number and the positions of the observed entries. We will demonstrate that with the aid of an example. Let one of the singular vectors be sparse. Assume, without loss of generality, that the third left singular vector, $\bm{u}_3$, is sparse with sparsity level $k=1$ and also that its nonzero component is the first one, i.e., $u_{31} \neq 0$. The rest of $\bm{u}_i$ and all $\bm{v}_i$ are assumed to be dense. Let us return to the SVD for a while, and specifically to the leftmost formula given in \eqref{ch10svd}. Observe that the matrix $M$ is written as the sum of $r$, $l_1 \times l_2$ matrices $\sigma_i \bm{u}_i \bm{v}_i^T$, $i=1,\ldots,r$. Thus, in this specific case where $\bm{u}_3$ is $k=1$ sparse, the matrix $\sigma_3 \bm{u}_3 \bm{v}_3^T$ has zeros everywhere except from its first row. In other words, the information that $\sigma_3 \bm{u}_3 \bm{v}_3^T$ brings to the formation of $M$ is concentrated to its first row only. This argument can also be viewed from another perspective; the entries of $M$ obtained from any row but the first one, do not provide any useful information with respect to the values of the free parameters $\sigma_3$, $\bm{u}_3$, $\bm{v}_3$. As a result, in this case, unless if one incorporates extra information about the sparse nature of the singular vector, the entries from the first row that are missed are not recoverable, since the number of parameters concerning this row is larger than the available number of data.

Intuitively, when a matrix has dense singular vectors is better rendered for exact completion, since each one among the observed entries carries information associated with all the $d_M$ parameters that fully describe it. To this end, a number of conditions, which evaluate the suitability of the singular vectors, have been established. The simplest one is given next \cite{candes_exact_mc_2009}:
\beqa \label{ch10:coher}
\norm{\bm{u}_i}_{\infty}\le \sqrt{\frac{\mu_B}{l_1}},~\norm{\bm{v}_i}_{\infty} \le \sqrt{\frac{\mu_B}{l_2}},~i=1,\ldots,r.
\eeqa
where $\mu_B$ is a bound parameter. In fact, $\mu_B$ is a measure of the coherence\footnote{This is a quantity different than the mutual-coherence already discussed in section \ref{ch10mutcoh}.} of matrix $U$ (and similarly of $V$),(vis-\`{a}-vis the standard basis), defined as follows:
\begin{equation}
\mu(U) \coloneqq \frac{l_1}{r}\max_{1 \le i \le l_1} \norm{\bm{P}_U \bm{e}_i}^2,
\end{equation}
where $\bm{P}_U$ defines the orthogonal projection to subspace $U$ and $\bm{e}_i$ is the $i$th vector of the canonical basis. It is easy to show that $\norm{\bm{P}_U \bm{e}_i}^2 = \norm{U^T \bm{e}_i}^2$. In essence, coherence is an index quantifying the extent to which the singular vectors are correlated with the standard basis, $\bm{e}_i$, $i=1,2,\ldots,l$. The smaller the $\mu_B$ is the less ``spiky'' the singular vectors are likely to be, and the corresponding matrix is better suited for exact completion. Indeed, assuming for simplicity a square matrix $M$, i.e. $l_1=l_2=l$, then if \textit{any one} among the singular vectors is sparse having a single nonzero component only, then, taking into account that $\bm{u}_i^T \bm{u}_i = \bm{v}_i^T \bm{v}_i = 1$, this value will have magnitude equal to one and the bound parameter will take its largest value possible, i.e., $\mu_B=l$. On the other hand, the smaller value that $\mu_B$ can get is $1$, something that occurs when the components of \textit{all} the singular vectors assume the same value (in magnitude). Note that in this case, due to the normalization, this common component value has magnitude $\frac{1}{l}$. Tighter bounds to the matrix coherence result via the more elaborate incoherence property \cite{candes_exact_mc_2009,recht_mc_2011} and the strong incoherence property \cite{candes_tao_mc_2010}. In all cases, the larger the bound parameter is the larger the number of known entries, which is required in order to guarantee uniqueness, becomes.

In section \ref{ch10:aplMC}, the aspects of uniqueness will be discussed in the context of a real life application.
\end{enumerate}

The problem described in \eqref{ch10:rankmin} is of limited practical interest since it is an NP-hard task. Thus, following the Compressed Sensing paradigm, it is replaced by a \textit{convexly} relaxed counterpart of it, i.e.,
\beqa
\min_{\hat{M} \in \Real^{l_1 \times l_2}} & & \norm{\bm{\sigma}_{\hat{M}}}_1 \nonumber \\
\text{s.t.} & &\hat{M}_{i,j}=M_{i,j}, \quad
(i,j) \in \Omega \label{ch10:nuclmin}
\eeqa
where $\norm{\bm{\sigma}_{\hat{M}}}_1$, i.e., the sum of the singular values, is referred to as \textit{nuclear norm} of the matrix $\hat{M}$, often denoted as $\norm{\hat{M}}_*$. The nuclear norm minimization was proposed in \cite{fazel_rank_2001} as a convex approximation of rank minimization, which can be cast as a semidefinite program.

\begin{theorem} \label{ch10:MatrixComplTheorem1}
Let $M$ be a $l_1 \times l_2$ matrix of rank $r$, which is a constant much smaller than $l = \max(l_1,l_2)$, obeying \eqref{ch10:coher}. Suppose that we observe $N$ entries of $M$ with locations sampled uniformly at random. Then there is a positive constant $C$ such that if
\beqa
N \ge C \mu_B^4 l \ln^2 l,
\eeqa
then $M$ is the unique solution to \eqref{ch10:nuclmin} with probability at leat $1-l^{-3}$.
\end{theorem}

There might be an ambiguity on how small the rank should be in order for the corresponding matrix to be characterized as ``low rank''. More rigorously, a matrix is said to be of low rank if $r=\mathcal{O}(1)$, which means that $r$ is a constant with no dependence (not even logarithmic), on $l$. Matrix completion is also possible for more general rank cases where instead of the mild coherence property of \eqref{ch10:coher}, the incoherence and the strong incoherence properties \cite{candes_exact_mc_2009,candes_tao_mc_2010,recht_mc_2011,gross_mc_2011} are mobilized in order to get similar theoretical guaranties.
The detailed exposition of these alternatives is out of the scope of this paper. In fact, Theorem \ref{ch10:MatrixComplTheorem1} embodies  the essence of the matrix completion task: with high probability, nuclear-norm minimization recovers all the entries of a low rank matrix $M$ with no error. More importantly, the number of entries $N$, which the convexly relaxed problem requires, is only by a logarithmic factor larger than the information theoretic limit; recall that the latter is equal to $C_0 l \ln l$. Moreover, similarly to Compressed Sensing, robust matrix completion in the present of noise is also possible as long as the request $\hat{M}_{i,j}=M_{i,j}$ in \eqref{ch10:rankmin} and \eqref{ch10:nuclmin} is replaced by $\norm{\hat{M}_{i,j}-M_{i,j}}_2 \le \epsilon$ \cite{candes_mcnoise_2010}. Furthermore, the notion of matrix completion has also been extended to tensors \cite{gandy_tensor_2011,signoretto_tensor_2011}.

\subsection{Robust PCA}
The developments on matrix completion theory led, more recently, to the formulation and solution of another problem of high significance. To this end, the notation $\norm{M}_1$, i.e., the $\ell_1$ norm of a matrix, is introduced and it is defined as the sum of the absolute values of its entries, i.e., $\norm{M}_1=\sum_{i=1}^{l_1} \sum_{j=1}^{l_2} |M_{i,j}|$. In other words, it acts on the matrix as if this were a long vector. Assume that $M$ is expressed as the sum of a low rank matrix, $L$, and a sparse matrix, $S$, i.e., $M=L+S$. The following convex minimization problem \cite{candes_rpca_2011,wright_rpca_2009,chandrasekaran_2011}, usually referred to as {\it principal component pursuit} (PCP),
\beqa
\min_{\hat{L} \in \Real^{l_1 \times l_2},~\hat{S} \in \Real^{l_1 \times l_2}} & & \norm{\bm{\sigma}_{M}}_1 + \lambda\norm{S}_1 \nonumber \\
\text{s.t.} & &\hat{L}+\hat{S}=M, \label{ch10:rpcamin}
\eeqa
is shown to recover {\it both} $L$ and $S$ according to the following theorem \cite{candes_rpca_2011}:

\begin{theorem}
The PCP recovers both $L$ and $S$ with probability at least $1-cl_1^{-10}$, where $c$ is a constant, provided that:
\begin{enumerate}
\item  the support set $\Omega$ of $S$ is uniformly distributed among all sets of
cardinality $N$,
\item the number, $k$, of nonzero entries of $S$ is relatively small, i.e., $k\le \rho l_1 l_2$, where $\rho$ is a sufficiently small positive constant,
\item $L$ obeys the incoherence property,
\item the regularization parameter, $\lambda$, is constant with value $\lambda=\frac{1}{\sqrt{l_2}}$,
\item $\rank(L) \le C \frac{l_2}{\ln^2 l_1}$, with $C$ being a constant.
\end{enumerate}
\end{theorem}
In other words, based on \textit{all} the entries of a matrix $M$, which is known that is the sum of two unknown matrices $L$ and $S$, with the first one being of low rank matrix and the second being sparse, then PCP recovers exactly, with probability almost 1, both $L$ and $S$, irrespective of how large the magnitude of the entries of $S$ are, provided \textit{that both $r$ and $k$ are sufficiently small}.

The applicability of the previous task is very broad. For example, PCP can be employed in order to find a low rank approximation of $M$. It is well known that the task of low rank approximation is closely related to the dimensionality reduction task, where the columns of $M$ are expressed in terms of the $r$ (principal components) columns of $U$, e.g., Chapter 6, \cite{Theod-10}. However, in contrast to the standard SVD or PCA approach, PCP is robust and insensitive to the presence of outliers, since these are naturally modeled, via the presence of $S$. For this reason, the above task is widely known as {\it robust PCA via nuclear norm minimization}. (More classical PCA techniques are known to be sensitive to outliers and a number of alternative approaches have in the past been proposed towards its robustification, e.g., \cite{Karhunen1995,Xu_robustpca,de_la_torre_robustpca_2003, Hubert_robustPCA2004,Hubert_robustPCA2005}).

When PCP serves as a robust PCA approach, the matrix of interest is $L$ and $S$ accounts for the outliers. However, PCP provides estimates for both $L$ and $S$. As it will be discussed in the next subsection, state-of-the-art applications are well accommodated when the focus of interest is turned into the sparse matrix $S$ itself.

\begin{remarks} \mbox{}
\begin{itemize}
\item Just as $\ell_1$ -minimization is the tightest convex relaxation of the combinatorial $\ell_0$-minimization problem in compressed sensing, the nuclear-norm minimization is the tightest convex relaxation of the NP-hard rank minimization problem; i.e., the nuclear ball $\{M \in \Real^{l_1 \times l_2}: \norm{M}_* \le 1 \}$ is the convex hull of the set of rank-one matrices with
spectral norm bounded by one. Besides the Nuclear norm, other heuristics have also been proposed, such as
the log-det heuristic \cite{fazel_rank_2004} and the max-norm \cite{foygel_2011}.

\item The nuclear norm, as a rank minimization approach, is the generalization of the trace-related cost, which is often used in the control community for the rank minimization of positive semidefinite matrices \cite{mesbahi_rank_1997}. Indeed, when the matrix is symmetric and positive semidefinite, the nuclear norm of $M$ is the sum of the eigenvalues and thus it is equal to the trace of $M$. Such problems arise when, for example, the rank minimization task refers to covariance matrices and positive semidefinite Toeplitz or Hankel matrices (see, e.g., \cite{fazel_rank_2004}).

\item Both matrix completion \eqref{ch10:nuclmin} and PCP \eqref{ch10:rpcamin} can be formulated as semidefinite programs and are solved via mobilizing interior-point methods. However, whenever the size of a matrix becomes large (e.g., $100 \times 100$), these methods are deemed to fail in practice due to excessive power and memory requirements. As a result, there is an increasing interest, which has propelled intensive research efforts, for the development of efficient methods to solve \eqref{ch10:nuclmin}, \eqref{ch10:rpcamin} or related approximations, which scale well with large matrices. Many of these methods revolve around the philosophy of the iterative soft and hard thresholding techniques, as discussed in previous sections. However, in the current low rank approximation setting, it is the singular values of the estimated matrix which are thresholded. As a result, in each iteration, the estimated matrix, after thresholding its singular values, tends to be of lower rank. The thesholding of the singular values is either imposed, such as in the case of the singular value thresholding (SVT) algorithm \cite{cai_svt_mc_2010} or it results as a solution of the regularized versions of \eqref{ch10:nuclmin} and \eqref{ch10:rpcamin} (see, e.g., \cite{toh_proxgrad_mc_2010,chen_mc_2011,lin_augmented_2010,ganesh_fast_rpcpa_2009,yuan_rpca_2009}). Moreover, algorithms inspired by greedy methods such as CoSaMP, have also been proposed \cite{lee_admira_2010,waters_rpca:_2011}.

\item Further developments on robust PCA led to improved versions \cite{ganesh_pcp_2010} allowing for exact recovery even though the number of nonzero entries of $S$ approaches $l_1l_2$ arbitrarily close, provided that the sign pattern of $S$ is random. Furthermore, even full columns are allowed to be corrupted \cite{xu_rpca_2012,mccoy_rpca_2011}. Moreover, fusions of PCP with matrix completion and Compressed Sensing are possible, in the sense that only a subset of the entries of $M$ is available and/or linear measurements of the matrix in a Compressed Sensing fashion can be used instead of matrix entries (see, e.g., \cite{waters_rpca:_2011,wright_compressive_rpca_2012}). Moreover, stable versions of PCP dealing with noise have also been investigated (see, e.g., \cite{zhou_stable_rpca_2010}).

\end{itemize}
\end{remarks}

\subsection{Applications of Matrix Completion and PCP}\label{ch10:aplMC}
The number of applications in which these techniques are involved is ever increasing and their extensive presentation is out of the scope of this paper. Next, some key applications will be selectively discussed since they reveal the potential of these methods and at the same time will assist the reader for a better understanding of the underline notions.
\subsubsection{Matrix Completion}
A typical application, where the matrix completion problem arises, is in the collaborative filtering task \cite{Su_collabfiltering_2009}, which is essential for building up successful recommender systems. Let us consider that a group of individuals provide their ratings concerning products, which they have enjoyed. Then a matrix with ratings can be filled where each row indexes a different individual and the columns index the products. As a popular example take the case where the products are different movies. Inevitably, the associated matrix will be partially filled, since it is not common that all customers have watched all the movies and submit ratings for all of them. Matrix completion comes to provide an answer, potentially in the affirmative, to the following question. Can we predict the ratings that the users would give to films that they have not seen yet? This is the task of a recommender system in order to encourage users to watch movies, which are likely to be of their preference. The exact objective of competition for the famous Netflix prize (\url{http://www.netflixprize.com/}) was the development of such a recommender system.

The aforementioned problem provides a good opportunity to build up our intuition about the matrix completion task. Fist, an individual's preferences or taste on movies are typically governed by a small number of factors, such as the gender, the actors they play in it, the continent of origin, etc. As a result, a matrix fully filled with ratings is expected to be low rank. Moreover, it is clear that each user need to have at least one movie rated in order to have any hope to fill out his/her ratings across all movies. The same is true for each movie. This requirement complies with the second requirement in section \ref{ch10:sec_MC}, concerning uniqueness, i.e., one needs to know at least one entry per row and column. Finally, imagine a single user who rates movies with criteria that are completely different to those used by the rest of the users. He/She could, for example, provide ratings at random or depending on, let us say, the first letter of the movie title. Such a scenario complies with the third point concerning the uniqueness in the matrix completion problem, as previously discussed. Unless all the ratings of the specific user are known, the matrix cannot get fully completed.

In the previous application, the matrix of interest can be characterized as approximately low rank. In other cases, such as in sensor network localization \cite{mao_localization_2007}, the rank of the matrix assumes an exact value. The goal of localization is to assign geographic coordinates to each node in the sensor network based on a square matrix, which contains the pairwise distances between the nodes \cite{biswas_sd_localization_2006,montanari_positioning_2010}. It turns out that this matrix is of very low rank, e.g., two or three, depending on whether the sensors are placed in the 2D or 3D space. As a result, matrix completion is possible using a limited number of distance measurements. The number of distance measurements is reduced either intentionally, in order to save power and/or due to the presence of irregularities and obstacles in the deployment area, which renders the communication among nodes impossible. Other applications of matrix completion includes system identification \cite{liu_mc_si_2010}, recovering structure from motion \cite{chen_sfm_2004} and multi-task learning \cite{argyriou_multi-task_2007}.

\subsubsection{Robust PCA/PCP}

In the collaborative filtering task, robust PCA offers an extra attribute compared to matrix completion, which can be proved very crucial in practice. The users are allowed to even tamper with some of the ratings without affecting the estimation of the low rank matrix. This seems to be the case whenever the rating process involves many individuals in an environment, which is not strictly controlled, since some of them occasionally are expected to provide ratings in an ad-hoc, or even malicious manner.

One of the first applications of PCP was in video surveillance systems \cite{candes_rpca_2011} and the main idea behind it appeared to be popular and extendable to a number of computer vision applications. Take the example of a camera recording a sequence of frames consisting of a merely static background and a foreground with few moving objects, e.g., vehicles and/or individuals. A common task in surveillance video is to extract from the background the foreground, in order,  for example, to detect any activity or to proceed with further processing such as face recognition. Suppose the successive frames are converted to vectors in lexicographic order and then are placed as columns in a matrix $M$. Due to the background, even though this may slightly vary due to changes in illumination, successive columns are expected to be highly correlated. As a result, the background can be modeled as an approximately low rank matrix $L$. On the other hand, the objects on the foreground, appear as ``anomalies'' concerning a fraction of pixels of each frame, i.e., a limited number of entries in each column of $M$. Moreover, due to the motion of the forground objects, the positions of these anomalies are likely to change from one column of $M$ to the next. Therefore, they can be modeled as a sparse matrix $S$. Note that in this application, the matrix of interest is the sparse matrix rather than the low rank one.

\section{Conclusions}

In this paper, we provided an overview of the major theoretical advances as well as the main trends in algorithmic developments in the area of sparsity-aware learning and compressed sensing. Both batch processing and online processing techniques were considered. A case study in the context of time-frequency analysis of signals was also presented. Our intent is to update the review from time to time, since this is a very hot research area with a momentum and speed that is sometimes difficult to follow up.

\clearpage
\pagebreak

\appendix

\section{Appendix}
\label{app:convex.analysis}
The stage of our discussion in this Appendix is the real Euclidean space
$\Real^l$, where $l$ is a positive integer. Although all of the following
arguments hold true even in the case where the $\Real^l$ is substituted by
the much more general Hilbert space setting, we confine ourselves here, for
the sake of simplicity, to the Euclidean space. Henceforth, the space
$\Real^l$ is considered to be equipped with an inner product, which, in the
present context, is denoted by $\innprod{\bm{\theta}_1} {\bm{\theta}_2}$,
$\forall \bm{\theta}_1, \bm{\theta}_2\in \Real^l$. A standard example of
such an inner product is the classical vector/dot one, defined by
$\innprod{\bm{\theta}_1} {\bm{\theta}_2} \coloneqq \bm{\theta}_1^T
\bm{\theta}_2$, $\forall \bm{\theta}_1, \bm{\theta}_2\in \Real^l$, where
the superscript $(\cdot)^T$ stands for vector transposition. Another
example of an inner product for the space $\Real^l$ is the following
\textit{weighted} one; $\innprod{\bm{\theta}_1} {\bm{\theta}_2} \coloneqq
\bm{\theta}_1^T W \bm{\theta}_2$, $\forall \bm{\theta}_1, \bm{\theta}_2\in
\Real^l$, where $W\in\Real^{l\times l}$ is any user-defined positive
definite matrix. In order not to spare the generality of the following
discussion, we let $\innprod{\cdot}{\cdot}$ stand for any user-defined
inner product on the linear space $\Real^l$. Given such an inner product,
the associated norm is induced according to the following rule:
$\norm{\cdot} \coloneqq \sqrt{\innprod{\cdot}{\cdot}}$. Excellent resources
for a deeper study on the extremely rich subject of convex analysis are
\cite{rockafellar.book, rockafellar.wets, Hiriart.Lemarechal.fundamentals,
  Bauschke.combettes.book}.

We start, now, with few notions of fundamental importance to convex
analysis.

\subsection{Closed convex sets and metric projection mappings}

\begin{definition}[Convex set, convex function]\label{def:convexity}
A non-empty subset $C$ of $\Real^l$ is called \textit{convex} if $\forall
\bm{\theta}_1, \bm{\theta}_2\in \Real^l$, and $\forall \lambda\in [0,1]$,
the following holds true: $\lambda \bm{\theta}_1 +
(1-\lambda)\bm{\theta}_2\in C$.

Moreover, a function $\mathcal{L}: \Real^l \rightarrow \Real$ is called
\textit{convex} if $\forall \bm{\theta}_1, \bm{\theta}_2\in \Real^l$, and
$\forall \lambda\in [0,1]$, $\mathcal{L}(\lambda \bm{\theta}_1 +
(1-\lambda) \bm{\theta}_2) \leq \lambda\mathcal{L}(\bm{\theta}_1) +
(1-\lambda)\mathcal{L}(\bm{\theta}_2)$. The function $\mathcal{L}$ is
called \textit{strictly convex} if $\forall \lambda\in (0,1)$ and $\forall
\bm{\theta}_1, \bm{\theta}_2\in \Real^l$, such that $\bm{\theta}_1\neq
\bm{\theta}_2$, we have $\mathcal{L}(\lambda \bm{\theta}_1 + (1-\lambda)
\bm{\theta}_2) < \lambda\mathcal{L}(\bm{\theta}_1) +
(1-\lambda)\mathcal{L}(\bm{\theta}_2)$.
\end{definition}

The \textit{epigraph} of a function $\mathcal{L}$ is defined as
the set
\begin{equation*}
\epi(\mathcal{L}) \coloneqq \bigl\{(\bm{\theta},r)\in \Real^l\times\Real:
\mathcal{L}(\bm{\theta})\leq r\bigr\}.
\end{equation*}
In other words, the epigraph of $\mathcal{L}$ is the set of all points of
$\Real^l\times\Real$ which belong to and lie above the graph of
$\mathcal{L}$. Notice, also, by the definition of convexity, that $\mathcal{L}$
is convex if and only if $\epi(\mathcal{L})$ is convex.

Given a real number $\xi$, the \textit{lower level set of $\mathcal{L}$ at
  height $\xi$} is defined as the set
\begin{equation*}
\lev{\xi}(\mathcal{L}) \coloneqq \bigl\{\bm{\theta}\in\Real^l:
\mathcal{L}(\bm{\theta})\leq \xi\bigr\}.
\end{equation*}
For the geometry behind the previous definitions, see
Fig.~\ref{fig:convex.function}.

\begin{figure}[!htbp]
\centering
\includegraphics{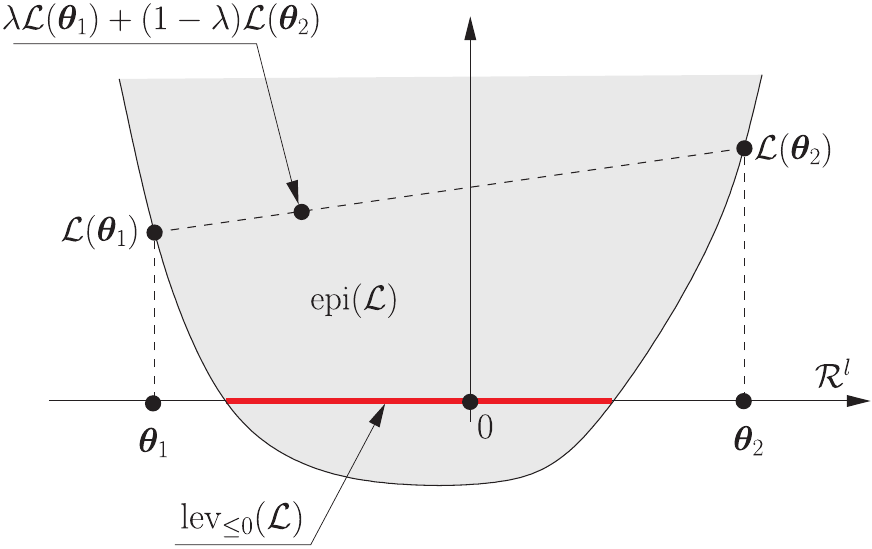}
\caption{A convex function $\mathcal{L}$, its epigraph, and the lower
  level set of $\mathcal{L}$ at height
  $0$.}\label{fig:convex.function}
\end{figure}

%
%

\begin{definition}[The metric projection mapping]\label{def:P_C}
Given a non-empty \textit{closed} convex set $C\subset\Real^l$, the
\textit{metric projection mapping onto $C$} is defined as the operator that
maps to each $\bm{\theta}\in\Real^l$ the \textit{unique}
$P_C(\bm{\theta})\in C$ such that
\begin{equation*}
\norm{\bm{\theta}-P_C(\bm{\theta})}= d(\bm{\theta},C).
\end{equation*}
In other words, the point $P_C(\bm{\theta})$ is the unique minimizer of the
function $\norm{\bm{\theta}-\bm{x}}$, $\bm{x}\in C$. Obviously, in
the case where $\bm{\theta}\in C$, then $P_C(\bm{\theta})= \bm{\theta}$.
\end{definition}

As an example, the metric projection mapping onto the {\it hyperslab}  is given next.

\begin{example}[Hyperslab]\label{ex:hyperslab}

\begin{figure}[!htbp]
\centering
\includegraphics[scale=1]{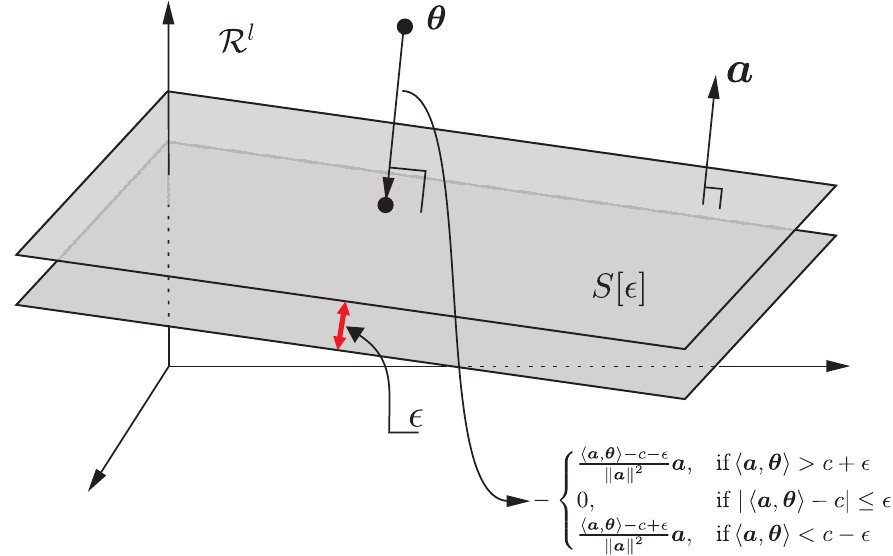}
\caption{A hyperslab and its associated projection
  mapping.}\label{fig:hyperslab}
\end{figure}

A hyperslab $S[\epsilon]$ is the closed convex subset of $\Real^l$
which is defined as
\begin{equation*}
S[\epsilon] \coloneqq \bigl\{\bm{x}\in\Real^l: |\innprod{\bm{a}}{\bm{x}} - c| \le \epsilon \bigr\},
\end{equation*}
for some nonzero $\bm{a}\in\Real^l$ and some $c\in \Real$. The projection
mapping $P_{S[\epsilon]}$ onto $S[\epsilon]$ is given as follows:
\begin{equation}
P_{S[\epsilon]}(\bm{\theta}) = \bm{\theta} -
\begin{cases} \frac{  \innprod{\bm{a}}{\bm{\theta}} - c - \epsilon}
  {\norm{\bm{a}}^2}\bm{a}, & \text{if} \innprod{\bm{a}}{\bm{\theta}} > c + \epsilon
,\\ 0, & \text{if}\ |\innprod{\bm{a}}{\bm{\theta}} - c |\leq  \epsilon
,\\ \frac{\innprod{\bm{a}}{\bm{\theta}} - c + \epsilon}
  {\norm{\bm{a}}^2}\bm{a}, & \text{if} \innprod{\bm{a}}{\bm{\theta}} < c - \epsilon.
\end{cases}
\label{project.hyperslab}
\end{equation}

For the related geometry, see Fig.~\ref{fig:hyperslab}. Notice that
$\bm{a}$ stands for the \textit{normal vector} defining the hyperplanes associated with the hyperslab.
\end{example}

\subsection{The subgradient}
\begin{figure}[!htbp]
\centering
\includegraphics{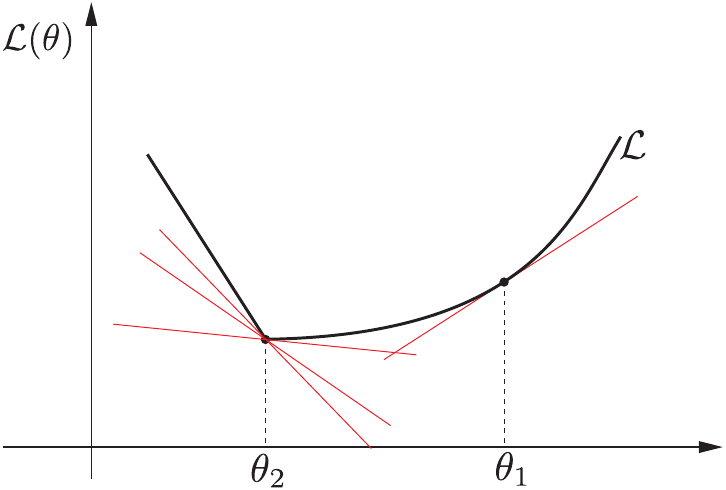}
\caption{The graph of $\mathcal{L}$ and supporting
  hyperplanes generated by the subgradients at points, $\theta_1$, $\theta_2$, where the function is differentiable and non-differentiable, respectively.}\label{fig:subgradient}
\end{figure}
\begin{definition}[Subgradient, Subdifferential]\label{def:subgradient} Given a convex function
  $\mathcal{L}$, defined on $\Real^l$, and a point $\bm{\theta}\in\Real^l$,
  the subgradient of $\mathcal{L}$ at $\bm{\theta}$ is defined as any vector, $\bm{h}$, such that
  \begin{equation}
  \innprod{\bm{h}}{\bm{x}- \bm{\theta}} + \mathcal{L}(\bm{\theta}) \le \mathcal{L}(\bm{x}), \forall \bm{x} \in \Real^l.
  \end{equation}
  If the function $\mathcal{L}$ is differentiable at $\bm{\theta}$ then the subgradient coincides with the (unique) gradient. As it is the case for the gradient, a subgradient defines a hyperplane. This hyperplane ``supports'' the epigraph of $\mathcal{L}$; that is, the epigraph is on the one side of this hyperplane (see Fig. \ref{fig:subgradient}). At $\theta_1$, the convex function is differentiable and there is only one subgradient, which coincides with the gradient. Thus at this point, there is a simple hyperplane that supports the epigraph. At $\theta_2$, the function is not differentiable. Hence there is an infinity of subgradients that define hyperplanes that support the epigraph. The set of all subgradients at a point $\bm{\theta}$ is known as the subdifferential and is denoted as $\partial \mathcal{L}$, i.e.,
  \begin{equation*}
\partial\mathcal{L}(\bm{\theta}) \coloneqq \bigl\{\bm{h}\in\Real^l:
\innprod{\bm{h}}{\bm{x}-\bm{\theta}} +
\mathcal{L}(\bm{x}) \leq \mathcal{L}(\bm{x}), \forall
\bm{x}\in\Real^l \bigr\}.
\end{equation*}
\end{definition}

\begin{figure}[!htbp]
\centering
\includegraphics{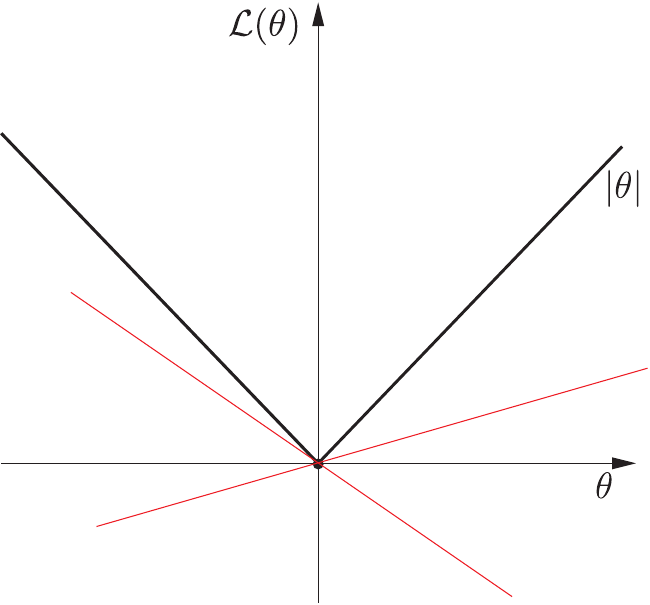}
\caption{The graph of the function $|\cdot|$, and the supporting
  hyperplanes generated by the subgradients of $|\cdot|$ at
  $\theta=0$.}\label{fig:l1.cost}
\end{figure}
Next, the subdifferential of $\mathcal{L}(\theta) \coloneqq
|\theta|$, $\theta\in\Real$ is given:
\begin{equation*}
\partial\mathcal{L}(\theta) = \begin{cases}
[-1,1], & \text{if}\ \theta=0,\\
\sign(\theta), & \text{if}\ \theta\neq 0,
\end{cases}
\end{equation*}
where $\sign(\cdot)$ stands for the sign of a real number. For the
geometry associated to this cost function see Fig.~\ref{fig:l1.cost}.

\clearpage
\fancyhead[LO]{\myfonts\rightmark}

\bibliographystyle{kostas_style}
\bibliography{ereferencebib}
\end{document}